\documentclass[useAMS,usenatbib]{mn2e} \usepackage{graphicx}

\title[Excited-state OH maser polarimetry]{Excited-state hydroxyl maser polarimetry:\\ Who ate all the $\pi$s?} %C2471

\author[Green et al.]
       {J. A. Green\thanks{E-mail: j.green@skatelescope.org}$^{1,2}$, J. L. Caswell$^{2}$\thanks{Deceased 2015 January 14.} and N. M. McClure-Griffiths$^{2,3}$\\
$^{1}$SKA Organisation, Jodrell Bank Observatory, Lower Withington, Macclesfield, SK11 9DL, UK\\
$^{2}$CSIRO Astronomy and Space Science, Australia Telescope National Facility, PO Box 76, Epping, NSW 1710, Australia\\
$^{3}$Research School of Astronomy \& Astrophysics, The Australian National University, Canberra ACT 2611 Australia\\
}
\date{Accepted XXXX . Received XXX; in original form XXXX}

\pagerange{\pageref{firstpage}--\pageref{lastpage}} \pubyear{XXXX}
\begin{document} \maketitle

\label{firstpage}

\begin{abstract}
We present polarimetric maser observations with the Australia Telescope Compact Array (ATCA) of excited-state hydroxyl (OH) masers. We observed 30 fields of OH masers in full Stokes polarization with the Compact Array Broadband Backend (CABB) at both the 6030 and 6035 MHz excited-state OH transitions, and the 6668-MHz methanol maser transition, detecting 70 sites of maser emission. Amongst the OH we found 112 Zeeman pairs, of which 18 exhibited candidate $\pi$ components. This is the largest single full polarimetric study of multiple sites of star formation for these frequencies, and the rate of 16\% $\pi$ components clearly indicates the $\pi$ component exists, and is comparable to the percentage recently found for ground-state transitions. This significant percentage of $\pi$ components, with consistent proportions at both ground- and excited-state transitions, argues against Faraday rotation suppressing the $\pi$ component emission. Our simultaneous observations of methanol found the expected low level of polarisation, with no circular detected, and linear only found at the $\le$10\% level for the brightest sources.
\end{abstract}


\section{Introduction}
Of the various masers found in high mass star forming regions, those of the hydroxyl (OH) molecule uniquely show highly polarized emission. It is well-established that the polarized properties of hydroxyl maser features arise from Zeeman splitting in magnetic fields of a few milliGauss. In the simple case, the presence of a magnetic field causes spectral line radiation to split into three frequencies, one with a positive frequency shift, one with a negative frequency shift, and one which remains at the original frequency. The two shifted components are denoted as $\sigma^{\pm}$ and the unshifted as $\pi$. The radiation generated from a $\pi$ component transition has an electric field which oscillates along the magnetic field axis, whilst the $\sigma$ components have electric fields which rotate perpendicular to the magnetic field axis. The orientation of the emission relative to the magnetic field dictates the pattern and nature of polarisation observed: if the radiation is seen exactly parallel to the magnetic field, only the $\sigma^{\pm}$ components are seen, circularly polarised, and not the linear $\pi$; if the radiation is seen exactly perpendicular to the magnetic field, all three components are seen, linearly polarised.  

The typical pairing of opposite circularly polarized $\sigma$ components provide a powerful means of mapping the magnetic fields in these regions \citep[e.g.][]{wright04b} and the ensemble of field measurements across several hundred star forming regions have the potential to delineate the magnetic field pattern in the spiral arms of our Galaxy \citep[e.g.][]{davies74,reid90,han07,green12magmo0}.

Zeeman patterns are often recognisable from single dish spectra \citep[e.g.][]{szymczak09}, and many identified this way have been validated by Very Long Baseline Interferometry (VLBI) measurements (with confirmation that either the postulated components of Zeeman splitting precisely coincide or at least trace the same field, with local conditions favouring one $\sigma$ component or the other). Investigations to date have primarily involved the OH transitions at 1665 and 1667 MHz  (the main-line transitions of the ground state), where the expected Zeeman pattern is especially simple: two (mainly) circularly polarised $\sigma$ components, straddling the unshifted linearly polarized $\pi$ component. While there is no doubt that the Zeeman interpretation is correct, there have been only a handful of features whose frequency and linear polarization may identify them as $\pi$ components \citep[see for example][]{gray03a,asanok10}. However, a more recent study of the ground-state OH transitions showed a small but significant percentage of sources with $\pi$ candidates \citep{caswell13}. This puzzling under-representation of linearly polarized $\pi$ components is an unsolved puzzle, with several suggested explanations including differential Faraday rotation within the emitting region destroying the $\pi$ component and a more general explanation not overtly dependent on Faraday rotation, but based on semi-quantitative competitive gain arguments \citep{gray95}.  Both explanations also lead to a reduced linearly polarized fraction in the $\sigma$ components. 

Transitions of the excited-state of OH near 6 GHz (with main lines at 6035 and 6030 MHz, hereafter referred to as `exOH') have similarly simple Zeeman patterns, and pairs of $\sigma$ components are often observed as expected, even in single dish observations.  For the large portion of Galactic plane visible from the southern sky, spectra of right and left hand circular polarization are available for more than 100 sources \citep[e.g.][]{caswell95c,caswell03}, but linear polarization measurements are almost non-existent. Full Stokes observations have rarely been made at this higher frequency, and so it is not clear whether the $\pi$ components are as rare as in the lower frequency ground-state transitions. Here we present new, full polarisation, observations to test if the linearly polarized $\pi$ component of the excited-state masers is as scarce as in the ground-state masers. We discuss the observations and detections in Section 2; then give a summary of the polarised properties and discuss Zeeman pairs and triplets and their implications for magnetic field direction in Section 3.

\section{Observations}\label{obs_section}
We targeted 30 fields of known exOH masers associated with high-mass star formation \citep{caswell03} with the Australia Telescope Compact Array (ATCA). Using the Compact Array Broadband Backend (CABB, \citealt{wilson11}), we simultaneously observed both exOH transitions at 6030.7485 MHz and 6035.0932 MHz, and the 6668.518 MHz methanol maser transition. CABB was configured to the CFB 1M$-$0.5 mode \citep{wilson11}, with six concatenated 1-MHz `zoom' bands centred at each of the exOH transitions (providing $\sim$170\,km\,s$^{-1}$ velocity coverage), and eight concatenated 1-MHz `zoom' bands at the methanol frequency (providing $\sim$200\,km\,s$^{-1}$ velocity coverage). The velocity channel separations were 0.024\,km\,s$^{-1}$ and 0.022\,km\,s$^{-1}$ for the exOH and methanol transitions, respectively. The observations were obtained 2011 August 29 to 2011 September 02 using the extended 6-km array configuration \citep[6B,][]{frater92}. Observations of each of the 30 target fields were made with an average of 50 minutes total integration time, split across five cuts, interspersed with calibrators, and spread across a 12 hour period of local sidereal time to enable good coverage of the UV plane. 

The primary (flux and bandpass) calibrator was 0823--500 and the five phase calibrators used were (dependent on source position): 1059--63, 1421--490, 1613--586, 1729--37 and 1730--130. The data were reduced and processed with the {\sc miriad} software package using standard techniques \citep{sault04}. The typical RMS noise was $\le$50mJy for the exOH transitions and $\le$150mJy for the methanol (with the higher noise for methanol due to the roll-off in performance of the feed system at the higher frequency). For the purposes of source detection the data were smoothed in velocity to channel widths of 0.1\,km\,s$^{-1}$, resulting in an rms noise of $\sim$20mJy. The synthesized beam has a Full Width at Half-Maximum (FWHM) in right ascension of $\sim$1.5 arcsec, and slightly larger in declination by a factor proportional to the cosecant of the declination of the targeted source. The rms position errors of the derived absolute positions are estimated to be $\sim$0.4 arcsec in each coordinate. This estimate is based on the phase calibrator being typically offset by 10$^{\circ}$ from the beam centre of the target pointing, and typical signal-to-noise ratios \citep[see][for further details]{caswell98}. Detections were made based on the presence of positionally coincident emission ($>$3$\sigma$) in three or more channels.  

\subsection{Detections}
From the 30 pointings we detected 70 sites of emission (Table \ref{resotable}), including 14 associated with Sgr B2 (see Section \ref{sgrb2section}). 
Of the 70 sites, any which had multiple spatially separate components, but corresponding Galactic longitudes and latitudes which were identical to three decimal places (i.e. positional separations of $>$0.4 arcsec, but  $<$3.6 arcsec), were given letters after their names to distinguish the components (e.g. 309.921+0.479a, 309.921+0.479b, etc). 7 of the 70 sites have these multiple components, bringing the total number of masers listed in Table \ref{resotable} to 84.
We include notes on all sources in Appendix \ref{appendixnotes}. 
%Of the 70 sites, any which had multiple spatially separate components (positional separations of $>$0.4 arcsec), but Galactic longitudes and latitudes which were identical to three decimal places (i.e. separations of $<$3.6 arc sec), and therefore had identical source names, where given letters after their names (e.g. 309.921+0.479a, 309.921+0.479b, etc). 7 of the 70 sites have these multiple components.
An example of the full spectra, with Stokes I, Q, and U together with right-hand circular (RHC) polarisation, left hand circular (LHC) polarisation and linear (LIN) polarisation is shown in Figure \ref{examplespectra}.
Spectra are shown in this manner for every site in Appendix B (online only). This includes the individual lettered components, as notable differences are seen in the spectra for the different spatial components (as clearly evidenced for example by the four components of 323.459--0.079). 

\begin{figure*}
\begin{center}
\renewcommand{\baselinestretch}{1.1}
\includegraphics[width=17cm]{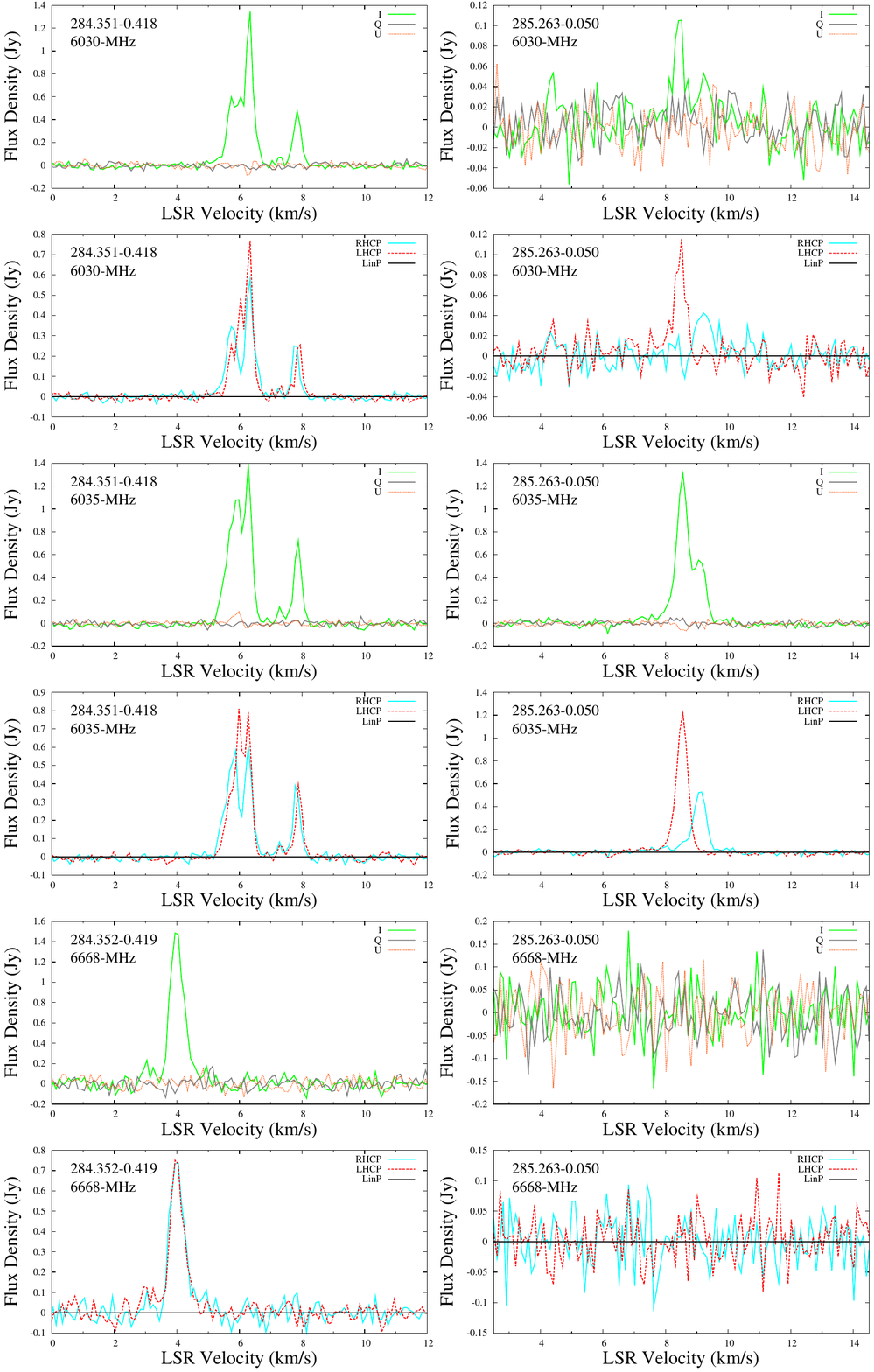}  
\caption{\small Example of the full spectra of detections, showing (from top to bottom) 6030-MHz OH, 6035-MHz OH and 6668-MHz methanol emission in both Stokes I, Q and U and RHC, LHC and LIN. Spectra are shown in this manner for every site in Appendix B (online only).}
\label{examplespectra}
\end{center}
\end{figure*}

\begin{figure*}
\begin{center}
\renewcommand{\baselinestretch}{1.1}
\includegraphics[width=17cm]{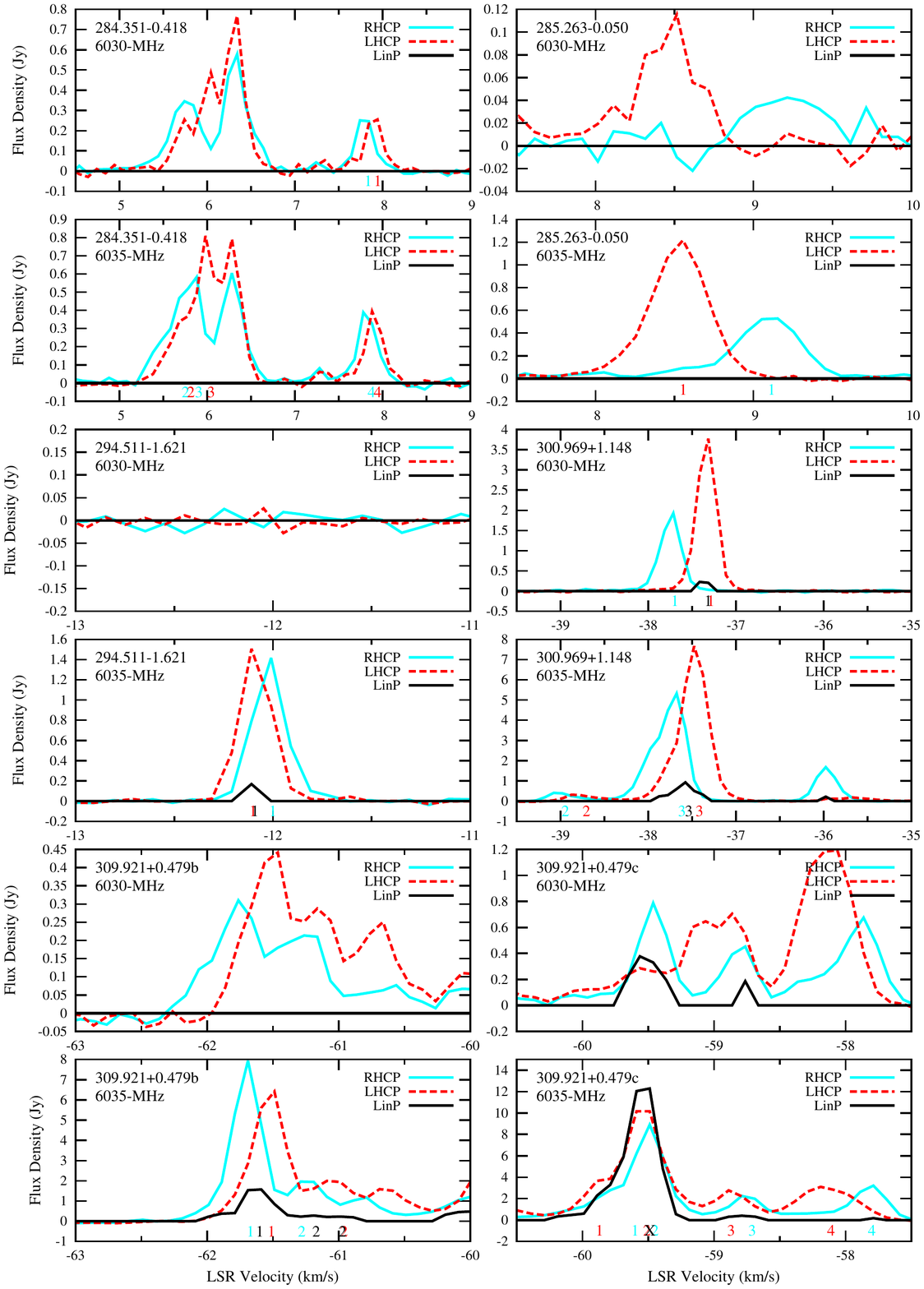}  
\caption{\small Example of the spectra of Zeeman pattern features, showing (from top to bottom) 6030-MHz OH and 6035-MHz OH, with identified patterns labelled with numbers corresponding to those listed in Table \ref{fullresultstable}. Only the velocity range relevant to the Zeeman pattern is displayed. Spectra are shown in this manner for every site in Appendix B (online only).}
\label{examplenewspectra}
\end{center}
\end{figure*}

\subsection{Zeeman pair identification}\label{zeeman_ident}
As discussed in the introduction, although VLBI measurements are desirable to determine Zeeman pairs with the highest confidence, many can be reliably identified from lower resolution observations \citep[see for example the discussion by][]{caswell13}. In the current data we initially identified Zeeman pairs with the RHC and LHC polarised components coincident to within 0.4 arcsec of the positions of fitted two-dimensional Gaussians to individual channel maps (and the LIN components for Zeeman triplets). We adopt the IEEE convention for polarisation handedness and define Stokes $V$ according to the IAU convention, RHC minus LHC. 

Zeeman pairs and triplets were then categorised based on a comparison of the least-squares ($\chi$$^{2}$) Gaussian fits of the spectral components. These standard Gaussian fits were made for peak velocity, FWHM and peak amplitude. The first two were used for comparison, but the amplitude was not, as it can vary depending on the local conditions for the maser spots, as noted in the earlier referenced VLBI studies. Pairs and triplets with fitted peak velocities separated by more than the individual errors of the components, typically less than or equal to the error on the velocity from the channelisation (half a channel width $\sim$0.01\,km\,s$^{-1}$), and which have comparable least-squares fitted FHWMs (within the errors, typically 2--3\% of the FWHM), were designated `A'. The pairs and triplets which may be spatially offset, opposite polarised, single features blended within the ATCA position accuracy, or those that are spectrally blended, were designated `B'. 
The results of these identifications, the polarised components of the Zeeman patterns of the exOH emission, are listed in Table \ref{fullresultstable}. An example of the spectra of identified Zeeman pattern features is given in Figure \ref{examplenewspectra}, with the complete set shown in Appendix B (online only)). 

An example of an `A' class Zeeman pair is shown in Figure \ref{examplenewspectra} for the source 284.351--0.418, at the 6030-MHz transition, where the RHC component has a fitted peak velocity of 7.79$\pm$0.01\,km\,s$^{-1}$  and the LHC component has a fitted peak velocity of 7.90$\pm$0.01\,km\,s$^{-1}$. An example of an `A' class Zeeman triplet is also shown in Figure \ref{examplenewspectra} for the source 300.969$+$1.148, at the 6035-MHz transition, where the RHC component has a fitted peak velocity of  --37.66$\pm$0.01\,km\,s$^{-1}$, the LIN component has a fitted peak velocity of --37.58$\pm$0.01\,km\,s$^{-1}$ and the LHC component has a fitted peak velocity of --37.46$\pm$0.01\,km\,s$^{-1}$. An example of a `B' class Zeeman pair is also shown in Figure \ref{examplenewspectra}, for the source 284.351--0.418 at the 6035-MHz transition, where the RHC component has a fitted peak velocity of 5.71$\pm$0.05\,km\,s$^{-1}$  and the LHC component has a fitted peak velocity of 5.77$\pm$0.05\,km\,s$^{-1}$.

The allocation of a LIN component to a $\sigma$, a $\pi$ or neither (denoted with an `X') takes into account several factors: the error of the fits to the peak velocity of the LIN and the RHC/LHC components; the FWHMs of the fitted components; the combined errors on the estimated unshifted velocity (the quadrature addition of the RHC and LHC errors); and a visual inspection. Ideally the LIN component is unambiguously associated with either a $\sigma$ component, or the unshifted $\pi$ velocity, but often requires interpretation. As examples, the LIN component at 6030-MHz for 300.969$+$1.148, is allocated to the $\sigma$ component, as it is within the error of the LHC feature, whilst the LIN component at 6035-MHz for the same source is allocated to the $\pi$ component, as it is outside the errors of the LHC and RHC features and within the error of the mid-point (unshifted velocity). An example of an unallocated LIN component (an `X') is 309.921+0.479c at the 6035 MHz transition, where the bright LIN feature is within the errors of both the LHC and RHC.

\subsection{Sgr B2}\label{sgrb2section} 
\begin{figure}
\begin{center}
\renewcommand{\baselinestretch}{1.1}
\includegraphics[width=7.5cm]{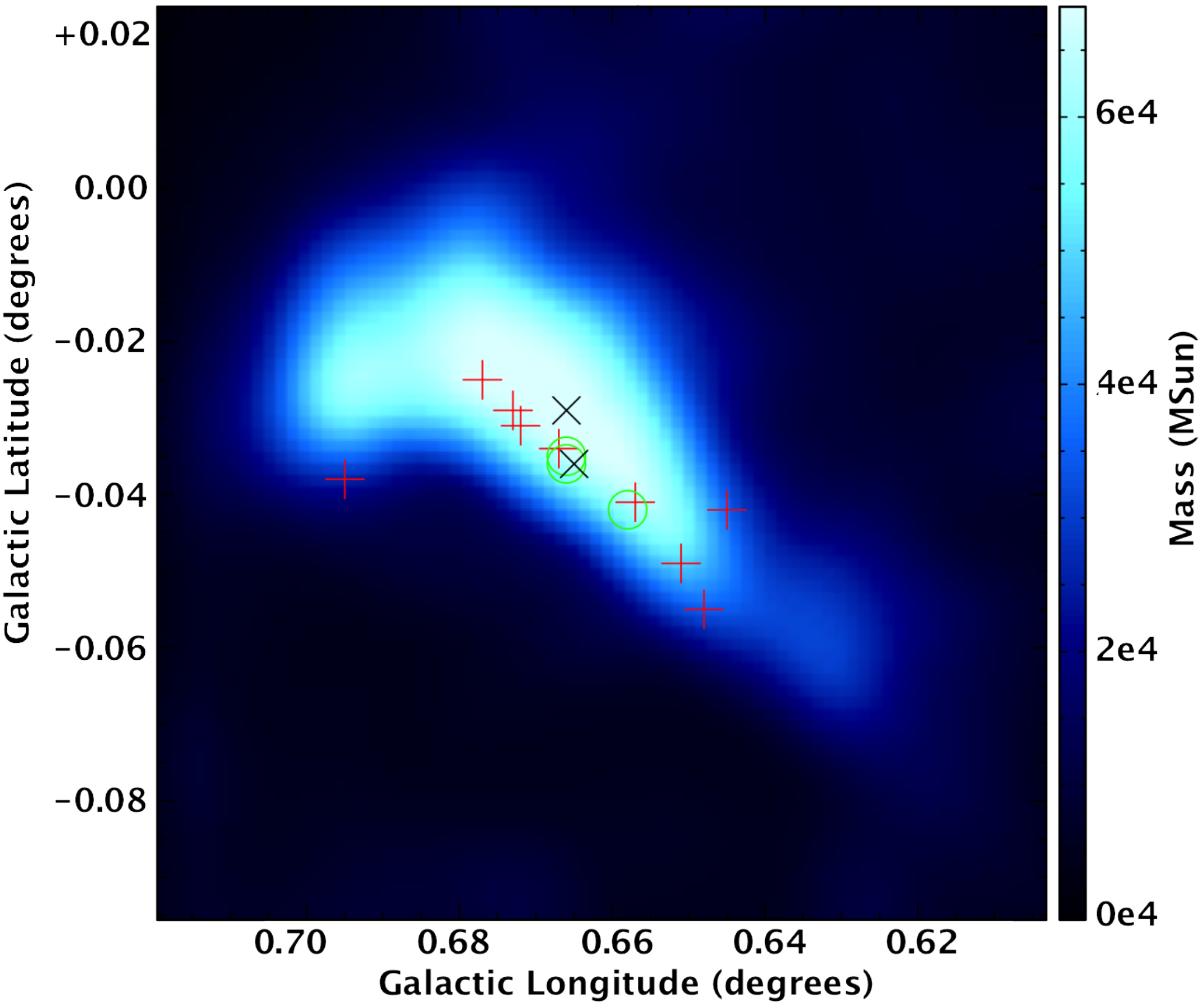}  
\caption{\small The sites of methanol and excited-state hydroxyl maser emission in Sgr-B2 overlaid on the estimate of the enclosed mass from \citet{longmore13}. Red plus is methanol only, green circles excited-state OH only, and black crosses methanol and excited-state OH. With the exception of two sites, the maser emission aligns with the peak in enclosed mass.}
\label{sgrb2fig}
\end{center}
\end{figure}

Here we make special note of the Sgr B2 complex (sources 0.645--0.042 through to 0.695--0.038), where we differentiate 14 sites of emission for this complex (Figure \ref{sgrb2fig}), including the 11 methanol sites listed by \citet{houghton95}. Three sites exhibit 6035-MHz emission, one of which is a new detection (000.665--0.036). Only one site exhibits 6030-MHz emission, 000.666--0.029, which is also the site of the strongest 6035-MHz emission (which has existing spectra in  \citealt{caswell95c} and  \citealt{caswell03}). Two of the three sites have LIN features and Zeeman pairs are also present, including an apparent shift in 6030-MHz emission at 000.666--0.029. With the exception of two of the sites, the Sgr B2 masers show a linear morphology across the region, aligned with the peak of enclosed mass estimated by \citet{longmore13} (Figure \ref{sgrb2fig}). This ridge of high-mass star formation corresponds with the leading edge of the complex in the orbital model of \citet{molinari11}, in the transition zone between $x_{1}$ and $x_{2}$ orbits within the Galactic bar (mimicking the high-mass star formation behaviour indicated by maser emission between the transition of the bar and the 3-kpc ring/arms, \citealt{green11a}). \\

\begin{table*} 
\centering 
\caption{\small Maser detections ($\ge$5$\sigma$ in two or more adjacent channels) as measured with the ATCA. Positions are given for the strongest Stokes I features at each transition, plus any additional features that are offset by more than 0.4$''$ (the positional accuracy of the observations). Previous source references are: HW95 = \citet{houghton95}; C95 = \citet{caswell95c}; C97 = \citet{caswell97}; C03 = \citet{caswell03}; C09 = \citet{caswell09a}; C11 = \citet{caswell11vlbi2}. 
The peak flux density in the fifth column is followed by a letter denoting the polarisation percentage of at least one emission feature: `c' $>$ 0\%, $<$ 50\% circularly polarised; `C' $>$ 50\% circularly polarised; `l' $>$ 0\%, $<$ 50\% linearly polarised; `L' $>$ 50\% linearly polarised. If no letter is present it is considered unpolarized. $^{1}$Primary beam correction to be applied to listed flux density to account for offset position of source from primary beam centre. Errors in velocities are dependent on signal to noise, but as the velocities are taken as the midpoint of channels, the maximum error is $\pm$0.5 channels (corresponding to $\sim$0.05\,km\,s$^{-1}$ for all three transitions).} 
\begin{tabular}{lcrcrlrrrl}
\hline
\multicolumn{1}{c}{Source Name} & \multicolumn{2}{c}{Equatorial Coordinates} & \multicolumn{1}{c}{Maser} & \multicolumn{1}{c}{Peak} && \multicolumn{1}{c}{Peak} & \multicolumn{2}{c}{Velocity} & \multicolumn{1}{l}{Previous}\\
\multicolumn{1}{c}{(~~~$l$,~~~~~~~$b$~~~)}    &       RA(J2000)        &       Dec(J2000)       & Transition & \multicolumn{1}{c}{Flux} &($\times$pbc$^{1}$)& \multicolumn{1}{c}{Velocity}   & \multicolumn{2}{c}{Range}      &Reference \\
\multicolumn{1}{c}{(~~~$^\circ$~~~~~~~$^\circ$~~~)} & (h~~m~~~s) & (~$^\circ$~~ $'$~~~~$''$) & \multicolumn{1}{c}{(MHz)} &  \multicolumn{1}{c}{(Jy)} && \multicolumn{1}{c}{(km\,s$^{-1}$ )}&   \multicolumn{2}{c}{(km\,s$^{-1}$ )}  & \\
\hline
284.351$-$0.418 & 10 24 10.69 &  $-$57 52 33.8 & 6030 & 1.2~C~~ & ($\times$1.00) & 6.3 & 5.0 & 8.5 &C95,C03 \\
 			 & 10 24 10.69 &  $-$57 52 33.9 & 6035 & 1.3~C~~ &($\times$1.00) & 6.3 & 5.0 & 8.5 &C95,C97,C03\\
284.352$-$0.419 & 10 24 10.89 &  $-$57 52 39.0 & 6668 & 1.4~~~~~ & ($\times$1.00)& 4.0 & 3.0 & 5.0 &C95,C97,C09\\
285.263$-$0.050 & 10 31 29.88 &  $-$58 02 18.3 & 6030 & 0.2~C~~ &($\times$1.00) & 8.4 & 8.0 & 9.0 &C03\\
 			 & 10 31 29.87 &  $-$58 02 18.5 & 6035 & 1.2~C~~ &($\times$1.00) & 8.6 & 7.5 & 10.0 &C95,C97,C03 \\
285.337$-$0.002 & 10 32 09.64 &  $-$58 02 05.0 & 6668 & 2.0~~~~~ & ($\times$3.85)& 0.7 & 0.0 & 2.0 &C09 \\
294.511$-$1.621 & 11 35 32.22 &  $-$63 14 43.2 & 6035 & 2.2~Cl~ &($\times$1.00) & $-$12.1 & $-$12.5 & $-$11.5 &C95,C97,C03 \\
 			 & 11 35 32.23 &  $-$63 14 42.8 & 6668 & 6.9~~~~~ & ($\times$1.00)& $-$9.9 & $-$14.0 & $-$4.0 &C97,C09 \\ 			 
300.969+1.148 & 12 34 53.22 &  $-$61 39 40.0 & 6030 & 3.6~Cl~ &($\times$1.00) & $-$37.3 & $-$38.5 & $-$36.5 &C95,C03\\
 			 & 12 34 53.22 &  $-$61 39 40.0 & 6035 & 8.8~Cl~ &($\times$1.00) & $-$37.6 & $-$39.5 & $-$35.0 &C95,C97,C03\\
 			 & 12 34 53.22 &  $-$61 39 40.0 & 6668 & 3.5~~~~~ &($\times$1.00) & $-$37.1 & $-$40.0 & $-$35.0&C97,C09 \\
309.921+0.479a & 13 50 41.78 &  $-$61 35 10.8 & 6668 & 62.8~~~l~ & ($\times$1.00)& $-$63.0 & $-$64.0 & $-$57.0 &C97,C09\\		
309.921+0.479b & 13 50 41.75 &  $-$61 35 10.4 & 6030 & 0.6~C~~ &($\times$1.00) & $-$61.5 & $-$62.5 & $-$57.0 &C95,C03\\
 			   & 13 50 41.75 &  $-$61 35 10.4 & 6035 & 10.1~Cl~  &($\times$1.00) & $-$61.6 & $-$62.5 & $-$56.0 &C95,C97,C03\\
 			   & 13 50 41.74 &  $-$61 35 10.6 & 6668 & 157.7~~~l~ & ($\times$1.00)& $-$62.4 & $-$64.0 & $-$57.0 &C97,C09\\			 
309.921+0.479c & 13 50 41.81 &  $-$61 35 10.0 & 6030 & 1.4~Cl~  &($\times$1.00) & $-$58.1 & $-$62.5 & $-$57.0 &C95,C03\\
 			   & 13 50 41.79 &  $-$61 35 10.1 & 6035 & 18.2~CL  &($\times$1.00) & $-$59.5 & $-$62.5 & $-$56.0 &C95,C97,C03\\
 			   & 13 50 41.78 &  $-$61 35 10.1 & 6668 & 669.4~~~l~ & ($\times$1.00)& $-$59.8 & $-$64.0 & $-$57.0 &C97,C09\\
311.596$-$0.398 & 14 06 18.34 &  $-$62 00 15.1 & 6035 & 1.3~CL &($\times$1.00) & 29.7 & 29.0 & 32.5 &C97,C03\\
311.643$-$0.380  & 14 06 38.74 &  $-$61 58 23.0 & 6035 & 0.5~C~~ & ($\times$1.47) & 33.8 & 33.0 & 34.5 &C95,C97,C03 \\
			     & 14 06 38.78 &  $-$61 58 23.2 & 6668 & 5.6~~~~~ & ($\times$1.47) & 32.5 & 33.0 & 34.5 &C97,C09 \\		     
323.459$-$0.079a & 15 29 19.36 &  $-$56 31 21.1 & 6030 & 0.5~C~~ &($\times$1.00) & $-$70.3 & $-$71.0 & $-$69.5 &C95,C03\\
 			       & 15 29 19.36 &  $-$56 31 21.1 & 6035 & 17.0~Cl~  & ($\times$1.00)& $-$70.3 & $-$71.0 & $-$69.5&C95,C97,C03 \\
323.459$-$0.079b & 15 29 19.33 &  $-$56 31 21.8 & 6030 & 0.4~C~~ & ($\times$1.00)& $-$68.4 & $-$69.0 & $-$65.0 &C95,C03\\
                                   & 15 29 19.35 &  $-$56 31 22.0 & 6035 & 3.6~Cl~ &($\times$1.00) & $-$68.5 & $-$69.0 & $-$65.0 &C95,C97,C03 \\
 			       & 15 29 19.33 &  $-$56 31 21.8 & 6668 & 1.3~~~~~ &($\times$1.00) & $-$68.1 & $-$68.5 & $-$66.0 &C97,C09\\
323.459$-$0.079c & 15 29 19.37 &  $-$56 31 22.2 & 6030 & 0.2~C~~  & ($\times$1.00)& $-$66.6& $-$69.0 & $-$65.0 &C95,C03\\
323.459$-$0.079d & 15 29 19.36 &  $-$56 31 22.7 & 6668 & 15.1~~~~~ & ($\times$1.00)& $-$67.0 & $-$68.5 & $-$66.0  &C97,C09\\
328.808+0.633a & 15 55 48.38 &  $-$52 43 06.7 & 6030 & 3.3~Cl~ & ($\times$1.00)& $-$46.1 & $-$47.0 & $-$45.5 &C95,C97,C03\\
 			   & 15 55 48.38 &  $-$52 43 06.6 & 6035 & 11.2~Cl~ & ($\times$1.00)& $-$46.1 & $-$47.5 & $-$45.0 &C95,C97,C03\\
 			   & 15 55 48.38 &  $-$52 43 06.5 & 6668 & 229.7~~~l~ &($\times$1.00) & $-$44.5 & $-$48.0 & $-$42.5 &C97,C09\\	   
328.808+0.633b & 15 55 48.52 & $-$52 43 06.1 &6035&4.7~C~~ &($\times$1.00) &$-$43.5 & $-$45.0 & $-$42.0&C95,C97,C03\\
                               & 15 55 48.49 & $-$52 43 06.5 & 6668 & 43.5~~~~~ &($\times$1.00)& $-$45.3&$-$47.0&$-$42.0&C97,C09\\
328.809+0.633    & 15 55 48.74 & $-$52 43 05.5 & 6668 & 36.2~~~l~ &($\times$1.00)& $-$44.2&$-$45.0&$-$43.0&C97,C09\\
329.339+0.148 & 16 00 33.12 &  $-$52 44 40.1 & 6035 & 0.2~C~~ &($\times$1.00) & $-$105.0 & $-$108.0 & $-$103.0 &C03\\
 			 & 16 00 33.12 &  $-$52 44 40.1 & 6668 & 17.0~~~l~ & ($\times$1.00)& $-$106.4 & $-$108.5 & $-$105.0 &C09\\		 
330.953$-$0.182 & 16 09 52.36 &  $-$51 54 57.6 & 6035 & 1.6~Cl~ &($\times$1.00) & $-$87.9 & $-$90.5 & $-$86.0 &C95,C03\\
 			     & 16 09 52.36 &  $-$51 54 57.7 & 6668 & 5.0~~~~~ &($\times$1.00) & $-$87.6 & $-$90.5 & $-$87.0 &C09\\			 
331.442$-$0.187 & 16 12 12.45 &  $-$51 35 10.3 & 6668 & 5.7~~~~~ & ($\times$16.67) & $-$88.6 & $-$93.0 & $-$84.0&C09 \\
331.496$-$0.078 & 16 11 59.13 & $-$51 28 13.2 & 6668 & 0.7~~~~~ &($\times$1.15) & $-$93.4 & $-$95.0 & $-$92.0 & $<$new site$>$ \\
331.511$-$0.102 & 16 12 09.71 &  $-$51 28 38.3 & 6030 & 0.6~C~~ & ($\times$1.00)& $-$89.1 & $-$91.0 & $-$88.5 &C95,C03 \\
 			     & 16 12 09.70 &  $-$51 28 38.2 & 6035 & 1.6~C~~ &($\times$1.00) & $-$89.1 & $-$92.0 & $-$88.5 &C95,C03\\
331.512$-$0.102 & 16 12 09.91 &  $-$51 28 37.2 & 6030 & 0.2~C~~ & ($\times$1.00)& $-$86.9 & $-$87.5 & $-$86.5 &C95,C03\\
                                & 16 12 09.91 &  $-$51 28 37.3 & 6035 & 0.7~C~~ & ($\times$1.00)& $-$87.0 & $-$87.5 & $-$86.5 &C95,C03\\
331.542$-$0.067 & 16 12 09.00 &  $-$51 25 48.0 & 6030 & 2.1~Cl~ & ($\times$1.00)& $-$85.9 & $-$86.5 & $-$85.0 &C95,C03\\
 			    & 16 12 09.00 &  $-$51 25 48.0 & 6035 & 11.7~Cl~ & ($\times$1.00)& $-$86.0 & $-$86.5 & $-$84.5 &C95,C03 \\
 			    & 16 12 08.99 &  $-$51 25 48.0 & 6668 & 4.2~~~~~ & ($\times$1.00)& $-$86.0 & $-$94.0 & $-$85.0 &C09\\ 
331.543$-$0.066 & 16 12 09.12 &  $-$51 25 45.6 & 6668 & 9.8~~~~~ & ($\times$1.00)& $-$84.5 & $-$85.0 & $-$79.0&C09 \\
331.556$-$0.121 & 16 12 27.19 &  $-$51 27 38.4 & 6668 & 23.8~~~~~ & ($\times$1.54) & $-$97.2 & $-$106.0 & $-$96.0&C09 \\
333.068$-$0.447 & 16 20 48.55 &  $-$50 38 36.4 & 6668 & 2.5~~~~~ & ($\times$2.38) & $-$54.4 & $-$57.0 & $-$49.5 &C09\\
333.121$-$0.434 & 16 20 59.70 &  $-$50 35 52.4 & 6668 & 9.6~~~~~ & ($\times$1.04) & $-$49.8 & $-$56.5 & $-$47.0 &C09\\
333.126$-$0.440 & 16 21 02.65 &  $-$50 35 55.2 & 6668 & 1.1~~~~~ & ($\times$1.03)& $-$47.2 & $-$48.0 & $-$42.0 &C09\\
333.128$-$0.440 & 16 21 03.30 &  $-$50 35 49.8 & 6668 & 4.2~~~~~ & ($\times$1.02) & $-$44.1 & $-$50.0 & $-$43.5&C09 \\
\hline
\end{tabular} 
\label{resotable}
\end{table*}
 
\begin{table*} 
\addtocounter{table}{-1}
\centering 
\caption{\small cont} 
\begin{tabular}{lcrcrlrrrl}
\hline
\multicolumn{1}{c}{Source Name} & \multicolumn{2}{c}{Equatorial Coordinates} & \multicolumn{1}{c}{Maser} & \multicolumn{1}{l}{Peak} && \multicolumn{1}{c}{Peak} & \multicolumn{2}{c}{Velocity} & \multicolumn{1}{l}{Previous}\\
\multicolumn{1}{c}{(~~~$l$,~~~~~~~$b$~~~)}    &       RA(J2000)        &       Dec(J2000)       & Transition & \multicolumn{1}{l}{Flux} &($\times$pbc$^{1}$)& \multicolumn{1}{c}{Velocity}   & \multicolumn{2}{c}{Range}      &Reference \\
\multicolumn{1}{c}{(~~~$^\circ$~~~~~~~$^\circ$~~~)} & (h~~m~~~s) & (~$^\circ$~~ $'$~~~~$''$) & \multicolumn{1}{c}{(MHz)} &  \multicolumn{2}{l}{(Jy)} & \multicolumn{1}{c}{(km\,s$^{-1}$ )}&   \multicolumn{2}{c}{(km\,s$^{-1}$ )}  & \\
\hline
333.135$-$0.432a & 16 21 02.97&  $-$50 35 13.5 & 6030 & 0.2~C~~ & ($\times$1.00)& $-$54.3 & $-$56.0 & $-$49.0  &C95,C03\\
 			       & 16 21 02.96&  $-$50 35 13.6 & 6035 & 0.6~C~~ & ($\times$1.00)& $-$54.5 & $-$57.0 & $-$48.5 &C95,C03 \\
			       & 16 21 03.00 &  $-$50 35 12.9 & 6668 & 0.8~~~~~ &($\times$1.00) & $-$54.0 & $-$56.0 & $-$51.0 &C09\\
333.135$-$0.432b & 16 21 02.99&  $-$50 35 10.7 & 6030 & 0.6~C~~ & ($\times$1.00)& $-$51.3 & $-$56.0 & $-$49.0  &C95,C03\\
 			       & 16 21 03.00&  $-$50 35 10.7 & 6035 & 6.4~Cl~ & ($\times$1.00)& $-$51.3 & $-$57.0 & $-$48.5 &C95,C03 \\
333.135$-$0.432c & 16 21 02.92&  $-$50 35 12.8 & 6030 & 0.3~C~~ & ($\times$1.00)& $-$50.1 & $-$56.0 & $-$49.0  &C95,C03\\
 			       & 16 21 02.92&  $-$50 35 12.9 & 6035 & 1.5~C~~ & ($\times$1.00)& $-$50.1 & $-$57.0 & $-$48.5 &C95,C03 \\
333.135$-$0.432d & 16 21 03.05 &  $-$50 35 11.9 & 6035 & 0.3~C~~ & ($\times$1.00)& $-$49.4 & $-$57.0 & $-$48.5 &C95,C03 \\
336.941$-$0.156    & 16 35 55.19 &  $-$47 38 45.5 & 6030 & 0.7~C~~ & ($\times$1.00)& $-$65.5 & $-$68.5 & $-$65.0  &C03\\
 			       & 16 35 55.19 &  $-$47 38 45.5 & 6035 & 2.3~Cl~ & ($\times$1.00)& $-$65.5 & $-$70.0 & $-$64.0  &C03\\
 			       & 16 35 55.19 &  $-$47 38 45.5 & 6668 & 18.1~~~l~ & ($\times$1.00)& $-$67.3 & $-$70.0 & $-$64.5 &C09\\
337.703$-$0.054 & 16 38 29.09 &  $-$47 00 43.5 & 6668 & 0.7~~~~~ &($\times$1.00) & $-$44.6 & $-$52.0 & $-$43.0&C09 \\			 
337.705$-$0.053a & 16 38 29.63 & $-$47 00 35.4   & 6035 & 0.7~C~~  &($\times$1.00) & $-$53.6  & $-$56.0 & $-$52.0&C95,C03\\
 			       & 16 38 29.62 &  $-$47 00 35.8 & 6668 & 121.4~~~l~ &($\times$1.00) & $-$54.7 & $-$56.5 & $-$48.0&C09 \\       
337.705$-$0.053b & 16 38 29.66 &  $-$47 00 36.0 & 6035 & 1.3~C~~ & ($\times$1.00)& $-$50.7 & $-$52.0 & $-$47.5  &C95,C03\\
& 16 38 29.70 &  $-$47 00 36.5 & 6668 & 5.9~~~l~ &($\times$1.00) & $-$44.1 & $-$45.0 & $-$43.0&C09 \\
339.884$-$1.259a & 16 52 04.62 &  $-$46 08 34.2 & 6030 & 11.1~Cl~ & ($\times$1.00)& $-$37.5 & $-$38.0 & $-$36.5  &C03\\
 			       & 16 52 04.62 &  $-$46 08 34.2 & 6035 & 47.0~Cl~ & ($\times$1.00)& $-$37.5 & $-$39.0 & $-$36.5  &C03\\
			       & 16 52 04.63 &  $-$46 08 34.2 & 6668 & 261.2~~~l~ &($\times$1.00) & $-$34.0 & $-$41.0 & $-$30.5&C09 \\
339.884$-$1.259b & 16 52 04.69 &  $-$46 08 34.6 & 6668 & 934.7~~~l~ &($\times$1.00) & $-$38.8 & $-$41.0 & $-$30.5&C09 \\
340.785$-$0.096a & 16 50 14.83 &  $-$44 42 26.5 & 6035 & 1.2~C~~ & ($\times$1.00)& $-$106.0 & $-$107.0 & $-$99.0 &C95,C03 \\
			       & 16 50 14.86 &  $-$44 42 26.5 & 6668 & 199.7~~~l~ & ($\times$1.00)& $-$108.0 & $-$111.5 & $-$85.0&C09 \\
340.785$-$0.096b & 16 50 14.82 &  $-$44 42 27.0 & 6035 & 3.5~C~~ & ($\times$1.00)& $-$102.1 & $-$107.0 & $-$99.0 &C95,C03 \\
			       & 16 50 14.83 &  $-$44 42 26.9 & 6668 & 16.2~~~l~ & ($\times$1.00)& $-$100.7 & $-$111.5 & $-$85.0&C09 \\
343.929+0.125 & 17 00 10.90 &  $-$42 07 19.4 & 6035 & 1.1~C~~ &($\times$1.00) & 13.6 & 11.5 & 15.5 &C95,C03 \\
 			& 17 00 10.90 &  $-$42 07 19.4 & 6668 & 6.7~~~~~ &($\times$1.00) & 14.5 & 8.5 & 19.0 &C09\\
345.003$-$0.223 & 17 05 10.89 &  $-$41 29 06.1 & 6668 & 42.8~~~l~ &($\times$1.00) & $-$23.0 & $-$31.0 & $-$21.0&C09\\
345.003$-$0.224 & 17 05 11.20 &  $-$41 29 06.8 & 6030 & 1.2~C~~ &($\times$1.00) & $-$26.0 & $-$29.5 & $-$25.0  &C95,C03\\
 			   & 17 05 11.21 &  $-$41 29 06.8 & 6035 & 3.1~CL & ($\times$1.00)& $-$26.0 & $-$31.0 & $-$24.5 &C95,C03 \\
  			   & 17 05 11.17 &  $-$41 29 06.9 & 6668 & 52.3~~~~~ & ($\times$1.00)& $-$26.2 & $-$31.5 & $-$21.5&C09 \\
345.009+1.792 & 16 56 47.58 &  $-$40 14 25.8 & 6030 & 0.3~C~~ &($\times$1.00) & $-$20.4 & $-$22.0 & $-$16.0  &C95,C03\\
 			 & 16 56 47.58 &  $-$40 14 25.9 & 6035 & 1.0~CL &($\times$1.00) & $-$21.4 & $-$24.5 & $-$15.0  &C95,C03\\
  			 & 16 56 47.58 &  $-$40 14 25.9 & 6668 & 199.2~~~l~ & ($\times$1.00)& $-$21.1 & $-$23.5 & $-$15.5 &C09\\
345.012+1.797 & 16 56 46.82 &  $-$40 14 08.9 & 6668 & 29.3~~~~~ &($\times$1.00) & $-$12.2 & $-$16.0 & $-$10.0&C09 \\
345.487+0.314 & 17 04 28.11 &  $-$40 46 24.9 & 6030 & 0.3~C~~ &($\times$1.00) & $-$22.2 & $-$22.5 & $-$21.5 &C95,C03 \\
 			& 17 04 28.11 &  $-$40 46 25.1 & 6035 & 2.7~Cl~ & ($\times$1.00)& $-$22.2 & $-$23.0 & $-$21.0 &C95,C03 \\
345.487+0.313 & 17 04 28.24 &  $-$40 46 29.1 & 6668 & 2.5~~~~~ &($\times$1.00) & $-$22.5 & $-$24.0 & $-$21.5 &C09\\
345.505+0.348 & 17 04 22.90 &  $-$40 44 22.2 & 6668 & 195.6~~~l~ & ($\times$1.32) & $-$17.8 & $-$23.0 & $-$10.5&C09 \\
345.698$-$0.090a & 17 06 50.60 &  $-$40 51 00.1 & 6030 & 2.0~C~~ & ($\times$1.00)& $-$4.9 & $-$9.0 & $-$2.5  &C95,C03\\
 			     & 17 06 50.60 &  $-$40 51 00.0 & 6035 & 2.2~C~~ &($\times$1.00) & $-$4.9 & $-$9.0 & $-$3.0 &C95,C03\\
345.698$-$0.090b			 & 17 06 50.64 &  $-$40 50 59.6 & 6035 & 4.1~C~~ &($\times$1.00) & $-$6.5 & $-$9.0 & $-$3.0 &C95,C03\\
347.628+0.148 & 17 11 51.02 &  $-$39 09 28.9 & 6030 & 0.9~C~~ & ($\times$1.00)& $-$96.7 & $-$97.5 & $-$96.0  &C95,C03\\
 			 & 17 11 51.02 &  $-$39 09 28.9 & 6035 & 6.1~C~~ & ($\times$1.00)& $-$96.7 & $-$98.5 & $-$94.0  &C95,C03\\
 			 & 17 11 51.02 &  $-$39 09 29.1 & 6668 & 15.0~~~~~ & ($\times$1.00)& $-$96.5 & $-$98.0 & $-$95.5&C09 \\
347.632+0.210 & 17 11 36.15 &  $-$39 07 06.7 & 6668 & 2.6~~~~~ & ($\times$2.04)& $-$91.9 & $-$94.0 & $-$90.5&C09 \\
351.417+0.645a & 17 20 53.37 &  $-$35 47 01.4 & 6030 & 11.5~Cl~ &($\times$1.00) & $-$10.3 & $-$11.5 & $-$10.0 &C95,C03,C11\\
 			   & 17 20 53.38 &  $-$35 47 01.4 & 6035 & 153.1~C~~ &($\times$1.00) & $-$10.3 & $-$15.5 & $-$5.0  &C95,C03,C11\\
 			   & 17 20 53.37 &  $-$35 47 01.4 & 6668 & 2350.1~~~l~ & ($\times$1.00)& $-$10.4 & $-$12.0 & $-$6.0&C09 \\   
351.417+0.645b & 17 20 53.42 &  $-$35 47 02.6 & 6030 & 0.7~C~~ & ($\times$1.00)& $-$7.8 & $-$8.2 & $-$7.0  &C95,C03,C11\\
  			   & 17 20 53.42 &  $-$35 47 02.6 & 6035 & 2.5~Cl~ & ($\times$1.00)& $-$7.7 & $-$9.0 & $-$7.0   &C95,C03,C11\\		   
351.417+0.645c & 17 20 53.45 &  $-$35 47 00.5 & 6035 & 4.1~CL &($\times$1.00) & $-$9.7 & $-$15.5 & $-$5.0  &C95,C03,C11\\
351.417+0.645d & 17 20 53.42 & $-$35 47 02.1 & 6030&0.6~C~~ &($\times$1.00) &$-$8.5&$-$9.0&$-$8.2&C95,C03,C11\\
& 17 20 53.43 &  $-$35 47 01.6 & 6035 & 6.8~Cl~ &($\times$1.00) & $-$8.7 & $-$15.5 & $-$5.0  &C95,C03,C11\\
			   & 17 20 53.44 &  $-$35 47 01.5 & 6668 & 81.3~~~~~ &($\times$1.00) & $-$8.6 & $-$9.0 & $-$6.0&C09 \\
351.417+0.645e & 17 20 53.44 &  $-$35 47 03.5 & 6668 & 100.2~~~l~ &($\times$1.00) & $-$6.6 & $-$9.0 & $-$6.0&C09 \\
351.417+0.646 & 17 20 53.18 &  $-$35 46 58.8 & 6668 & 966.8~~~l~ & ($\times$1.00)& $-$11.2 & $-$12.0 & $-$7.0&C09 \\
351.445+0.660 & 17 20 54.62 &  $-$35 45 08.3 & 6668 & 106.6~~~l~ & ($\times$1.20)& $-$9.3 & $-$10.5 & $-$2.0 &C09\\
353.410$-$0.360 & 17 30 26.19 &  $-$34 41 45.3 & 6030 & 0.9~C~~ & ($\times$1.00)& $-$21.9 & $-$23.0 & $-$20.0 &C95,C03,C11 \\
 			     & 17 30 26.19 &  $-$34 41 45.3 & 6035 & 14.5~Cl~ & ($\times$1.00)& $-$20.7 & $-$23.0 & $-$19.0 &C95,C03,C11 \\
 			     & 17 30 26.18 &  $-$34 41 45.5 & 6668 & 83.8~~~l~ & ($\times$1.00)& $-$20.4 & $-$23.0 & $-$18.5 &C09\\
355.343+0.148 & 17 33 28.80	& $-$32 47 59.6 &6668&0.9~~~~~ &($\times$1.00)&5.8&4.5&7.0&C97,C09\\

\hline
\end{tabular} 
\label{resotable}
\end{table*}

\begin{table*} 
\addtocounter{table}{-1}
\centering 
\caption{\small cont} 
\begin{tabular}{lcrcrlrrrl}
\hline
\multicolumn{1}{c}{Source Name} & \multicolumn{2}{c}{Equatorial Coordinates} & \multicolumn{1}{c}{Maser} & \multicolumn{1}{l}{Peak} && \multicolumn{1}{c}{Peak} & \multicolumn{2}{c}{Velocity} & \multicolumn{1}{l}{Previous}\\
\multicolumn{1}{c}{(~~~$l$,~~~~~~~$b$~~~)}    &       RA(J2000)        &       Dec(J2000)       & Transition & \multicolumn{1}{l}{Flux} &($\times$pbc$^{1}$)& \multicolumn{1}{c}{Velocity}   & \multicolumn{2}{c}{Range}      &Reference \\
\multicolumn{1}{c}{(~~~$^\circ$~~~~~~~$^\circ$~~~)} & (h~~m~~~s) & (~$^\circ$~~ $'$~~~~$''$) & \multicolumn{1}{c}{(MHz)} &  \multicolumn{2}{l}{(Jy)} & \multicolumn{1}{c}{(km\,s$^{-1}$ )}&   \multicolumn{2}{c}{(km\,s$^{-1}$ )}  & \\
\hline
355.344+0.147 & 17 33 29.05 &  $-$32 47 58.7 & 6030 & 1.2~C~~ & ($\times$1.00)& 17.8 & 17.0 & 19.5&C95,C03\\
 			 & 17 33 29.05 &  $-$32 47 58.8 & 6035 & 3.4~C~~ & ($\times$1.00)& 17.9 & 17.0 & 19.5 &C95,C97,C03\\
 			 & 17 33 29.06 &  $-$32 47 58.7 & 6668 & 7.4~~~~~ & ($\times$1.00)& 19.9 & 18.5 & 20.5&C97,C09\\
355.346+0.149 & 17 33 28.91 &  $-$32 47 49.1 & 6668 & 6.0~~~~~ & ($\times$1.00)& 10.4 & 8.0 & 12.0 &C03,C97\\
000.645$-$0.042 & 17 47 18.65 &  $-$28 24 25.0 & 6668 & 40.3~~~l~ & ($\times$1.11)& 49.5 & 46.0 & 53.0 &HW95 \\
000.648$-$0.055 & 17 47 22.25 &  $-$28 24 39.1 & 6668 & 4.8~~~~~ &($\times$1.23) & 49.8 & 49.0 & 52.0 &HW95 \\
000.651$-$0.049 & 17 47 21.11 &  $-$28 24 18.4 & 6668 & 12.2~~~~~ &($\times$1.11) & 47.8 & 46.0 & 49.0 &HW95 \\
000.657$-$0.041 & 17 47 20.05 &  $-$28 23 46.9 & 6668 & 2.9~~~~~ & ($\times$1.04)& 51.1 & 48.0 & 56.0 &HW95\\
000.658$-$0.042 & 17 47 20.48 & $-$28 23 45.5 & 6035 & 0.2~C~~ &($\times$1.04) & 68.4 &65.0&70.0&$<$new site$>$\\
000.665$-$0.036 & 17 47 20.10 &  $-$28 23 13.0 & 6035 & 0.3~C~~  & ($\times$1.00)& 60.6 & 60.0 & 61.0 &$<$new site$>$\\
 			    & 17 47 20.12 &  $-$28 23 12.9 & 6668 & 4.9~~~~~ &($\times$1.01) & 60.7 & 58.0 & 62.0 &HW95\\
000.666$-$0.029 & 17 47 18.64 &  $-$28 22 54.7 & 6030 & 0.6~C~~  & ($\times$1.00)& 70.2 & 68.5 & 73.0 &C95,C03\\
 			    & 17 47 18.64 &  $-$28 22 54.7 & 6035 & 9.5~Cl~  & ($\times$1.00)& 72.1 & 68.5 & 73.5 &C95,C97,C03\\
 			    & 17 47 18.64 &  $-$28 22 54.7 & 6668 & 32.0~~~~~ & ($\times$1.00)& 70.4 & 68.0 & 73.0&HW95,C97\\
000.666$-$0.035 & 17 47 20.10 &  $-$28 23 06.4 & 6030 & 0.2~C~~ & ($\times$1.00)& 65.3 & 60.0 & 68.0 &C97,C03\\
                                & 17 47 20.15 &  $-$28 23 06.3 & 6035 & 1.2~C~~ & ($\times$1.00)& 67.2 & 60.0 & 68.0 &C97,C03\\
000.666$-$0.036 & 17 47 20.10 &  $-$28 23 08.7 & 6030 & 0.2~C~~  & ($\times$1.00)& 68.2 & 60.0 & 68.0 &$<$new site$>$\\
                                & 17 47 20.12 &  $-$28 23 08.7 & 6035 & 0.7~C~~  & ($\times$1.00)& 62.1 & 60.0 & 68.0 &$<$new site$>$\\                                                             
000.667$-$0.034 & 17 47 19.86 &  $-$28 23 00.9 & 6668 & 0.6~~~~~ & ($\times$1.00)& 55.0 & 49.0 & 56.0&HW95 \\
000.672$-$0.031 & 17 47 20.04 &  $-$28 22 41.4 & 6668 & 6.8~~~~~ &($\times$1.01) & 58.3 & 55.0 & 59.0 &HW95\\
000.673$-$0.029 & 17 47 19.56 &  $-$28 22 33.0 & 6668 & 0.5~~~~~ & ($\times$1.01)& 60.5 & 60.0 & 66.5 &HW95\\
000.677$-$0.025 & 17 47 19.28 &  $-$28 22 14.9 & 6668 & 4.8~~~~~ & ($\times$1.02)& 73.4 & 70.0 & 77.0 &HW95\\
000.695$-$0.038 & 17 47 24.74 &  $-$28 21 43.7 & 6668 & 19.6~~~~~ &($\times$1.18) & 68.6 & 64.0 & 75.0 &HW95\\
010.959+0.023 & 18 09 39.31 &  $-$19 26 26.6 & 6035 & 0.4~C~~  &($\times$3.13) & 24.8 & 24.0 & 25.5 &$<$new site$>$\\
 			 & 18 09 39.31 &  $-$19 26 26.6 & 6668 & 3.4~~~~~ &($\times$4.17) & 24.6 & 23.5 & 25.5 &C09 \\
011.034+0.062 & 18 09 39.84 &  $-$19 21 20.1 & 6035 & 1.0~Cl~  &($\times$1.00) & 24.0 & 21.0 & 25.0 &C95,C03\\
 			 & 18 09 39.83 &  $-$19 21 20.6 & 6668 & 0.4~~~~~ & ($\times$1.00)& 20.6 & 15.0 & 21.0 &C03\\			 
011.903$-$0.102 & 18 12 02.69 &  $-$18 40 23.5 & 6668 & 7.7~~~~~ & ($\times$1.32)& 33.8 & 33.0 & 37.0 &C97,C09 \\
011.904$-$0.141 & 18 12 11.45 &  $-$18 41 27.7 & 6035 & 2.1~CL & ($\times$1.00)& 42.9 & 41.0 & 44.0 &C95,C97,C03\\
 			     & 18 12 11.45 &  $-$18 41 27.7 & 6668 & 46.0~~~l~ & ($\times$1.00)& 43.2 & 39.5 & 44.5 &C09,C97 \\
011.934$-$0.150 & 18 12 17.15 &  $-$18 40 08.8 & 6668 & 0.9~~~~~ & ($\times$1.20)& 33.3 & 31.5 & 36.0 &$<$new site$>$\\
011.936$-$0.150 & 18 12 17.28 &  $-$18 40 01.7 & 6668 & 1.6~~~~~  &($\times$1.23) & 48.4 & 47.0 & 50.0 &C09,C97\\
015.034$-$0.677 & 18 20 24.80 &  $-$16 11 35.1 & 6030 & 3.1~CL & ($\times$1.00)& 21.5 & 21.0 & 22.0 &C95,C03\\
 			    & 18 20 24.80 &  $-$16 11 35.1 & 6035 & 30.1~Cl~ & ($\times$1.00)& 21.4 & 20.5 & 24.5 &C95,C97,C03 \\
 			    & 18 20 24.80 &  $-$16 11 35.1 & 6668 & 48.8~~~l~ & ($\times$1.00)& 21.3 & 20.5 & 24.0  &C97,C09\\
\hline
\end{tabular} 
\label{resotable}
\end{table*}

\begin{table*}
\begin{minipage}{180mm}
\small
\centering
\caption{\small Polarised excited-state hydroxyl features associated with Zeeman splitting. Components identified from least square fitted Gaussian components (see Section \ref{zeeman_ident}). The formal detection limit is three channels with emission above five $\sigma_{\rm rms}$, but where appropriate, weaker components evident in the spectra are listed in italics. Splitting factors are 0.079 \,km\,s$^{-1}$ mG$^{-1}$ for 6030 MHz and  0.056 \,km\,s$^{-1}$ mG$^{-1}$ for 6035 MHz. All Zeeman associations are spatially coincident within the observational errors, but a reliability indicator is given in the last column to identify those which we believe to be the most reliable (`A') and those that may be spatially offset opposite polarised single features blended within the ATCA position accuracy or are spectrally blended (`B'). $^{1}$For the Zeeman splitting type `P' denotes Zeeman pair, `T' denotes possible Zeeman triplet, `X' denotes a linearly polarized component which could not readily be attributed to a Zeeman pair/triplet.
 Linearly polarised components believed to represent elliptical polarisation are identified as $\sigma^{\pm}$ components. Pairs are individually numbered for each source.   Field strengths are listed after the first component of the pair/triplet with errors based on the accuracy of the Gaussian fitted components. The commonly adopted field direction convention is chosen here with RHC at a lower velocity than LHC representing a negative magnetic field strength, directed towards us.}
\begin{tabular}{l r r l r c c c r c}
\\
\hline
\multicolumn{1}{l}{Source Name}& \multicolumn{1}{c}{Freq} & \multicolumn{1}{c}{$V_{\rm LSR}$} &\multicolumn{1}{c}{Pol}& \multicolumn{1}{c}{$S_p$} & \multicolumn{5}{c}{Zeeman Splitting} \\
\ (~~~$l$,~~~~~~~$b$~~~)   &                       &&& &Type$^{1}$ & Number  & Component& Strength &Reliability\\
\ (~~~$^\circ$~~~~~~~$^\circ$~~~)    &  \multicolumn{1}{c}{(MHz)} & \multicolumn{1}{c}{(\,km\,s$^{-1}$) }&  & (Jy) & && &(mG)  \\
\hline
284.351$-$0.418  & 6030 & 7.79 & RHC & 0.28 & P & $Z_1$ & $\sigma$ & $-$1.4$\pm$0.2 & A \\
 & 6030 & 7.90 & LHC & 0.28 & P & $Z_1$ & $\sigma$ &   &  \\
 & 6035 & 5.71 & RHC & 0.42 & P & $Z_2$ & $\sigma$ & $-$1.1$\pm$1.3 & B \\
 & 6035 & 5.77 & LHC & 0.36 & P & $Z_2$ & $\sigma$ &   &  \\
 & 6035 & 5.86 & RHC & 0.27 & P & $Z_3$ & $\sigma$ & $-$2.5$\pm$0.4 & B \\
 & 6035 & 6.00 & LHC & 0.62 & P & $Z_3$ & $\sigma$ &   &  \\
 & 6035 & 7.83 & RHC & 0.39 & P & $Z_4$ & $\sigma$ & $-$1.3$\pm$0.3 & A \\
 & 6035 & 7.90 & LHC & 0.39 & P & $Z_4$ & $\sigma$ &   &  \\
285.263$-$0.050 & 6035 & 8.53 & LHC & 1.19 & P & $Z_1$ & $\sigma$ & +10.0$\pm$0.2 & A \\
 & 6035 & 9.09 & RHC & 0.53 & P & $Z_1$ & $\sigma$ &   &  \\
294.511$-$1.621 & 6035 & $-$12.11  & LHC & 1.56 & P & $Z_1$ & $\sigma$ &    +1.3$\pm$0.1 & B\\
 & 6035 & $-$12.11 & LIN & 0.17 & P & $Z_1$ & $\sigma$ & &  \\
 & 6035 & $-$12.02 & RHC & 1.43 & P & $Z_1$ & $\sigma$ &   &  \\
300.969+1.148 & 6030 & $-$37.74 & RHC & 1.91 & P & $Z_1$ & $\sigma$ & $-$5.2$\pm$0.1 & A \\
 & 6030 & $-$37.36 & LIN & 0.27 & P & $Z_1$ & $\sigma$ &   &  \\
 & 6030 & $-$37.33 & LHC & 3.82 & P & $Z_1$ & $\sigma$ &   &  \\
 & 6035 & $-$38.99 & RHC & 0.39 & P & $Z_2$ & $\sigma$ & $-$4.3$\pm$0.6 & A \\
 & 6035 & $-$38.75 & LHC & 0.29 & P & $Z_2$ & $\sigma$ &   &  \\
 & 6035 & $-$37.66 & RHC & 4.46 & T & $Z_3$ & $\sigma$ & $-$3.6$\pm$0.3 & A \\
 & 6035 & $-$37.58 & LIN & 0.87 & T & $Z_3$ & $\pi$ &   &  \\
 & 6035 & $-$37.46 & LHC & 7.64 & T & $Z_3$ & $\sigma$ &   &  \\
309.921+0.479b & 6035 & $-$61.70 & RHC & 6.88 & P/T & $Z_1$ & $\sigma$ & $-$2.9$\pm$0.2 & B \\
 & 6035 & $-$61.63 & LIN & 1.60 & P/T & $Z_1$ & $\sigma$/$\pi$ &   &  \\
 & 6035 & $-$61.54 & LHC & 5.58 & P/T & $Z_1$ & $\sigma$ &   &  \\
 & 6035 & $-$61.31 & RHC & 1.56 & P/T & $Z_2$ & $\sigma$ & $-$5.7$\pm$1.3 & B \\
 & 6035 & $-$61.20 & LIN & 0.30 &  P/T & $Z_2$  & $\sigma$/$\pi$ &   &  \\
 & 6035 & $-$61.00 & LIN & 0.25 &  P/T & $Z_2$ & $\sigma$ &   &  \\
 & 6035 & $-$60.99 & LHC & 1.66 & P/T & $Z_2$ & $\sigma$ &   &  \\
309.921+0.479c & 6035 & $-$59.90 & LHC & 3.29 & P & $Z_1$ & $\sigma$ & +4.8$\pm$0.4 & B \\
 & 6035 & $-$59.63 & RHC & 3.05 & P & $Z_1$ & $\sigma$ &   &  \\
 & 6035 & $-$59.54 & LHC & 10.67 & P & $Z_2$ & $\sigma$ & +1.1$\pm$0.7 & B \\
 & 6035 & $-$59.53 & LIN & 13.53 & X& &&   &  \\
 & 6035 & $-$59.48 & RHC & 6.27 & P & $Z_2$ & $\sigma$ &   &  \\
 & 6035 & $-$58.90 & LHC & 2.45 & P & $Z_3$ & $\sigma$ & +2.9$\pm$0.4 & B \\
 & 6035 & $-$58.74 & RHC & 2.09 & P & $Z_3$ & $\sigma$ &   &  \\
 & 6035 & $-$58.14 & LHC & 3.15 & P & $Z_4$ & $\sigma$ & +5.5$\pm$0.4 & B \\
 & 6035 & $-$57.83 & RHC & 3.10 & P & $Z_4$ & $\sigma$ &   &  \\
311.596$-$0.398 & 6035 & 29.65 & LHC & 1.10 & P & $Z_1$ & $\sigma$ & +0.9$\pm$0.3 & B \\
 & 6035 & 29.70 & RHC & 0.33 & P & $Z_1$ & $\sigma$ &   &  \\
 & 6035 & 29.70 & LIN & 0.86 & X& &&   &  \\
 & 6035 & 30.13 & LHC & 0.82 & P & $Z_2$ & $\sigma$ & +1.1$\pm$0.3 & B \\
 & 6035 & 30.17 & LIN & 0.52 & X& &&   &  \\
 & 6035 & 30.19 & RHC & 0.51 & P & $Z_2$ & $\sigma$ &   &  \\
311.643$-$0.380 & 6035 & 33.80 & LHC & 0.43 & P & $Z_1$ & $\sigma$ & +4.1$\pm$0.4 & B \\
 & 6035 & 34.03 & RHC & 0.18 & P & $Z_1$ & $\sigma$ &   &  \\
323.459$-$0.079a & 6030 & $-$70.50 & LHC & 0.49 & P & $Z_1$ & $\sigma$ & +4.1$\pm$0.1 & A \\
 & 6030 & $-$70.18 & RHC & 0.50 & P & $Z_1$ & $\sigma$ &   &  \\
 & 6035 & $-$70.51 & LHC & 8.17 & T & $Z_2$ & $\sigma$ & +4.1$\pm$0.1 & A \\
 & 6035 & $-$70.48 & LIN & 0.51 & T & $Z_2$ & $\sigma$ &   &  \\
 & 6035 & $-$70.37 & LIN & 0.38 & T & $Z_2$ & $\pi$ &   &  \\
 & 6035 & $-$70.28 & RHC & 13.92 & T & $Z_2$ & $\sigma$ &   &  \\
 \hline
\end{tabular} 
\label{fullresultstable}
\end{minipage}
\end{table*}

\begin{table*}
\addtocounter{table}{-1}
\begin{minipage}{180mm}
\small
\centering
\caption{\small Cont.}
\begin{tabular}{l r r l r c c c r c}
\\
\hline
\multicolumn{1}{l}{Source Name}& \multicolumn{1}{c}{Freq} & \multicolumn{1}{c}{$V_{\rm LSR}$} &\multicolumn{1}{c}{Pol}& \multicolumn{1}{c}{$S_p$} & \multicolumn{5}{c}{Zeeman Splitting} \\
\ (~~~$l$,~~~~~~~$b$~~~)   &                       &&& &Type$^{1}$ & Number  & Component& Strength &Reliability\\
\ (~~~$^\circ$~~~~~~~$^\circ$~~~)    &  \multicolumn{1}{c}{(MHz)} & \multicolumn{1}{c}{(\,km\,s$^{-1}$) }&  & (Jy) & && &(mG)  \\
\hline 
323.459$-$0.079b & 6030 & $-$68.42 & LHC & 0.19 & P & $Z_1$ & $\sigma$ & +0.4$\pm$0.2 & A \\
 & 6030 & $-$68.39 & RHC & 0.25 & P & $Z_1$ & $\sigma$ &   &  \\
 & 6035 & $-$68.67 & LHC & 1.07 & P & $Z_2$ & $\sigma$ & +3.2$\pm$0.5 & B \\
 & 6035 & $-$68.65 & LIN & 0.49 & P & $Z_2$ & $\sigma$ &   &  \\
 & 6035 & $-$68.49 & RHC & 2.35 & P & $Z_2$ & $\sigma$ &   &  \\
 & 6035 & $-$68.45 & LIN & 0.50 & P & $Z_2$ & $\sigma$ &   &  \\
 & 6035 & $-$68.30 & LHC & 1.96 & T & $Z_3$ & $\sigma$ & +3.9$\pm$0.3 & B \\
 & 6035 & $-$68.19 & LIN & 0.27 & T & $Z_3$ & $\pi$ &   &  \\
 & 6035 & $-$68.08 & RHC & 1.37 & T & $Z_3$ & $\sigma$ &   &  \\
 & 6035 & $-$66.09 & LHC & 0.34 & P & $Z_4$ & $\sigma$ & +0.9$\pm$0.8 & B \\
 & 6035 & $-$66.04 & RHC & 0.68 & P & $Z_4$ & $\sigma$ &   &  \\
328.808+0.633a & 6030 & $-$46.53 & RHC & 0.63 & T & $Z_1$ & $\sigma$ & $-$5.1$\pm$0.1 & B \\
 & 6030 & $-$46.20 & LIN & 0.25 & T & $Z_1$ & $\pi$ &   &  \\
 & 6030 & $-$46.13 & LHC & 3.09 & T & $Z_1$ & $\sigma$ &   &  \\
 & 6035 & $-$46.41 & RHC & 2.31 & T & $Z_2$ & $\sigma$ & $-$4.6$\pm$0.3 & B \\
 & 6035 & $-$46.20 & LIN & 1.45 & T & $Z_2$ & $\pi$ &   &  \\
 & 6035 & $-$46.15 & LHC & 10.15 & T & $Z_2$ & $\sigma$ &   &  \\
 & 6035 & $-$45.88 & RHC & 4.28 & T & $Z_3$ & $\sigma$ & $-$2.9$\pm$0.4 & B \\
 & 6035 & $-$45.81 & LIN & 0.63 & T & $Z_3$ & $\pi$ &   &  \\
 & 6035 & $-$45.72 & LHC & 3.07 & T & $Z_3$ & $\sigma$ &   &  \\
 & 6035 & $-$45.64 & LIN & 0.54 & X &  &  &   &  \\
 & 6035 & $-$45.11 & LIN & 0.64 & X &  &  &   &  \\
328.808+0.633b & 6035 & $-$44.08 & RHC & 0.93 & P & $Z_1$ & $\sigma$ & $-$0.9$\pm$0.4 & B \\
 & 6035 & $-$44.03 & LHC & 0.38 & P & $Z_1$ & $\sigma$ &   &  \\
 & 6035 & $-$43.46 & RHC & 2.35 & P & $Z_2$ & $\sigma$ & $-$0.4$\pm$0.2 & B \\
 & 6035 & $-$43.44 & LHC & 1.81 & P & $Z_2$ & $\sigma$ &   &  \\
329.339+0.148 & 6035 & $-$105.54 & LHC & 0.14 & P & $Z_1$ & $\sigma$ & +7.9$\pm$0.8 & B \\
 & 6035 & $-$105.10 & RHC & 0.15 & P & $Z_1$ & $\sigma$ &   &  \\
 & 6035 & $-$104.83 & LHC & 0.11 & P & $Z_2$ & $\sigma$ & +11.4$\pm$0.6 & B \\
 & 6035 & $-$104.19 & RHC & 0.16 & P & $Z_2$ & $\sigma$ &   &  \\
330.953$-$0.182 & 6035 & $-$89.00 & RHC & 0.12 & P & $Z_1$ & $\sigma$ & $-$2.0$\pm$0.8 & B \\
 & 6035 & $-$88.89 & LHC & 0.22 & P & $Z_1$ & $\sigma$ &   &  \\
 & 6035 & $-$88.08 & RHC & 0.50 & T & $Z_2$ & $\sigma$ & $-$3.4$\pm$0.2 & B \\
 & 6035 & $-$87.93 & LIN & 0.54 & T & $Z_2$ & $\pi$ &   &  \\
 & 6035 & $-$87.89 & LHC & 1.44 & T & $Z_2$ & $\sigma$ &   &  \\
331.511$-$0.102 & 6030 & $-$89.60 & RHC & 0.16 & P & $Z_1$ & $\sigma$ & $-$0.4$\pm$0.9 & B \\
 & 6030 & $-$89.57 & LHC & 0.21 & P & $Z_1$ & $\sigma$ &   &  \\
 & 6030 & $-$89.15 & LHC & 0.15 & P & $Z_2$ & $\sigma$ & +0.9$\pm$0.4 & B \\
 & 6030 & $-$89.08 & RHC & 0.24 & P & $Z_2$ & $\sigma$ &   &  \\
 & 6035 & $-$89.25 & LHC & 0.75 & P & $Z_3$ & $\sigma$ & +1.4$\pm$0.4 & B \\
 & 6035 & $-$89.17 & RHC & 0.88 & P & $Z_3$ & $\sigma$ &   &  \\
331.512$-$0.102 & 6030 & $-$87.05 & LHC & 0.08 & P & $Z_1$ & $\sigma$ & +0.6$\pm$0.7 & B \\
 & 6030 & $-$87.00 & RHC & 0.10 & P & $Z_1$ & $\sigma$ &   &  \\
 & 6035 & $-$87.02 & LHC & 0.30 & P & $Z_2$ & $\sigma$ & +0.7$\pm$0.9 & B \\
 & 6035 & $-$86.98 & RHC & 0.31 & P & $Z_2$ & $\sigma$ &   &  \\
331.542$-$0.067 & 6030 & $-$86.07 & LHC & 0.91 & T & $Z_1$ & $\sigma$ & +2.3$\pm$0.1 & A \\
 & 6030 & $-$85.97 & LIN & 0.16 & T & $Z_1$ & $\pi$ &   &  \\
 & 6030 & $-$85.89 & RHC & 2.04 & T & $Z_1$ & $\sigma$ &   &  \\
 & 6030 & $-$85.63 & LHC & 0.69 & T & $Z_2$ & $\sigma$ & +2.0$\pm$0.1 & A \\
 & 6030 & $-$85.57 & LIN & 0.84 & T & $Z_2$ & $\pi$ &   &  \\
 & 6030 & $-$85.47 & RHC & 1.58 & T & $Z_2$ & $\sigma$ &   &  \\
 & 6035 & $-$86.06 & LHC & 3.70 & T & $Z_3$ & $\sigma$ & +2.1$\pm$0.2 & A \\
 & 6035 & $-$86.04 & LIN & 1.07 & T & $Z_3$ & $\pi$ &   &  \\
 & 6035 & $-$85.94 & RHC & 9.22 & T & $Z_3$ & $\sigma$ &   &  \\
 & 6035 & $-$85.58 & LIN & 0.67 & X &  &  &   &  \\
 & 6035 & $-$85.53 & LHC & 3.16 & T & $Z_4$ & $\sigma$ & +1.6$\pm$0.3 & A \\
 & 6035 & $-$85.49 & LIN & 2.08 & T & $Z_4$ & $\pi$ &   &  \\
 & 6035 & $-$85.44 & RHC & 5.05 & T & $Z_4$ & $\sigma$ &   &  \\
 & 6035 & $-$85.32 & LIN & 0.84 & X &  &  &   &  \\
\hline
\end{tabular} 
\label{fullresultstable}
\end{minipage}
\end{table*}

\begin{table*}
\addtocounter{table}{-1}
\begin{minipage}{180mm}
\small
\centering
\caption{\small Cont.}
\begin{tabular}{l r r l r c c c r c}
\\
\hline
\multicolumn{1}{l}{Source Name}& \multicolumn{1}{c}{Freq} & \multicolumn{1}{c}{$V_{\rm LSR}$} &\multicolumn{1}{c}{Pol}& \multicolumn{1}{c}{$S_p$} & \multicolumn{5}{c}{Zeeman Splitting} \\
\ (~~~$l$,~~~~~~~$b$~~~)   &                       &&& &Type$^{1}$ & Number  & Component& Strength &Reliability\\
\ (~~~$^\circ$~~~~~~~$^\circ$~~~)    &  \multicolumn{1}{c}{(MHz)} & \multicolumn{1}{c}{(\,km\,s$^{-1}$) }&  & (Jy) & && &(mG)  \\
\hline 
333.135$-$0.432b & 6030 & $-$51.34 & RHC & 0.31 & P & $Z_1$ & $\sigma$ & $-$1.6$\pm$0.5 & B \\
 & 6030 & $-$51.21 & LHC & 0.26 & P & $Z_1$ & $\sigma$ &   &  \\
 & 6035 & $-$51.39 & RHC & 3.44 & P & $Z_2$ & $\sigma$ & $-$2.0$\pm$0.3 & B \\
 & 6035 & $-$51.36 & LIN & 0.24 & P & $Z_2$ & $\sigma$ &   &  \\
 & 6035 & $-$51.28 & LHC & 2.60 & P & $Z_2$ & $\sigma$ &   &  \\
333.135$-$0.432c & 6035 & $-$50.61 & RHC & 1.31 & P & $Z_1$ & $\sigma$ & $-$2.9$\pm$0.4 & B \\
 & 6035 & $-$50.45 & LHC & 1.01 & P & $Z_1$ & $\sigma$ &   &  \\
336.941$-$0.156 & 6030 & $-$67.23 & LHC & 0.21 & P & $Z_1$ & $\sigma$ & +2.2$\pm$0.6 & B \\
 & 6030 & $-$67.06 & RHC & 0.12 & P & $Z_1$ & $\sigma$ &   &  \\
337.705$-$0.053a & 6035 & $-$54.27 & LHC & 0.28 & P & $Z_1$ & $\sigma$ & +4.1$\pm$0.9 & B \\
 & 6035 & $-$54.04 & RHC & 0.11 & P & $Z_1$ & $\sigma$ &   &  \\
 & 6035 & $-$53.63 & LHC & 0.54 & P & $Z_2$ & $\sigma$ & +8.9$\pm$1.0 & B \\
 & 6035 & $-$53.13 & RHC & 0.26 & P & $Z_2$ & $\sigma$ &   &  \\
337.705$-$0.053b & 6035 & $-$51.30 & LHC & 0.56 & P & $Z_1$ & $\sigma$ & +9.6$\pm$0.3 & B \\
 & 6035 & $-$50.76 & RHC & 0.82 & P & $Z_1$ & $\sigma$ &   &  \\
 & 6035 & $-$50.66 & LHC & 0.46 & P & $Z_2$ & $\sigma$ & +8.7$\pm$0.3 & B \\
 & 6035 & $-$50.17 & RHC & 0.57 & P & $Z_2$ & $\sigma$ &   &  \\
 & 6035 & $-$49.88 & LHC & 0.54 & P & $Z_3$ & $\sigma$ & +7.9$\pm$0.4 & B \\
 & 6035 & $-$49.44 & RHC & 0.51 & P & $Z_3$ & $\sigma$ &   &  \\
339.884$-$1.259a & 6030 & $-$37.56 & RHC & 10.96 & T & $Z_1$ & $\sigma$ & $-$3.8$\pm$0.1 & A \\
 & 6030 & $-$37.53 & LIN & 1.08 & T & $Z_1$ & $\pi$ &   &  \\
 & 6030 & $-$37.28 & LIN & 1.41 & T & $Z_1$ & $\sigma$ &   &  \\
 & 6030 & $-$37.26 & LHC & 10.37 & T & $Z_1$ & $\sigma$ &   &  \\
 & 6035 & $-$37.71 & LIN & 0.89 &  X&  &  &   &  \\
 & 6035 & $-$37.58 & RHC & 27.41 & T & $Z_2$ & $\sigma$ & $-$3.4$\pm$0.3 & A \\
 & 6035 & $-$37.43 & LIN & 5.39 & T & $Z_2$ & $\pi$ &   &  \\
 & 6035 & $-$37.39 & LHC & 25.16 & T & $Z_2$ & $\sigma$ &   &  \\
340.785$-$0.096a & 6035 & $-$106.44 & RHC & 1.20 & P & $Z_1$ & $\sigma$ & $-$8.0$\pm$0.4 & B \\
 & 6035 & $-$105.99 & LHC & 1.04 & P & $Z_1$ & $\sigma$ &   &  \\
 & 6035 & $-$105.57 & RHC & 0.43 & P & $Z_2$ & $\sigma$ & $-$4.6$\pm$0.9 & B \\
 & 6035 & $-$105.31 & LHC & 0.65 & P & $Z_2$ & $\sigma$ &   &  \\
340.785$-$0.096b & 6035 & $-$102.17 & LHC & 2.17 & P & $Z_1$ & $\sigma$ & +2.5$\pm$0.1 & A \\
 & 6035 & $-$102.03 & RHC & 2.19 & P & $Z_1$ & $\sigma$ &   &  \\
 & 6035 & $-$101.69 & LHC & 0.61 & P & $Z_2$ & $\sigma$ & +2.3$\pm$0.8 & A \\
 & 6035 & $-$101.56 & RHC & 0.55 & P & $Z_2$ & $\sigma$ &   &  \\
 & 6035 & $-$100.63 & LHC & 0.21 & P & $Z_3$ & $\sigma$ & +3.0$\pm$0.9 & A \\
 & 6035 & $-$100.46 & RHC & 0.19 & P & $Z_3$ & $\sigma$ &   &  \\
343.929+0.125 & 6035 & 12.61 & LHC & 0.37 & P & $Z_1$ & $\sigma$ & +3.8$\pm$0.3 & A \\
 & 6035 & 12.82 & RHC & 0.35 & P & $Z_1$ & $\sigma$ &   &  \\
 & 6035 & 13.58 & LHC & 1.11 & P & $Z_2$ & $\sigma$ & +6.8$\pm$0.2 & A \\
 & 6035 & 13.96 & RHC & 0.61 & P & $Z_2$ & $\sigma$ &   &  \\
 & 6035 & 14.19 & LHC & 0.30 & P & $Z_3$ & $\sigma$ & +6.2$\pm$0.4 & A \\
 & 6035 & 14.54 & RHC & 0.24 & P & $Z_3$ & $\sigma$ &   &  \\
345.003$-$0.224 & 6030 & $-$28.42 & LHC & 0.08 & P & $Z_1$ & $\sigma$ & +3.8$\pm$1.2 & A \\
 & 6030 & $-$28.12 & RHC & 0.12 & P & $Z_1$ & $\sigma$ &   &  \\
 & 6030 & $-$27.50 & LHC & 0.19 & P & $Z_2$ & $\sigma$ & +4.9$\pm$0.7 & A \\
 & 6030 & $-$27.11 & RHC & 0.13 & P & $Z_2$ & $\sigma$ &   &  \\
 & 6030 & $-$26.09 & LHC & 0.70 & P & $Z_3$ & $\sigma$ & +4.4$\pm$0.2 & A \\
 & 6030 & $-$25.74 & RHC & 0.93 & P & $Z_3$ & $\sigma$ &   &  \\
 & 6035 & $-$28.37 & LHC & 0.26 & P & $Z_4$ & $\sigma$ & +4.1$\pm$0.9 & A \\
 & 6035 & $-$28.14 & RHC & 0.51 & P & $Z_4$ & $\sigma$ &   &  \\
 & 6035 & $-$27.42 & LHC & 1.00 & T & $Z_5$ & $\sigma$ & +3.9$\pm$0.5 & A \\
 & 6035 & $-$27.33 & LIN & 0.32 & T & $Z_5$ & $\pi$ &   &  \\
 & 6035 & $-$27.20 & RHC & 0.84 & T & $Z_5$ & $\sigma$ &   &  \\
 & 6035 & $-$26.98 & LHC & 0.56 & P & $Z_6$ & $\sigma$ & +3.6$\pm$0.6 & A \\
 & 6035 & $-$26.78 & RHC & 0.70 & P & $Z_6$ & $\sigma$ &   &  \\
 & 6035 & $-$26.53 & LHC & 1.17 & P & $Z_7$ & $\sigma$ & +3.9$\pm$0.3 & A \\
 & 6035 & $-$26.31 & RHC & 1.05 & P & $Z_7$ & $\sigma$ &   &  \\
 & 6035 & $-$26.02 & LHC & 1.75 & P & $Z_8$ & $\sigma$ & +4.3$\pm$1.0 & A \\
 & 6035 & $-$25.78 & RHC & 1.86 & P & $Z_8$ & $\sigma$ &   &  \\
 \hline
\end{tabular} 
\label{fullresultstable}
\end{minipage}
\end{table*}

\begin{table*}
\addtocounter{table}{-1}
\begin{minipage}{180mm}
\small
\centering
\caption{\small Cont.}
\begin{tabular}{l r r l r c c c r c}
\\
\hline
\multicolumn{1}{l}{Source Name}& \multicolumn{1}{c}{Freq} & \multicolumn{1}{c}{$V_{\rm LSR}$} &\multicolumn{1}{c}{Pol}& \multicolumn{1}{c}{$S_p$} & \multicolumn{5}{c}{Zeeman Splitting} \\
\ (~~~$l$,~~~~~~~$b$~~~)   &                       &&& &Type$^{1}$ & Number  & Component& Strength &Reliability\\
\ (~~~$^\circ$~~~~~~~$^\circ$~~~)    &  \multicolumn{1}{c}{(MHz)} & \multicolumn{1}{c}{(\,km\,s$^{-1}$) }&  & (Jy) & && &(mG)  \\
\hline 

345.009+1.792 & 6035 & $-$17.83 & LHC & 0.26 & P & $Z_1$ & $\sigma$ & +1.6$\pm$0.3 & B \\
 & 6035 & $-$17.74 & RHC & 0.36 & P & $Z_1$ & $\sigma$ &   &  \\
345.487+0.314 & 6035 & $-$22.19 & LHC & 1.98 & P & $Z_1$ & $\sigma$ & +0.5$\pm$0.2 & B \\
 & 6035 & $-$22.16 & RHC & 0.96 & P & $Z_1$ & $\sigma$ &   &  \\
345.698$-$0.090a & 6030 & $-$4.96 & LHC & 1.14 & P & $Z_1$ & $\sigma$ & +1.1$\pm$0.1 & A \\
 & 6030 & $-$4.87 & RHC & 1.05 & P & $Z_1$ & $\sigma$ &   &  \\
 & 6030 & $-$3.58 & LHC & 0.17 & P & $Z_2$ & $\sigma$ & +0.6$\pm$0.7 & B \\
 & 6030 & $-$3.53 & RHC & 0.16 & P & $Z_2$ & $\sigma$ &   &  \\
 & 6035 & $-$4.92 & LHC & 1.14 & P & $Z_3$ & $\sigma$ & +1.3$\pm$0.3 & A \\
 & 6035 & $-$4.85 & RHC & 1.14 & P & $Z_3$ & $\sigma$ &   &  \\
345.698$-$0.090b & 6035 & $-$6.50 & RHC & 2.32 & P & $Z_1$ & $\sigma$ & $-$0.5$\pm$0.1 & A \\
 & 6035 & $-$6.47 & LHC & 2.31 & P & $Z_1$ & $\sigma$ &   &  \\
347.628+0.148 & 6030 & $-$97.04 & LHC & 0.67 & P & $Z_1$ & $\sigma$ & +4.6$\pm$0.1 & A \\
 & 6030 & $-$96.68 & RHC & 1.06 & P & $Z_1$ & $\sigma$ &   &  \\
 & 6035 & $-$96.95 & LHC & 2.02 & P & $Z_2$ & $\sigma$ & +6.4$\pm$4.6 & B \\
 & 6035 & $-$96.59 & RHC & 4.56 & P & $Z_2$ & $\sigma$ &   &  \\
351.417+0.645a & 6030 & $-$10.85 & LIN & 1.03 &  X &  &  &   &  \\
 & 6030 & $-$10.74 & RHC & 8.57 & T & $Z_1$ & $\sigma$ & $-$3.8$\pm$0.1 & A \\
 & 6030 & $-$10.56 & LIN & 0.98 & T & $Z_1$ & $\pi$ &   &  \\
 & 6030 & $-$10.35 & LIN & 1.27 & T & $Z_1$ & $\sigma$ &   &  \\
 & 6030 & $-$10.34 & LHC & 12.26 & T & $Z_1$ & $\sigma$ &   &  \\
 & 6035 & $-$11.22 & RHC & 13.20 & P  & $Z_2$  &$\sigma$  & $-$4.5$\pm$0.3  & B  \\
 & 6035 & $-$10.97 & LHC & 16.81 &  P & $Z_2$  &  $\sigma$&   &  \\
 & 6035 & $-$10.58 & RHC & 124.41 & T & $Z_3$ & $\sigma$ & $-$4.8$\pm$0.1 & A \\
 & 6035 & $-$10.44 & LIN & 14.43 & T & $Z_3$ & $\pi$ &   &  \\
 & 6035 & $-$10.31 & LHC & 153.00 & T & $Z_3$ & $\sigma$ &   &  \\
 & 6035 & $-$10.23 & LIN & 7.40 & X &  &  &   &  \\
351.417+0.645b & 6030 & $-$8.22 & RHC & 0.22 & P & $Z_1$ & $\sigma$ & $-$4.3$\pm$0.3 & A \\
 & 6030 & $-$7.88 & LHC & 0.19 & P & $Z_1$ & $\sigma$ &   &  \\
 & 6030 & $-$7.80 & RHC & 0.58 & P & $Z_2$ & $\sigma$ & $-$5.2$\pm$0.1 & A \\
 & 6030 & $-$7.39 & LHC & 0.58 & P & $Z_2$ & $\sigma$ &   &  \\
 & 6035 & $-$8.23 & RHC & 0.67 & P & $Z_3$ & $\sigma$ & $-$5.7$\pm$0.7 & A \\
 & 6035 & $-$7.91 & LHC & 0.85 & P & $Z_3$ & $\sigma$ &   &  \\
 & 6035 & $-$7.73 & RHC & 2.30 & P & $Z_4$ & $\sigma$ & $-$5.2$\pm$0.1 & A \\
 & 6035 & $-$7.44 & LHC & 2.39 & P & $Z_4$ & $\sigma$ &   &  \\
351.417+0.645c & 6035 & $-$5.22 & LIN & 0.30 & P & $Z_1$ & $\sigma$ &  +0.7$\pm$3.6 & B \\
 & 6035 & $-$5.22 & LHC & 0.40 & P & $Z_1$ & $\sigma$ & &  \\
 & 6035 & $-$5.18 & RHC & 0.45 & P & $Z_1$ & $\sigma$ &   &  \\
351.417+0.645d & 6030 & $-$8.80 & RHC & 0.56 & P & $Z_1$ & $\sigma$ & $-$4.1$\pm$0.3 & A \\
 & 6030 & $-$8.48 & LHC & 0.60 & P & $Z_1$ & $\sigma$ &   &  \\
 & 6035 & $-$8.79 & RHC & 3.86 & T & $Z_2$ & $\sigma$ & $-$3.6$\pm$0.4 & A \\
 & 6035 & $-$8.68 & LIN & 0.67 & T & $Z_2$ & $\pi$ &   &  \\
 & 6035 & $-$8.59 & LHC & 4.30 & T & $Z_2$ & $\sigma$ &   &  \\
353.410$-$0.360 & 6030 & $-$22.46 & RHC & 0.83 & P & $Z_1$ & $\sigma$ & $-$6.3$\pm$0.1 & A \\
 & 6030 & $-$21.96 & LHC & 0.90 & P & $Z_1$ & $\sigma$ &   &  \\
 & 6035 & $-$22.57 & RHC & 3.86 & P & $Z_2$ & $\sigma$ & $-$10.2$\pm$0.3 & A \\
 & 6035 & $-$22.28 & RHC & 5.53 & P & $Z_3$ & $\sigma$ & $-$10.4$\pm$0.3 & A \\
 & 6035 & $-$22.00 & LHC & 5.60 & P & $Z_2$ & $\sigma$ &   &  \\
 & 6035 & $-$21.70 & LHC & 3.37 & P & $Z_3$ & $\sigma$ &   &  \\
 & 6035 & $-$20.79 & RHC & 5.52 & P & $Z_4$ & $\sigma$ & $-$3.6$\pm$0.1 & B \\
 & 6035 & $-$20.65 & LHC & 10.43 & P & $Z_4$ & $\sigma$ &   &  \\
355.344+0.147 & 6030 & 17.78 & RHC & 1.03 & P & $Z_1$ & $\sigma$ & $-$5.2$\pm$0.2 & A \\
 & 6030 & 18.19 & LHC & 0.89 & P & $Z_1$ & $\sigma$ &   &  \\
 & 6030 & 18.36 & RHC & 0.33 & P & $Z_2$ & $\sigma$ & $-$4.9$\pm$0.3 & A \\
 & 6030 & 18.75 & LHC & 0.32 & P & $Z_2$ & $\sigma$ &   &  \\
 & 6035 & 17.74 & RHC & 2.01 & P & $Z_3$ & $\sigma$ & $-$5.4$\pm$0.1 & A \\
 & 6035 & 18.04 & LHC & 2.12 & P & $Z_3$ & $\sigma$ &   &  \\
 & 6035 & 18.43 & RHC & 0.58 & P & $Z_4$ & $\sigma$ & $-$4.6$\pm$0.3 & A \\
 & 6035 & 18.69 & LHC & 0.59 & P & $Z_4$ & $\sigma$ &   &  \\
\hline
\end{tabular} 
\label{fullresultstable}
\end{minipage}
\end{table*}

\begin{table*}
\addtocounter{table}{-1}
\begin{minipage}{180mm}
\small
\centering
\caption{\small Cont.}
\begin{tabular}{l r r l r c c c r c}
\\
\hline
\multicolumn{1}{l}{Source Name}& \multicolumn{1}{c}{Freq} & \multicolumn{1}{c}{$V_{\rm LSR}$} &\multicolumn{1}{c}{Pol}& \multicolumn{1}{c}{$S_p$} & \multicolumn{5}{c}{Zeeman Splitting} \\
\ (~~~$l$,~~~~~~~$b$~~~)   &                       &&& &Type$^{1}$ & Number  & Component& Strength &Reliability\\
\ (~~~$^\circ$~~~~~~~$^\circ$~~~)    &  \multicolumn{1}{c}{(MHz)} & \multicolumn{1}{c}{(\,km\,s$^{-1}$) }&  & (Jy) & && &(mG)  \\
\hline 
000.665$-$0.036 & 6035 & 60.69 & RHC & 0.17 & P & $Z_1$ & $\sigma$ & $-$0.4$\pm$1.0 & B \\
 & 6035 & 60.71 & LHC & 0.12 & P & $Z_1$ & $\sigma$ &   &  \\
000.666$-$0.029 & 6030 & 70.20 & RHC & 0.60 & P & $Z_1$ & $\sigma$ & $-$5.3$\pm$0.2 & A \\
 & 6030 & 70.62 & LHC & 0.43 & P & $Z_1$ & $\sigma$ &   &  \\
 & 6030 & 72.16 & LHC & 0.25 & P & $Z_2$ & $\sigma$ & +4.2$\pm$0.3 & A \\
 & 6030 & 72.49 & RHC & 0.25 & P & $Z_2$ & $\sigma$ &   &  \\
 & 6035 & 69.12 & RHC & 0.39 & P & $Z_3$ & $\sigma$ & $-$4.5$\pm$1.2 & A \\
 & 6035 & 69.37 & LHC & 0.41 & P & $Z_3$ & $\sigma$ &   &  \\
 & 6035 & 70.83 & RHC & 2.18 & P & $Z_4$ & $\sigma$ & $-$3.8$\pm$0.4 & B \\
 & 6035 & 70.85 & LIN & 0.38 & P & $Z_4$ & $\sigma$ &   &  \\
 & 6035 & 71.06 & LIN & 0.41 & P & $Z_4$ & $\sigma$ &   &  \\
 & 6035 & 71.07 & LHC & 2.56 & P & $Z_4$ & $\sigma$ &   &  \\
 & 6035 & 72.04 & LIN & 0.93 & X &  &  &   &  \\
 & 6035 & 72.16 & LHC & 7.81 & P & $Z_5$ & $\sigma$  & $+$4.8$\pm$0.3  & B  \\
 & 6035 & 72.43 & RHC & 3.69 & P & $Z_5$ & $\sigma$ &   &  \\
000.666$-$0.035 & 6035 & 63.36 & RHC & 0.32 & P & $Z_1$ & $\sigma$ & $-$4.5$\pm$0.5 & A \\
 & 6035 & 63.61 & LHC & 0.34 & P & $Z_1$ & $\sigma$ &   &  \\
 & 6035 & 64.66 & RHC & 0.34 & P & $Z_2$ & $\sigma$ & $-$7.5$\pm$0.3 & A \\
 & 6035 & 65.08 & LHC & 0.52 & P & $Z_2$ & $\sigma$ &   &  \\
 & 6035 & 67.13 & RHC & 0.77 & P & $Z_3$ & $\sigma$ & $-$8.0$\pm$3.0 & A \\
 & 6035 & 67.58 & LHC & 0.80 & P & $Z_3$ & $\sigma$ &   &  \\
000.666$-$0.036 & 6035 & 62.18 & LHC & 0.40 & P & $Z_1$ & $\sigma$ & +4.3$\pm$1.0 & B \\
 & 6035 & 62.42 & RHC & 0.32 & P & $Z_1$ & $\sigma$ &   &  \\
010.959+0.023 & 6035 & 24.75 & RHC & 0.41 & P & $Z_1$ & $\sigma$ & $-$6.1$\pm$0.4 & B \\
 & 6035 & 25.09 & LHC & 0.22 & P & $Z_1$ & $\sigma$ &   &  \\
011.034+0.062 & 6035 & 23.23 & RHC & 0.52 & P & $Z_1$ & $\sigma$ & $-$6.1$\pm$0.7 & B \\
 & 6035 & 23.57 & LHC & 0.46 & P & $Z_1$ & $\sigma$ &   &  \\
011.904$-$0.141 & 6035 & 41.61 & LHC & 0.49 & P & $Z_1$ & $\sigma$ & +1.6$\pm$0.4 & B \\
 & 6035 & 41.70 & RHC & 0.44 & P & $Z_1$ & $\sigma$ &   &  \\
 & 6035 & 42.88 & LIN & 0.97 & X &  &  &   &  \\
015.034$-$0.677 & 6030 & 21.49 & LHC & 2.27 & T & $Z_1$ & $\sigma$ & +0.5$\pm$0.1 & B \\
 & 6030 & 21.52 & LIN & 1.51 & T & $Z_1$ & $\pi$ &   &  \\
 & 6030 & 21.53 & RHC & 1.28 & T & $Z_1$ & $\sigma$ &   &  \\
 & 6035 & 21.41 & LHC & 18.16 & P & $Z_2$ & $\sigma$ & +0.9$\pm$0.1 & B \\
 & 6035 & 21.46 & RHC & 16.63 & P & $Z_2$ & $\sigma$ &   &  \\
 & 6035 & 21.47 & LIN & 6.95 & P & $Z_2$ & $\sigma$ &   &  \\
 & 6035 & 22.54 & LIN & 2.04 & X  &  &  &   &  \\
 & 6035 & 22.60 & RHC & 8.79 & P & $Z_3$ & $\sigma$ & $-$0.2$\pm$0.1 & B \\
 & 6035 & 22.61 & LHC & 10.06 & P & $Z_3$ & $\sigma$ &   &  \\
 & 6035 & 22.71 & LIN & 1.91 &  X&  &  &   &  \\
 & 6035 & 23.28 & RHC & 0.58 & P & $Z_4$ & $\sigma$ & $-$5.4$\pm$1.0 & B \\
 & 6035 & 23.58 & LHC & 1.33 & P & $Z_4$ & $\sigma$ &   &  \\
\hline
\end{tabular} 
\label{fullresultstable}
\end{minipage}
\end{table*}

%%%%%%%%%%%%%%%%%%%%%%%%%%%%%%%%%%%%%%%%%%%%%%%%%%%%%%%%%%%%%%%%%%%%%%%%%%%%%%%%%%

\section{Polarised characteristics}\label{polarisation_section}
The polarization characteristics of the 70 sites of exOH and methanol maser emission are shown in Tables \ref{resotable} and \ref{fullresultstable}. There are 10 sites of exclusively exOH emission, 32 sites of exclusively methanol emission, and 28 sites with both exOH and methanol emission. 

Of the 38 exOH emission sites, 20 exhibit detectable linearly polarised emission ($>$0.10\,Jy), and seven of these sources have a linear polarisation percentage greater than 50\% (309.921+0.479, 311.596--0.398, 345.003-0.224, 345.009+1.792, 351.417+0.654, 11.904--0.141 and 15.034--0.677). All 38 sites of exOH emission exhibit detectable circularly polarised emission, with typical percentages ranging from 50 to 100\%.

Of the 60 sites of methanol maser emission, 22 exhibit detectable linear polarisation. 16 of these have associated exOH, and all but one (340.785--0.096) show linearly polarised features in both species. There is no indication of correspondence between the existence of linear features in methanol and the strength of linear features in exOH (only three of the 16 have exOH features with $>$50\,\% linear polarisation). None of the methanol sites exhibit detectable circular polarisation.

\subsubsection{Velocity range of emission}
The velocity range of polarised features at the sites of maser emission show a small spread in velocities for the exOH, with the median of 6030-MHz and 6035-MHz 3.3\,km\,s$^{-1}$ and 4.5\,km\,s$^{-1}$, respectively. Methanol demonstrates a slightly larger median spread of 5.3\,km\,s$^{-1}$ and, with the exception of  the source 351.417+0.645, is the only transition to have features spread over more than 10\,km\,s$^{-1}$ in velocity. The velocity spreads are fully consistent with those seen in previous studies \citep{caswell95c,caswell10mmb1,green10mmb2,caswell11mmb3,green12mmb4}.

\subsection{Zeeman pairs and triplets}\label{zeeman_section}
\begin{figure}
\begin{center}
\renewcommand{\baselinestretch}{1.1}
\includegraphics[width=8.5cm]{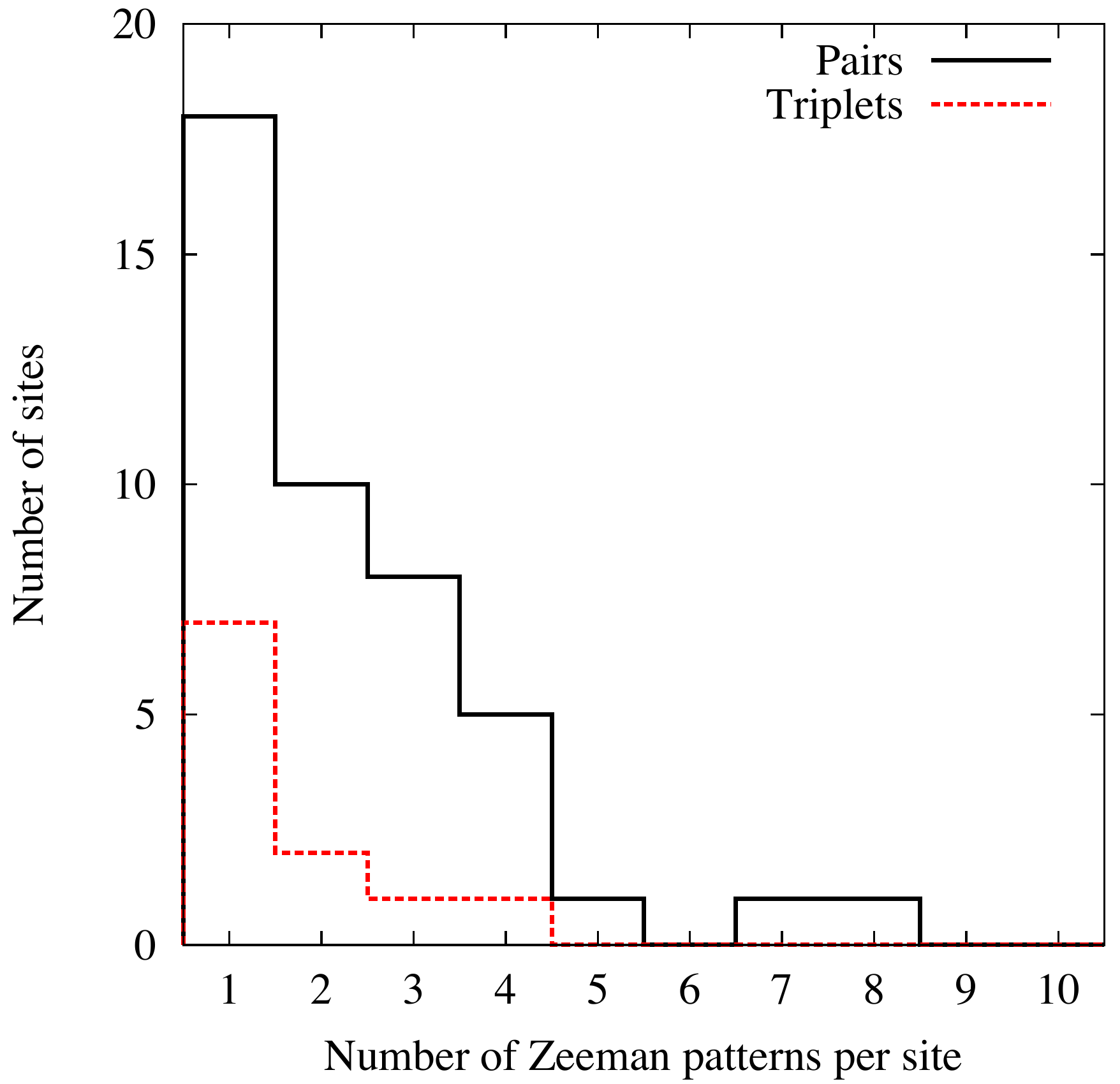}  
\caption{\small Number of Zeeman pairs (solid black) and triplets (dashed red) per site of maser emission.}
\label{patternbreakdown}
\end{center}
\end{figure}

Following the criterion outlined in Section \ref{obs_section}, we identify 112 Zeeman patterns in the exOH emission (55 class `A', 57 class `B', see Table  \ref{fullresultstable}); of these, 18 are Zeeman triplet candidates (16\% of total Zeeman identifications)\footnote{Two of the Zeeman pairs may actually be triplets (identified by `P/T' in Table \ref{fullresultstable}), but could not be ascertained with a high enough degree of confidence, and so are treated as pairs.}. The number of pair and triplet identifications per site is shown in Figure\,\ref{patternbreakdown}, with a strong tendency for only one identification per maser site (both distributions peaking in the lowest bin), although 50\% of the sites with Zeeman pairs have more than two pairs identified.

Previously there has been only one ground-state OH Zeeman triplet identification, first suggested from observations with the MERLIN array by \citet{busaba02} of W75N; this identification was later shown by \citet{fish06b}, with Very Long Baseline Array (VLBA) observations, to be features spatially separated by 110 AU. Our number of lower resolution triplets is an upper limit, with all potentially being resolved with higher resolution VLBI measurements, however we can confidently conclude that the linearly polarised $\pi$ component has a similar propensity at the high frequency maser transitions as at the ground-state transitions (recently demonstrated by \citealt{caswell13,caswell14}, as detailed in Section \ref{iso_section}). Especially well defined triplets include 300.969+1.148 and 323.459--0.079a at 6035 MHz, and 339.884--1.259a and 351.417+0.645a at both transitions, the latter pair identified as `text book' examples, exhibiting the simple split spectra and coincident features (with the corresponding difference in splitting factor).

For the 18 Zeeman triplets, three have both $\pi$ and $\sigma$ LIN components identified (323.459--0.079a, 339.884--1.259a, 351.417+0.645a, see Table \ref{fullresultstable}), and two of these (323.459--0.079a and 351.417+0.645a) clearly show the signature of orthogonal polarization in the Stokes Q and U in the spectra in Appendix B (online only) (it is suggested for 339.884--1.259a as well but is spectrally blended).

\subsubsection{Ratio of component peak flux densities}
\begin{figure}
\begin{center}
\renewcommand{\baselinestretch}{1.1}
\includegraphics[width=8.5cm]{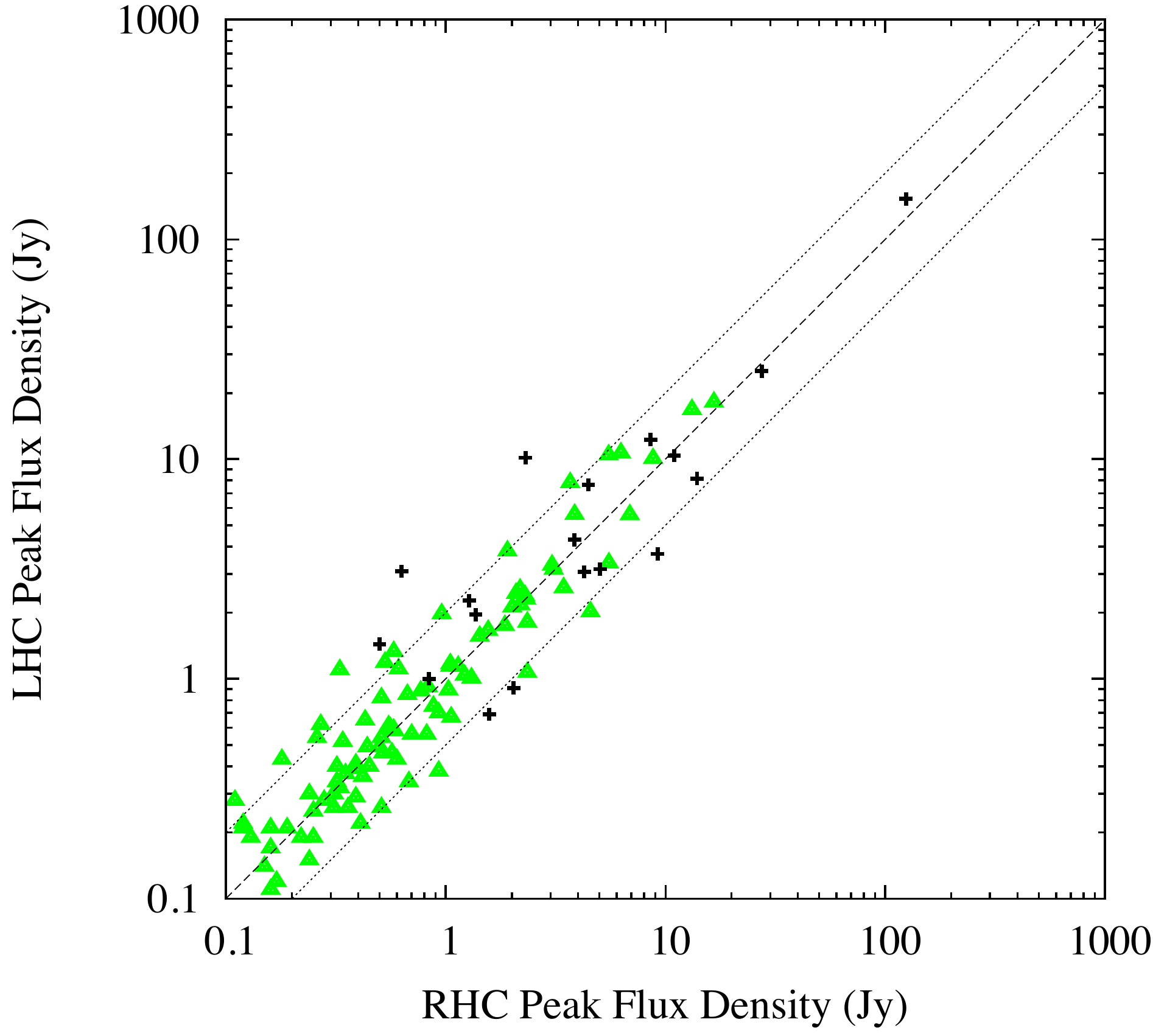}  
\caption{\small Ratios in peak flux density for paired $\sigma$ components for Zeeman pairs (green triangles) and Zeeman triplets (black crosses). Dashed line shows 1:1 ratio, dotted lines a factor of two variation.}
\label{FluxRatios}
\end{center}
\end{figure}

The ratios of the peak flux densities of the RHC and LHC components for Zeeman pairs and triplets are displayed in Figure \ref{FluxRatios}, and demonstrate a median of 1.01 and 1.15 for the pairs and triplets respectively. The ratios range from 0.3 to 2.4, with only $\sim$13\% of the pairs (12 of the 94) showing a ratio of fluxes greater than two. In comparison, $\sim$33\% of the triplets (6 of the 18) show a ratio of fluxes greater than two. 
The RHC and LHC components of the triplet candidates tend to be brighter than those of the pairs, with the median of the averaged RHC and LHC components for the triplets being 4.1 Jy, compared with 0.6 Jy for the pairs. The peak flux density of the $\pi$ component of the triplets varies from 3 to 85\% of the averaged circular component flux, but has a median value of 16\%. Hence, if the flux density ratios between $\sigma$s and $\pi$s were comparable for our pair candidates, the average flux density of 0.6 Jy for the pairs would imply the $\pi$ component would have an average peak flux density of $\sim$0.13\,Jy, just below our 3$\sigma$ detection limit. This therefore implies that we may only be detecting the brightest of the Zeeman triplets.

\subsubsection{Isolated $\pi$ components}\label{iso_section}
\citet{caswell13,caswell14} identify isolated $\pi$ components, based on the criterion of strongly linearly polarized features at both the 1665- and 1667-MHz ground-state transitions with no nearby (spatially or spectrally) circularly polarised components (see for example 339.282+0.136 in figure 1 of \citealt{caswell14}). Within the current sample of exOH maser transitions there were no examples of isolated $\pi$ emission. There are however 14 linearly polarized components which are not readily attributable to Zeeman patterns, i.e. strong linear polarization with nearby unassociated circular polarization (denoted with `X's in Table \ref{fullresultstable}, a good example is 309.921+0.479c, apparent in the spectrum in  Appendix B (online only)).

\subsection{Implied magnetic fields}
\begin{figure}
\begin{center}
\renewcommand{\baselinestretch}{1.1}
\includegraphics[width=8cm]{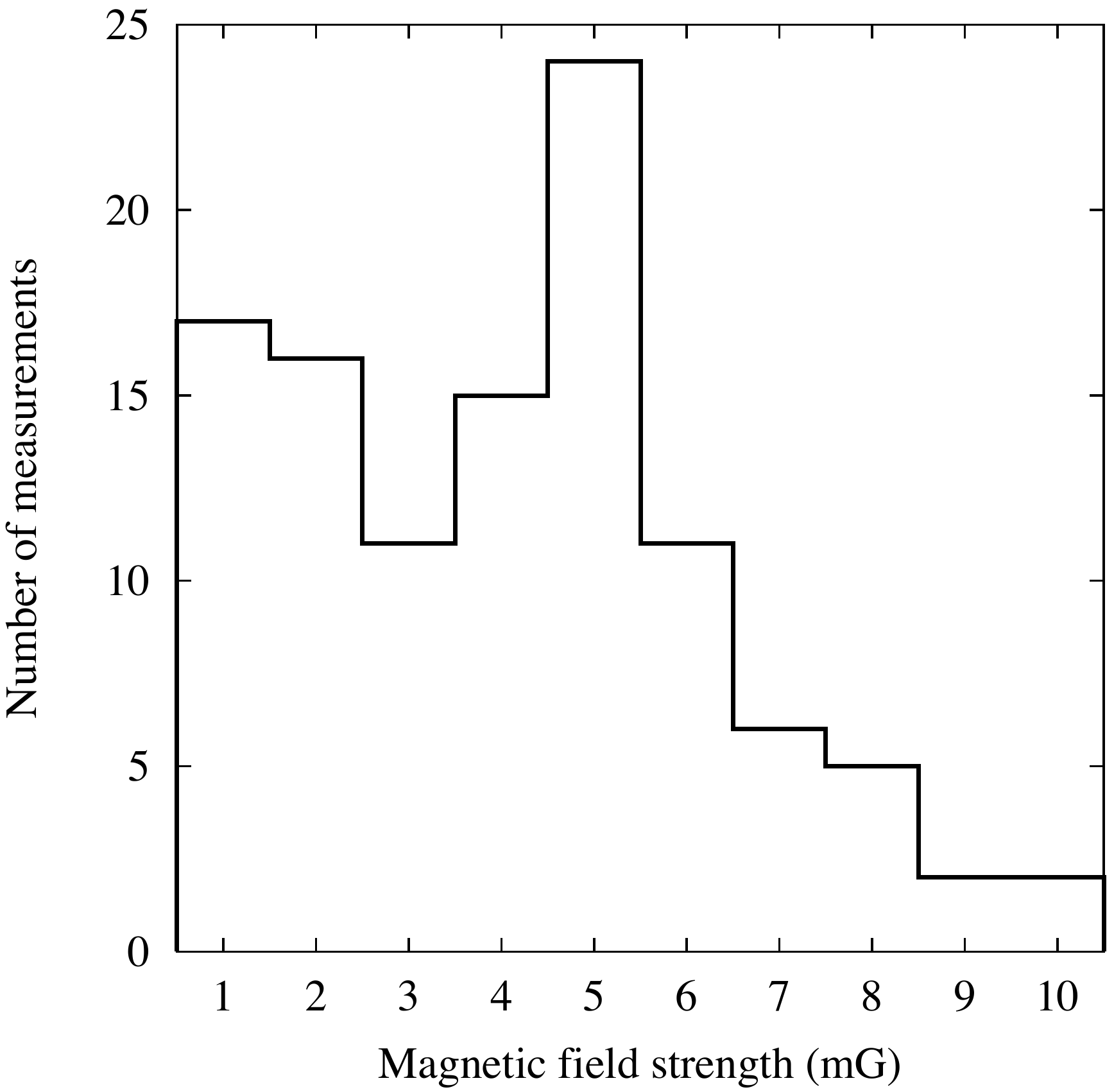}  
\caption{\small Absolute magnetic field strength determined from Zeeman pairs at both 6030 and 6035 MHz combined. Three sources have strengths exceeding 10mG.}
\label{magfieldstrengths}
\end{center}
\end{figure}

From the Zeeman pairs and triplets we can estimate the magnetic field strength. We adopt splitting factors of: 0.079 \,km\,s$^{-1}$ mG$^{-1}$ for 6030 MHz and  0.056 \,km\,s$^{-1}$ mG$^{-1}$ for 6035 MHz \citep{yen69,zuckerman72}. Consequently, we find the magnetic field strengths varying between $-$10.4 and +11.4 mG, with absolute field strengths varying between 0.2\,mG and 11.4\,mG, with a median of 3.9\,mG. The distribution is shown in Figure\,\ref{magfieldstrengths}, with over 50\% of the measurements below 3\,mG. Focussing purely on the subset of Zeeman triplet candidates we find the absolute field strengths range from 0.5\,mG to 5\,mG with a median of 3.7\,mG. We therefore find no strong dependence of the magnetic field strength on the presence of a $\pi$ component.

\begin{figure}
\begin{center}
\renewcommand{\baselinestretch}{1.1}
\includegraphics[width=8.5cm]{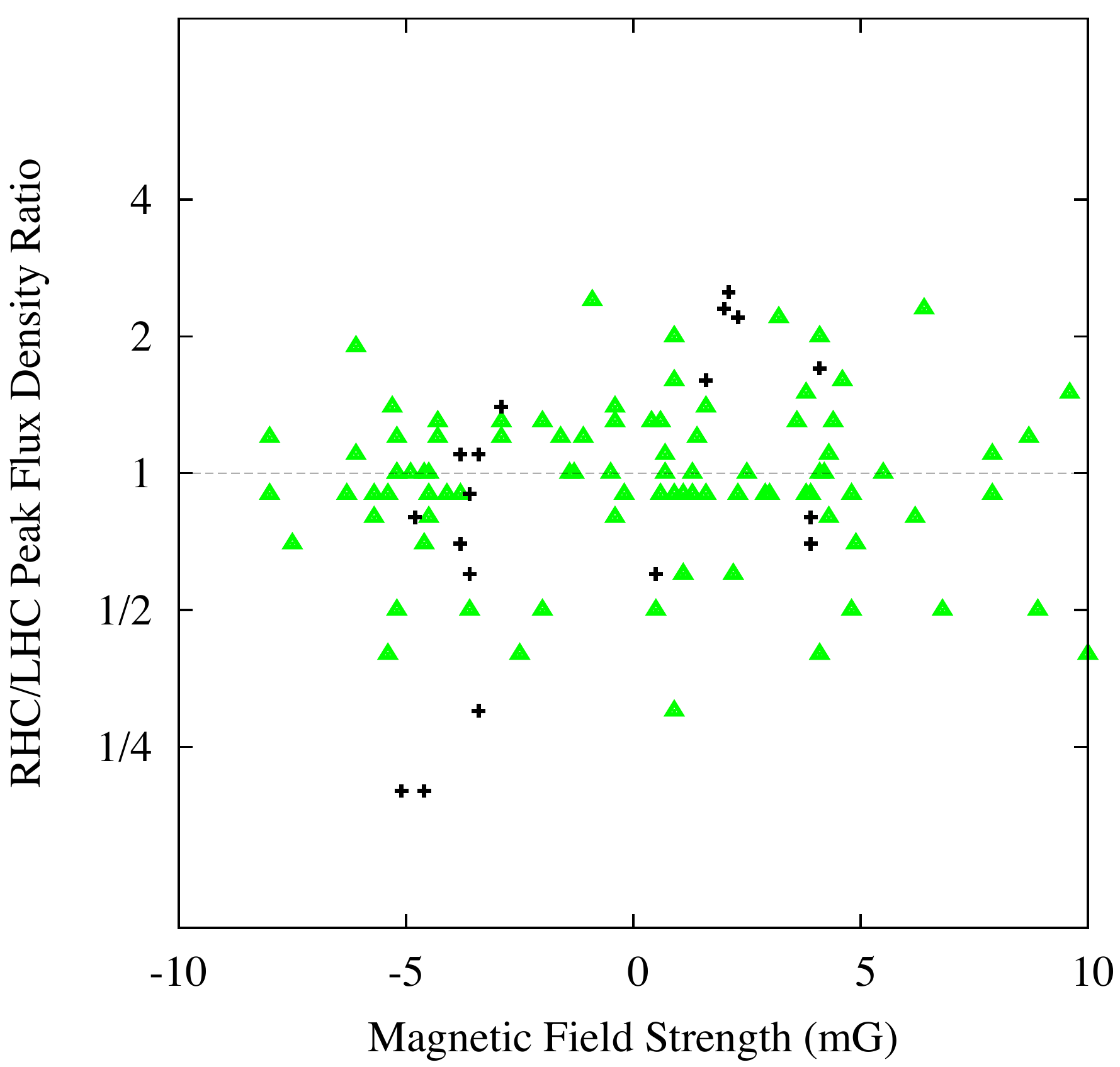}  
\caption{\small Variation of magnetic field strength with ratio of RHC and LHC peak flux densities for Zeeman pairs (green triangles) and Zeeman triplets (black crosses). No trend is visible for the Zeeman pairs, but the Zeeman triplets show a marginal trend (a least-squares linear fit has an error of 40\%) for positive fields to have a larger ratio (i.e. stronger RHC component), negative fields a smaller ratio (i.e. stronger LHC component).}
\label{fieldvsratios}
\end{center}
\end{figure}

Figure\,\ref{fieldvsratios} demonstrates that there is no clear relationship between the magnetic field strength and the flux density ratio of the RHC and LHC components of the Zeeman pairs discussed in the previous section. The plot suggests a marginal trend for the Zeeman triplets (a least-squares linear fit has an error of 40\%), with positive fields producing a larger ratio (five of eight triplets have brighter RHC) and negative fields producing a smaller ratio (seven of 10 triplets have brighter LHC). Although the low statistics prevent any significance being applied, it is interesting to note that for a positive magnetic field, directed away from us (with the RHC at the higher velocity), the component that is equivalently shifted away from us (redshifted), is the brighter of the two.

We found five candidates for internal field reversals (331.511--0.102, 345.698--0.090, 351.417+0.645, 0.666-0.029, and 15.034--0.677), although two (331.511--0.102, 351.417+0.645) are consistent to within the errors. There are also nine Zeeman pairs associated with the Sgr B2 complex, six of which indicate a negative field of 4 to 8 mG strength, three a positive field of $\sim$4\,mG.

\subsubsection{Large-scale magnetic field direction}
\begin{figure}
\begin{center}
\renewcommand{\baselinestretch}{1.1}
\includegraphics[width=7.5cm]{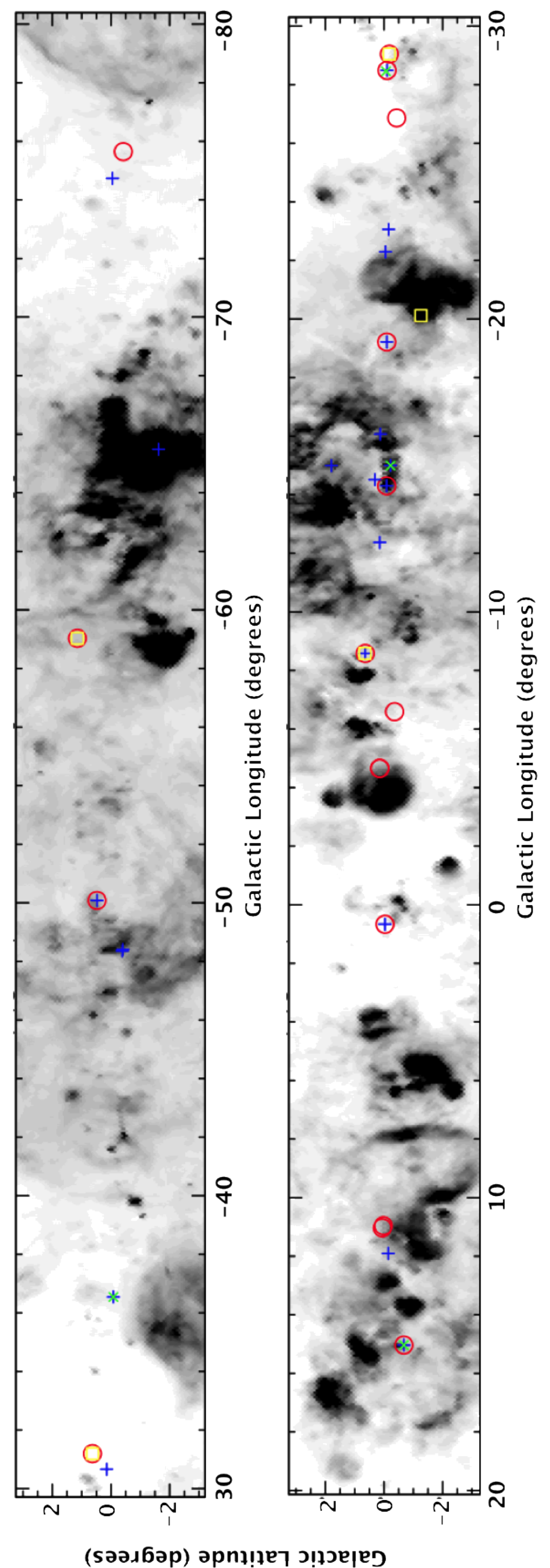}  
\caption{\small Distribution of maser field directions overlaid on the composite H$\alpha$ emission of \citet{finkbeiner03} using the Southern H-Alpha Sky Survey Atlas data of \citet{gaustad01}. The H$\alpha$ emission indicates regions of ionized hydrogen.  Blue crosses are positive fields from Zeeman pairs, rotated green crosses are positive fields from Zeeman triplets, red circles negative fields from Zeeman pairs, yellow boxes negative fields from Zeeman triplets.}
\label{lbbplot}
\end{center}
\end{figure}

\begin{figure}
\begin{center}
\renewcommand{\baselinestretch}{1.1}
\includegraphics[width=8.5cm]{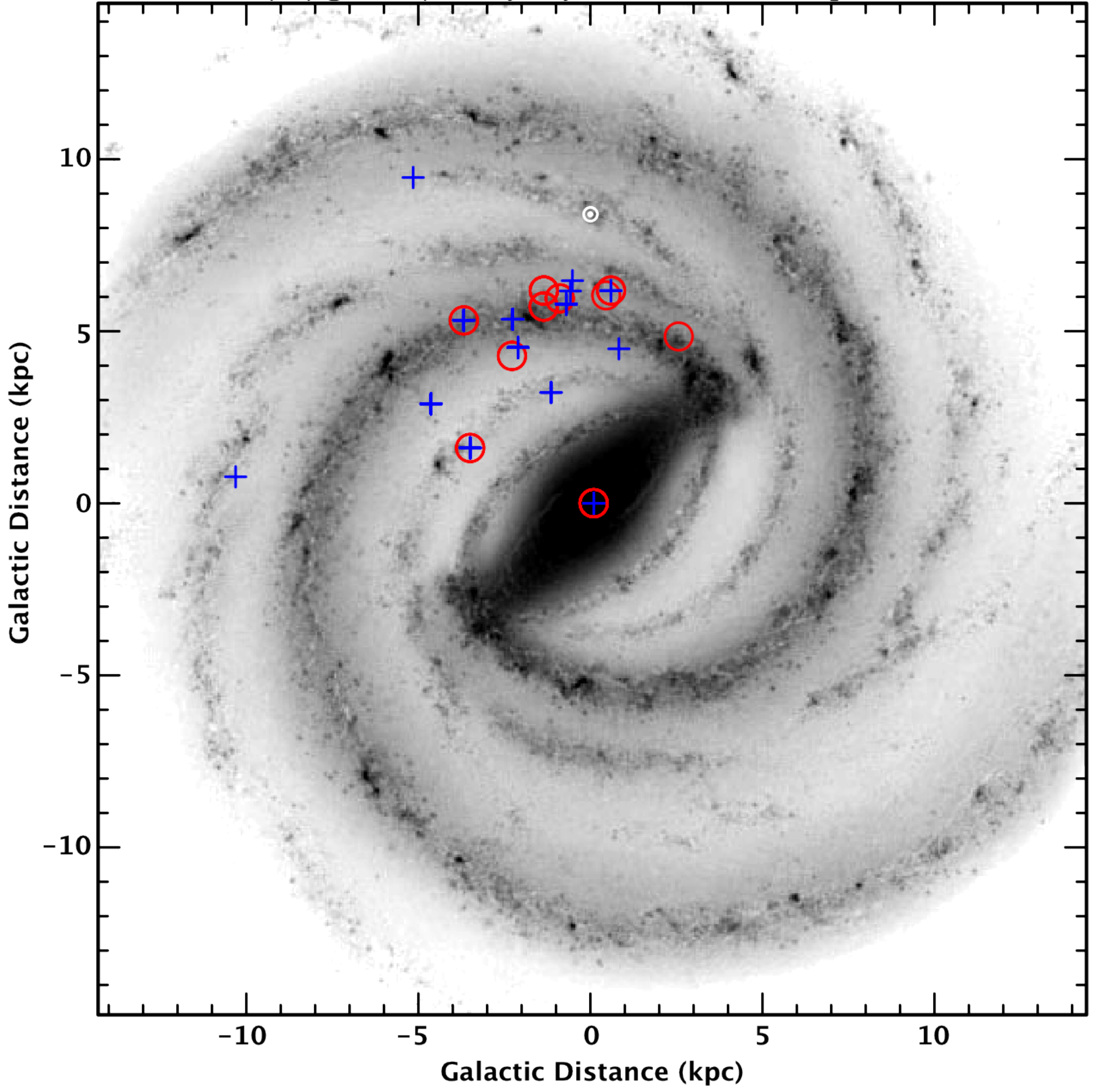}  
\caption{\small Magnetic field direction (blue pluses corresponding to positive magnetic fields, away from the Sun,  and red circles negative magnetic fields, towards the Sun) as inferred from the current Zeeman splitting measurements overlaid on the informed artist impression of the Milky Way (R. Hurt: NASA/JPL- Caltech/SSC).}
\label{MWTopDown}
\end{center}
\end{figure}

In Figure \ref{lbbplot} we show the magnetic field orientations in the plane of the Galaxy, in comparison to H-$\alpha$ emission (indicative of local compact H{\sc ii} regions).  Recent polarimetric observations of ground-state OH maser emission \citep{green12magmo0} have reinforced the potential for tracing large-scale magnetic fields through maser emission. Whilst the ground-state OH masers are typically found on the edges of ultra-compact H{\sc ii} regions, their exOH cousins are found closer to (or coincident with) the methanol emission, typically located nearer to the forming high-mass star(s). This means unlike the ground-state masers, the excited-state masers may be more strongly influenced by the small-scale magnetic field structure of the star(s), rather than any larger scale field. For 23 of the 70 sites (67 of the Zeeman pairs), a distance was available (using \citealt{green11b}, references therein) and we utilise that to show the magnetic field orientations with respect to their Galactic location (Figure\,\ref{MWTopDown}). There is no clear correspondence of field direction with spiral arm features. 

\subsection{Effect of internal Faraday rotation on $\pi$ emission}
We find a small, but significant, fraction of Zeeman patterns exhibit $\pi$ components, similar to the ground-state transitions \citep{caswell13,caswell14}. As such, we revisit the concept of suppressing the $\pi$ component through Faraday rotation.  This concept speculates that Faraday rotation within the maser medium may contribute to the low incidence of observed $\pi$ components). 
The electric field vector of the emission will rotate in the presence of a magnetic field parallel to the direction of propagation. As stated in the Introduction, the $\pi$ component is seen (linearly polarised) when the field is perpendicular and is not seen when the field is exactly parallel. Thus Faraday rotation will only impact on cases where the angle between the field and propagation is {\it between} 0$^{\circ}$ and 90$^{\circ}$, and will have the highest impact when the $\pi$ component is least likely to be observed.
From maser theory \citep[e.g.][]{elitzur92} the expected path length is of the order of 10$^{12}$\,m (with the ultra-compact H{\sc ii} region containing it approximately 10$^{15}$\,m or more). The magnetic fields that are traced by the masers are expected to be of the order of a few mG and the electron density of an ultra-compact H{\sc ii} region is approximately 10\,cm$^{-3}$. Using the standard equation for Faraday rotation \citep[see][]{brentjens05}, this equates to a rotation measure of the order of 25 rad\,m$^{-2}$. This is insufficient to depolarise the $\pi$ component signal within a channel ($\sim$500\,Hz). Coupled with the orientation considerations, this makes depolarisation of the $\pi$ component due to Faraday rotation very unlikely\footnote{As an aside, the Zeeman patterns with and without $\pi$ components can be compared with the H-$\alpha$ emission in Figure \ref{lbbplot} and we see no correspondence between the presence of a $\pi$ component and the intensity of H-$\alpha$ (indicative, where not heavily extincted, of the local ultra-compact H{\sc ii} regions).}. 

\subsection{Magnetic field orientation and magnetic beaming}
\citet{gray94} postulated that the lack of $\pi$ components was due to preferential beaming of maser emission parallel to the magnetic field. This was partially revisited recently by \citet{green14} with respect to a quantum mechanical derivation of the Zeeman splitting spectrum in relation to field orientation. Equations 2 and 3 in  \citet{green14} (reproduced from \citealt{goldreich73part2}) demonstrate the dependence of the components with respect to the angle between the maser propagation and the magnetic field: a purely tangential field produces no circular $\sigma$ components; and a purely parallel field produces no linear $\pi$ component. 

To consider in a simple sense what this equates to observationally, we assume that the case of a purely parallel field (and therefore intrinsically no $\pi$ emission) is equivalent in our data to the unresolved ATCA beam, which at the exOH frequencies, subtends a solid angle of 0.002 steradians. 
In the preliminary analysis of \citet{gray94} there is severe suppression of emission 16 degrees (a corresponding solid angle of 0.24 steradians) from the magnetic field direction, in the simple case that the direction is extremely uniform. We can therefore only exclude $\sim$1\% (0.002/0.24) of sources as having no intrinsic $\pi$ emission (with an expectation that the majority of sources exhibit LIN, but at a low level) and our result of 16\% of Zeeman patterns exhibiting $\pi$s is fully consistent.

\section{Summary} 
We made targeted observations of 30 fields of high-mass star formation in full Stokes polarization with the Australia Telescope Compact Array and detected 70 sites of maser emission, including 14 associated with Sgr B2. Of the polarised excited-state hydroxyl features associated with these sources, we identified 94 Zeeman pair candidates and 18 Zeeman triplet candidates, a 16\% propensity for $\pi$ components in Zeeman patterns. We thus demonstrate that $\pi$ components do exist (and no one has in fact eaten all the $\pi$s). This result agrees with the recently demonstrated similar rate of occurrence at the lower frequency maser transitions. We argue that depolarisation through Faraday rotation is not significantly affecting the maser emission (specifically, not reducing the linear polarisation in the lower frequency ground-state OH transition relative to the excited-state) and that our results are consistent with magnetic maser beaming. We identify two sources as `text book' examples of Zeeman triplets, 339.884--1.259a and 351.417+0.645a, with coincident splitting at both frequencies (with the corresponding difference in splitting factor) and linear polarisation in $\pi$ and $\sigma$ components (with orthogonal polarisation angles demonstrated in the Stokes Q and U emission).

\section*{Acknowledgments}
This paper is dedicated to the memory of James (`Jim') Caswell, friend, colleague and mentor.\\

\noindent We thank the anonymous referee for their comments, which enhanced and clarified the paper.
The authors thank the staff of the Australia Telescope Compact Array. The Australia Telescope Compact Array is part of the Australia Telescope which is funded by the Commonwealth of Australia for operation as a National Facility managed by CSIRO.\\

\bibliographystyle{mn2e} \bibliography{UberRef}

\appendix
\section{Individual source notes}\label{appendixnotes}%%%%%%%%%%%%%%%%%%%%%%%%%%%%%%%%%%%%%%%%%%%%%%%%%%%%%%%%%%%%%%%%%%%%%%%%
Here we discuss the relevant history and properties of individual sources, including reference to other transitions of maser emission. We also note any peculiarities in the spectra for the sources (such as side-lobe responses). The full spectra are presented in Appendix B (online only). To note, 331.135--0.431, a strong source of 6035-MHz emission with existing spectra in  \citet{caswell95c} and  \citet{caswell03} was not observed (not targeted) in the current study.\\

{\it 284.351--0.418 and 284.352--0.419} This source has strong 6035-MHz emission and existing spectra in  \citet{caswell95c} and  \citet{caswell03}. The first site is that of the  6030/6035-MHz emission, the second that of the 6668-MHz emission. Zeeman splitting is seen at both the 6030- and 6035-MHz transitions, with an orientation consistent with a negative field direction. The corresponding 1612-MHz and 1665-MHz have a similar negative field direction \citep[as seen in the `MAGMO' survey of][]{green12magmo0} and a history of variability \citep{caswell13}.\\
 
{\it 285.263--0.050} This source is a strong 6035-MHz site, and existing spectra in \citet{caswell95c} and  \citet{caswell03} show variations to stronger and weaker emission respectively. For the first time we also see weak 6030-MHz emission. Zeeman splitting demonstrates a large positive field, comparable to that reported in \citet{caswell95c} and that seen with the ground-state transitions \citep{green12magmo0,caswell13}. The clear Zeeman pair seen at 6035 MHz, indicating a field strength of 10mG, appears replicated at 6030 MHz, but below the formal detection limit of the current study (the RHC component is $\le$1 $\sigma$). The same peak flux density ratio between RHC and LHC components is seen at both frequencies (1:2.5). This site does not exhibit methanol emission.\\

{\it 285.337--0.002} This 6668-MHz methanol maser was  detected $\sim$5$'$ offset from the nearby, but unrelated source 285.263--0.050. It exhibits no detectable exOH emission.\\
 
{\it 294.511--1.621} This source has strong 6035-MHz emission, although weaker than at several earlier epochs \citep{caswell95c,caswell03}. The high spectral resolution of the current observations make the narrowly split Zeeman pair clear. LIN emission is detected, and could be a $\pi$ component; however, the velocity resolution is inadequate to separate it spectrally from the LHC component. This source is also reported at ground-state in \citet{green12magmo0}, where there are numerous LIN features with comparable velocities.  In comparison to the narrow exOH emission, the known associated methanol is seen across $\sim$10\,km\,s$^{-1}$ of the spectrum.\\

{\it 300.969+1.148}  This source has Long Baseline Array (LBA) maps of the ground state transitions  \citep{caswell10vlbi}, and the accompanying single dish measurements of 6035-MHz emission in circular polarization show strong evidence of the $\sigma$ components of Zeeman splitting. The spectrum is similar to earlier reports \citep{caswell03}, but emission is now stronger. There is a clear Zeeman pair at 6030 MHz, and corresponding behaviour at 6035 MHz, with some LIN present in both spectra (with the 6035-MHz LIN identified as a $\pi$ component). The field orientation is in agreement with the ground-state Zeeman pairs \citep{caswell13}. This source was the monitor source for the MMB survey \citep{green09a} with well characterised methanol and exOH emission.\\
 
{\it 309.921+0.479a,b,c}  This source is a reasonably strong maser at 6035 and 6030 MHz, with the intensity increasing since a minimum in 2001 \citep{caswell03}. When observed by  \citet{knowles76}, this source showed the best example of linear polarization, and our present observations suggest a slight increase, with a dominant LIN feature at --59.5\,km\,s$^{-1}$ (attributed to site `c').  Despite changes over a 35-year interval (evident from the circular polarization spectra), it is still clearly a good candidate for $\pi$ emission, with the coincidence in velocity of LIN features at both the 6030- and 6035-MHz transitions a strong indicator (as argued for the coincidence of 1665- and 1667-MHz transitions in \citealt{caswell13}). The distribution of RHC and LHC components suggest both positive and negative field Zeeman pairs. Several of the pairs identified in the current work differ from previous, and a Zeeman triplet is likely in this source, despite the failure to pass the criteria outlined in Section \ref{zeeman_ident}. For the brightest 6035-MHz LIN feature, our negative Stokes Q and positive Stokes U suggests a position angle between 45$^{\circ}$ and 90$^{\circ}$, whereas \citet{knowles76} appears to be --20$^{\circ}$. This source exhibits bright methanol maser emission.\\  
 
{\it 311.596--0.398} This 6035-MHz maser also exhibits maser emission at the even more highly excited 13.441-GHz transition \citep{caswell04d}. The 6035-MHz emission is stable with several narrow LIN features, but the complexity of the spectrum prevents Zeeman triplet identification. This source has strong linear polarisation at the ground-state transitions \citep{caswell13}. There is no emission seen at 6030 MHz or at 6668 MHz. The positive LSR velocity of this source indicates a large unambiguous kinematic distance.\\

{\it 311.643--0.380} A site of both 6035-MHz exOH and methanol emission, which also has associated ground-state OH emission with high LIN \citep{caswell13}. A simple Zeeman pair is identified in the 6035-MHz emission.\\
  
{\it 323.459--0.079a,b,c,d}  This source has LBA maps of the ground state transitions  \citep{caswell10vlbi}, and the accompanying single dish measurements of exOH emission in circular polarization show strong evidence of the $\sigma$ components of Zeeman splitting. Four sites are detected in the current observations, two sites of exOH, one with methanol, and one with both exOH and methanol, all within the known H{\sc ii} region, and in accord with the spot distribution presented in \citet{caswell97}. The source has demonstrated stable emission at both 6030 and 6035 MHz with several clear Zeeman pairs at both transitions.  This source also demonstrates LIN emission at 6035 MHz, with two weak $\pi$ components identified, most notably in the triplet of 323.459--0.079a, which also has weak LIN emission accompanying the $\sigma$ components. It is clear from the Stokes Q and U that these LIN features show the expected variation in polarisation position angle between the LIN of the $\sigma$ components and the $\pi$.\\

{\it 328.808+0.633a,b and 328.809+0.633} The first two of these three sites exhibit emission at both exOH and methanol transitions, whilst the other site exhibits emission at the methanol transition only (6035-MHz emission in the spectrum of 328.809+0.633 in Figure \ref{spectra} is a side-lobe response to 328.808+0.633a,b). These sites have associated 13.441-GHz exOH maser emission \citep{caswell04d}.  The spectra of the exOH emission are comparable to that reported in 2001 by \citet{caswell03}. 328.808+0.633a has multiple weak 6035-MHz LIN features (one of them also at 6030 MHz with similar velocity and position angle), which are likely $\pi$ components (three Zeeman triplets are identified). The Stokes Q and U spectra show variation in polarisation position angle across the LIN features.\\

{\it 329.339+0.148} This 6035-MHz maser exhibits maser emission at the more highly excited 13.441-GHz transition, which has a similar spectral structure of offset LHC and RHC components \citep{caswell04d}. The emission has diminished since that reported by \citet{caswell03}. Two Zeeman pairs are identified. Associated methanol seen with $\sim$10\% linear polarisation.\\
 
{\it 330.953--0.182}  This source has LBA maps of the ground state transitions  \citep{caswell10vlbi}, and the accompanying single dish measurements of 6035-MHz emission in circular polarization \citep{caswell03} show evidence of the $\sigma$ components of Zeeman splitting. We identify three Zeeman pairs; the strongest centred at --87.9\,km\,s$^{-1}$ indicates a magnetic field of --3.4 mG (LHCP at more positive velocity) and is in fact identified as a Zeeman triplet; the other two, centred at --88.8\,km\,s$^{-1}$ and --89.8\,km\,s$^{-1}$ are weaker; they also indicate negative fields of comparable strength (--2 and --4 mG respectively). We also note that the polarization position angle of the LIN component is close to +22.5$^{\circ}$ (U and Q both positive), similar to that of the 1665-MHz LIN component of \citet{caswell13}. The site exhibits methanol maser emission across a comparable range of LSR velocities.\\

{\it 331.442--0.187} This known site of only methanol maser emission was detected significantly offset ($\sim$400$''$, 0.95 of the Full Width at Half Maximum) from the phase centre of observations of 331.511--0.102. The 6030- and 6035-MHz emission shown in the spectrum is simply a side-lobe response of 331.542--0.066.\\

{\it 331.496--0.078} This is a new detection of methanol maser emission close to the MMB detection limit, not recognisable in the MMB survey, perhaps due to variability.\\

{\it 331.511--0.102 and 331.512--0.102}  Both of these sites have emission at both of the exOH transitions, but neither have methanol emission (the methanol emission present in the spectra is a side-lobe response of 331.442--0.187).\\
  
{\it 331.542--0.067 and 331.543--0.066} The first of these is a site of both exOH emission and methanol, and additionally  exhibits maser emission at the more highly excited 13.441-GHz transition \citep{caswell04d}. The second, separated by only 2$''$, is a site of purely methanol maser emission. 331.542--0.067 has LIN features at 6030 and 6035 MHz.\\

{\it 331.556--0.121} This is a site of  methanol maser emission only, detected offset by $\sim$3$'$ from 331.511--0.102.  \\

{\it 333.068--0.447} This site of methanol emission  may exhibit weak ($<$0.15\,Jy) exOH at --56\,km\,s$^{-1}$. The emission seen in the spectrum between --52\,km\,s$^{-1}$ and --49\,km\,s$^{-1}$ is a side-lobe of 333.135--0.432.\\

{\it 333.121--0.434, 333.126--0.440 and 333.128--0.440} This is a trio of methanol only sites. Again, the emission seen in the spectrum between --52\,km\,s$^{-1}$ and --49\,km\,s$^{-1}$ is a side-lobe of 333.135--0.432.\\

{\it 333.135--0.432a,b,c,d}  This source has both 6030-MHz and strong 6035-MHz emission and existing spectra in  \citet{caswell95c} and  \citet{caswell03}. The emission at --55\,km\,s$^{-1}$ has decreased since the Parkes observations of  \citet{caswell95c}. The first site (`a') exhibits known weak methanol maser emission.\\

{\it 336.941--0.156}  This site has emission from both exOH transitions and methanol. It also exhibits maser emission at the more highly excited 13.441-GHz transition \citep{caswell04d}. The 6035-MHz exOH and methanol show some LIN features.\\ 
   
{\it 337.703--0.054 and 337.705--0.053a,b}  This source has LBA maps of the ground state transitions  \citep{caswell10vlbi}, and the accompanying single dish measurements of 6035-MHz emission in circular polarization shows strong evidence of the $\sigma$ components of Zeeman splitting \citep{caswell03}. The emission at 6030 is very weak ($<$0.15\,Jy). The 6035-MHz emission has several RHC and LHC pairs, with the RHCs shifted to  higher velocities, which is opposite to 1665- and 1667-MHz emission (both from the LBA maps and the overall spectral appearance), as first reported by \citet{caswell03} and most recently by \citet{caswell13}. 337.703--0.054 is the marginally offset methanol emission.\\
  
{\it 339.884--1.259a,b} The first site has strong 6030- and 6035-MHz emission, and the second site represents the marginally offset methanol emission. The exOH site has existing spectra in  \citet{caswell95c} and  \citet{caswell03}. The emission at both transitions is now larger than that previously reported, by factors of five and nearly two respectively.  There are clear coincident LIN features at both 6030 and 6035 MHz, especially evident at the 6030-MHz transition with the larger splitting. Although the LIN features are spectrally blended, the Stokes Q and U spectra show a change in polarisation position angle as would be expected from the LIN of the two $\sigma$s and the intermediary $\pi$, thereby confirming the triplet identification. This source can be considered a `text-book example' of a Zeeman triplet, with the simple split spectra and coincident features at both transitions (with the corresponding difference in splitting factor).\\ 
  
{\it 340.785--0.096a,b} This source has strong 6035-MHz emission, but no 6030-MHz emission, and existing spectra in \citet{caswell95c} and  \citet{caswell03}. The weak feature ($<$0.2\,Jy) at --91\,km\,s$^{-1}$ is still present. The spectrum suggests a field reversal between two major separated velocity ranges of features, with those near --106\,km\,s$^{-1}$ showing RHC shifted to lower velocities and those near --102\,km\,s$^{-1}$, showing RHC shifted to higher velocities. The 1665- and 1667-MHz emission matches this behaviour.
The methanol emission (site `b') is also wide. It is not clear which velocity represents the systemic, but the weaker emission is at red shift and thus might be a red-shifted outflow. Notably water maser emission is very weak (detected with a peak flux density of 1 Jy in 2003, but below 0.2 Jy in 2004, \citealt{breen10b}), with a peak at a more negative velocity, near --120\,km\,s$^{-1}$.  It appears this source is certainly evolved and thus not expected to be blue-shifted dominant.\\
   
{\it 343.929+0.125}  This source has strong 6035-MHz emission, associated methanol, and existing spectra in  \citet{caswell95c} and  \citet{caswell03}. The spectrum has three Zeeman pairs, which are clearer than  \citet{caswell95c} (and additionally clearer than the ground-state 1665 emission). \\
    
{\it 345.003--0.224 and 345.003--0.223}  This close pair  exhibits exOH and methanol at spatially separate sites. The exOH maser also exhibits maser emission at the more highly excited 13.441-GHz transition \citep{caswell04d}. Both the 6030- and 6035-MHz transitions exhibit clear Zeeman pairs, as previously noted by  \citet{caswell95c} and  \citet{caswell03}.\\
    
{\it 345.009+1.792}  This source has strong 6035-MHz emission and existing spectra in  \citet{caswell95c} and  \citet{caswell03}. The emission has greatly changed, with consistently RHC at a larger velocity, for 6030- and 6035-MHz emission.  The emission at 1665 and 1667 MHz may agree, but the spectrum is complicated.\\

{\it 345.012+1.797} This is a site of only methanol emission, offset by 20$''$ from the previous OH site (345.010+1.792).\\
    
{\it 345.487+0.314 and 345.487+0.313}  The first site (`a') has both 6030- and 6035-MHz emission, with existing spectra in  \citet{caswell95c} and  \citet{caswell03}. Both exOH transitions show similar polarised properties, each showing RHC 50\% or less relative to LHC  and a similar marginal shift of RHC and LHC polarised components. The second site (`b') is the slightly offset (by 3$''$) position with weak methanol emission. \\

{\it 345.505+0.348} This is a nearby site of strong methanol maser emission. It is also a site of strong 1665-MHz emission, and was originally believed to be located at 345.488+0.315 \citep{caswell95ab}.\\
    
{\it 345.698--0.090a,b}  This source has strong 6035-MHz emission and existing spectra in  \citet{caswell95c} and  \citet{caswell03}. The methanol spectrum shows evidence of a corresponding feature at --6.7\,km\,s$^{-1}$, however this could not be positioned with the current data. The brightest 6030-MHz emission, at a velocity of --4.9\,km\,s$^{-1}$, comprises a close Zeeman pair which are matched by a close 6035-MHz Zeeman pair (with comparable implied field strengths).\\
    
{\it 347.628+0.148}  This source has strong 6035-MHz emission and existing spectra in  \citet{caswell95c} and  \citet{caswell03}. There is a clear matching Zeeman pair (at both 6030 and 6035 MHz) compared to the complex ground-state emission spectrum. There is weak LIN emission at the 6035-MHz transition, which may be $\pi$ emission, but is not formally identified.\\

{\it 347.632+0.210} This is the offset site of methanol emission (with the excited state OH emission present in the spectra reduced to 20 per cent at this offset of two arcmin).\\

{\it 351.417+0.645a,b,c,d,e}  Part `e' corresponds to the purely methanol site listed in \citet{caswell09a} as 351.417+0.646. This complex (also known as NGC6334F) is of special interest as it has recently been studied at 6030 and 6035 MHz with the Long Baseline Array \citep{caswell11vlbi2}.  In that study, all 56 detected features appeared to be $\sigma$ components of Zeeman pairs, but the observations were not designed to record linear polarization, so there was little information on possible $\pi$ components or on the ellipticity of polarization displayed by the $\sigma$ components (even VLBI observations in the northern sky of the much studied W3(OH) maser site have not included linear polarization measurements, e.g. \citealt{etoka05} and \citealt{fish07c}, although more recent measurements of ON1 by \citealt{fish10} now show successful linear polarization with the European VLBI Network). In the current study we detect LIN at both 6030 and 6035 MHz, with Zeeman triplets at sites `a' and `d', with the former exhibiting the expected change in polarisation angle (seen in Stokes Q and U) with LIN features associated with both the $\sigma$ components in addition to the $\pi$. As with 339.884--1.259a, 351.417+0.645a can be considered a `text-book example' of a Zeeman triplet, with the simple split spectra and coincident features at both transitions (with the corresponding difference in splitting factor).\\

{\it 351.445+0.660} This is a site of only 6668-MHz methanol emission.\\

{\it 353.410--0.360}  This source is of special interest as it was recently studied with the LBA \citep{caswell11vlbi2}. The 6030-MHz emission has dropped by almost 90\%, whilst 6035-MHz emission has remained stable. Most 6035-MHz features show linear polarisation and this is especially notable in the brightest 6035 MHz feature, with $\sim$30\% LIN. The three 6035-MHz Zeeman pairs likely have $\pi$ emission spectrally blended with LIN of the $\sigma$s (and so not formally identified as triplets in this study, c.f. Section \ref{zeeman_ident}).\\
    
{\it 355.343+0.148, 355.344+0.147 and 355.346+0.149}  This close trio of sources has strong 6035-MHz emission at the first two sites and existing spectra in  \citet{caswell95c} and  \citet{caswell03}. There is less than 3\% linear polarisation at 6035MHz. Methanol emission is offset at 355.346+0.149. Magnetic field strength is comparable to the measurement of the ground-state transition (Caswell et al. 2013).\\

{\it 10.959+0.023} This site of methanol and newly detected exOH was detected $\sim$5$'$ offset from the pointing centred on 11.034+0.062. There is no previous recorded ground-state OH emission for this site, however the recent `MAGMO' survey of \citet{green12magmo0} indicates emission exists. The large offset implies a strong source of emission (peak intensity of $\sim$1.5\,Jy). The methanol is a known MMB source \citep{green10mmb2}, with expected reduction of intensity at the large position offset.\\

{\it 11.034+0.062}  The spectrum of this source indicates it may represent a good candidate for a possible strong $\pi$ component between two $\sigma$ components; however, a Zeeman triplet could not be identified in the current data due to the complexity of the spectrum. The exOH emission appears stable, with spectral properties comparable to the spectra presented in \citet{caswell03}. Similarly, the weak ($<$0.4\,Jy) methanol emission is comparable to previous observations \citep{green10mmb2}.\\

{\it 11.903--0.102 and 11.904--0.141} The first of this pair of closely spaced sources is a site of only methanol maser emission \citep{caswell09a,green10mmb2}. The second is a known site of both methanol and exOH emission, and also exhibits emission at the 13.441-GHz transition \citep{caswell04d,fish07d}. The 6035-MHz emission demonstrates a positive velocity shift of RHC components relative to LHC (opposite to that seen in the higher frequency 13-GHz transition), and has strong linear polarisation ($\sim$40\%).\\

{\it 11.934--0.150 and 11.936--0.150} This is a pair of methanol only sites (the 6035-MHz emission in this direction arises from a weak side-lobe of 11.904--0.141). The wider methanol spectrum of 11.936--0.150 shows a negative side-lobe response of the bright source 11.904--0.141 (between 40\,km\,s$^{-1}$ and 45\,km\,s$^{-1}$), and a response of 11.934--0.150 (at 34\,km\,s$^{-1}$).\\
    
{\it 15.034--0.677}  This source (associated with M17) has two strong 6035-MHz features, and a weaker 6030-MHz feature, consistent with existing spectra in  \citet{caswell95c} and  \citet{caswell03}. \citet{knowles76} described the 6035-MHz feature at 21\,km\,s$^{-1}$ as about 20\% LIN with a position angle of  5$^{\circ}$; and the other feature as having no significant LIN. In the current data, at 6035 MHz we find $\sim$15\% LIN, and a position angle of about +120$^{\circ}$ in the brightest feature and 10\% in the second brightest feature. At 6030 MHz, we find $\sim$30\% LIN and a similar position angle. The LIN emission at 6030 MHz is identified as a $\pi$ component, and it is likely that the 6035-MHz feature at the corresponding velocity is also a $\pi$ component. The methanol emission is similar in total intensity to that found in the MMB survey \citep{green10mmb2}, and our polarimetric observations reveal weak LIN ($\le$5\%).\\

\section{Spectra}
The full Stokes I, Q and U and circular polarisation spectra with full velocity coverage are shown in Figure \ref{spectra} and the restricted spectra of the polarised features identified in Table~2 are shown in Figure \ref{newspectra}, with Zeeman pairs and triplets labelled accordingly.

\begin{figure*}
\begin{center}
\renewcommand{\baselinestretch}{1.1}
\includegraphics[width=17cm]{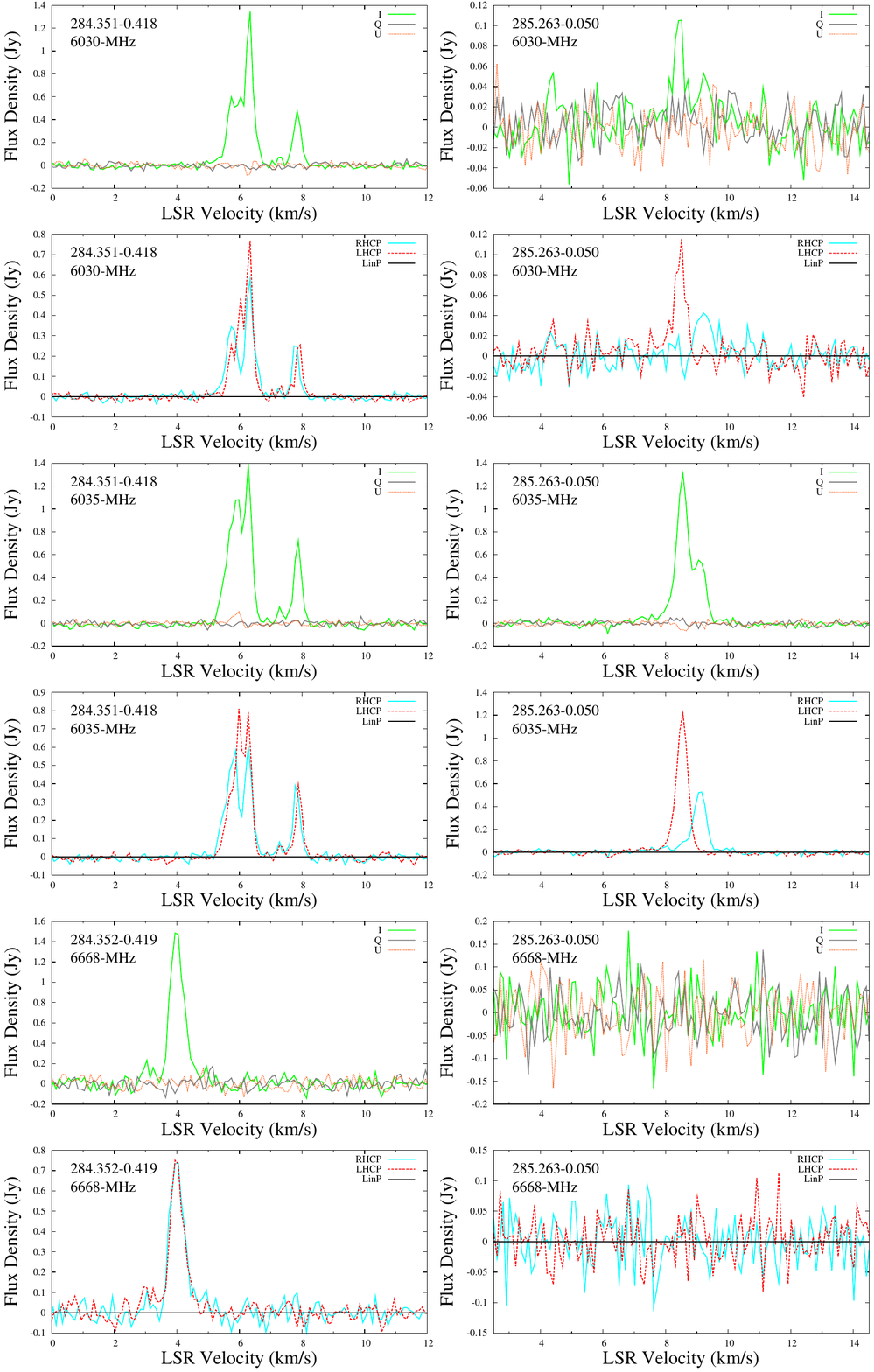}  
\caption{\small Full spectra of detections, showing (from top to bottom) 6030-MHz OH, 6035-MHz OH and 6668-MHz methanol emission in both Stokes I, Q and U and RHC, LHC and LIN.}
\label{spectra}
\end{center}
\end{figure*}

\begin{figure*}
\begin{center}
\addtocounter{figure}{-1}
\renewcommand{\baselinestretch}{1.1}
\includegraphics[width=17cm]{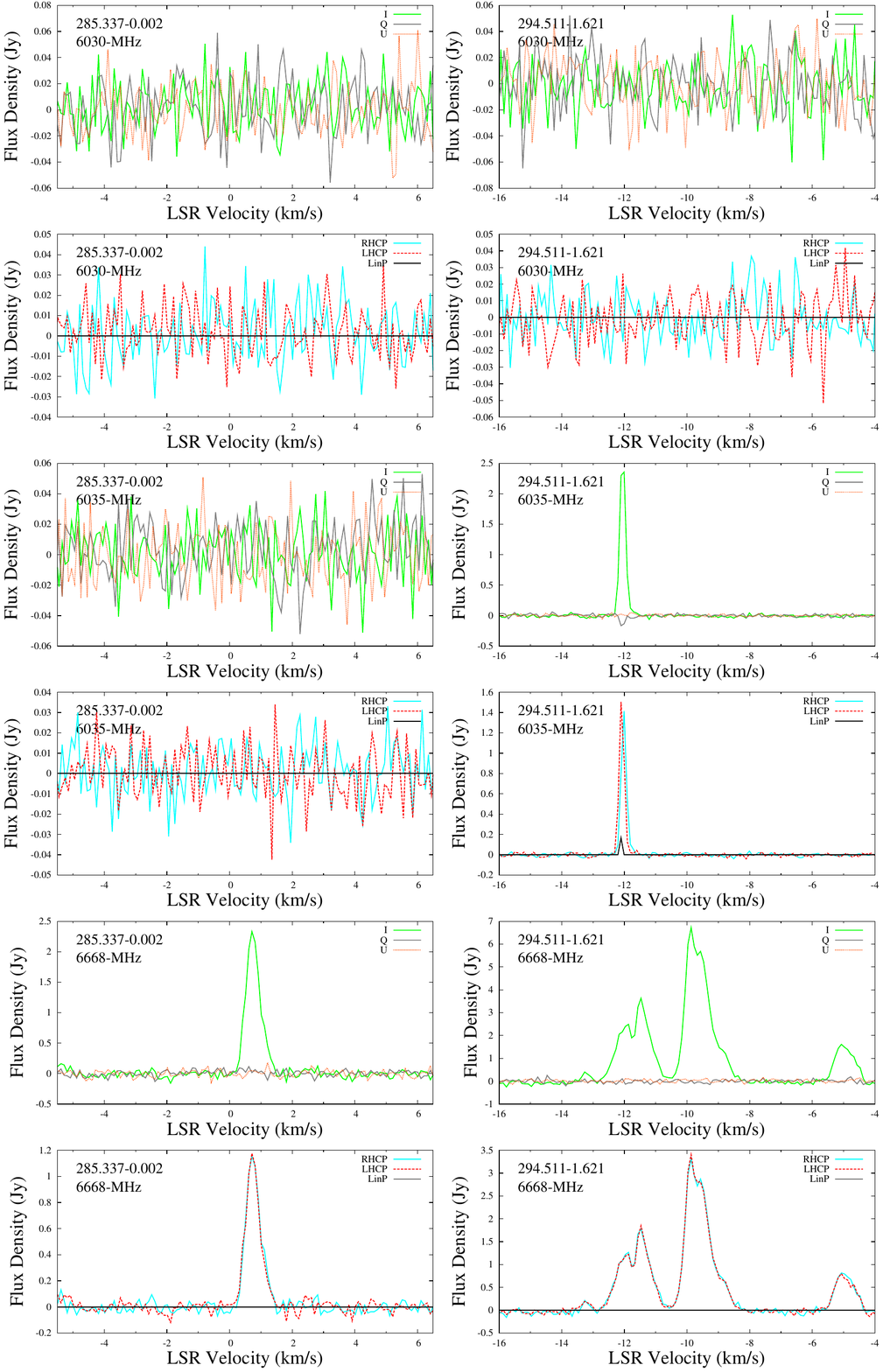}  
\caption{\small continued.}
\label{spectra}
\end{center}
\end{figure*}

\begin{figure*}
\begin{center}
\addtocounter{figure}{-1}
\renewcommand{\baselinestretch}{1.1}
\includegraphics[width=17cm]{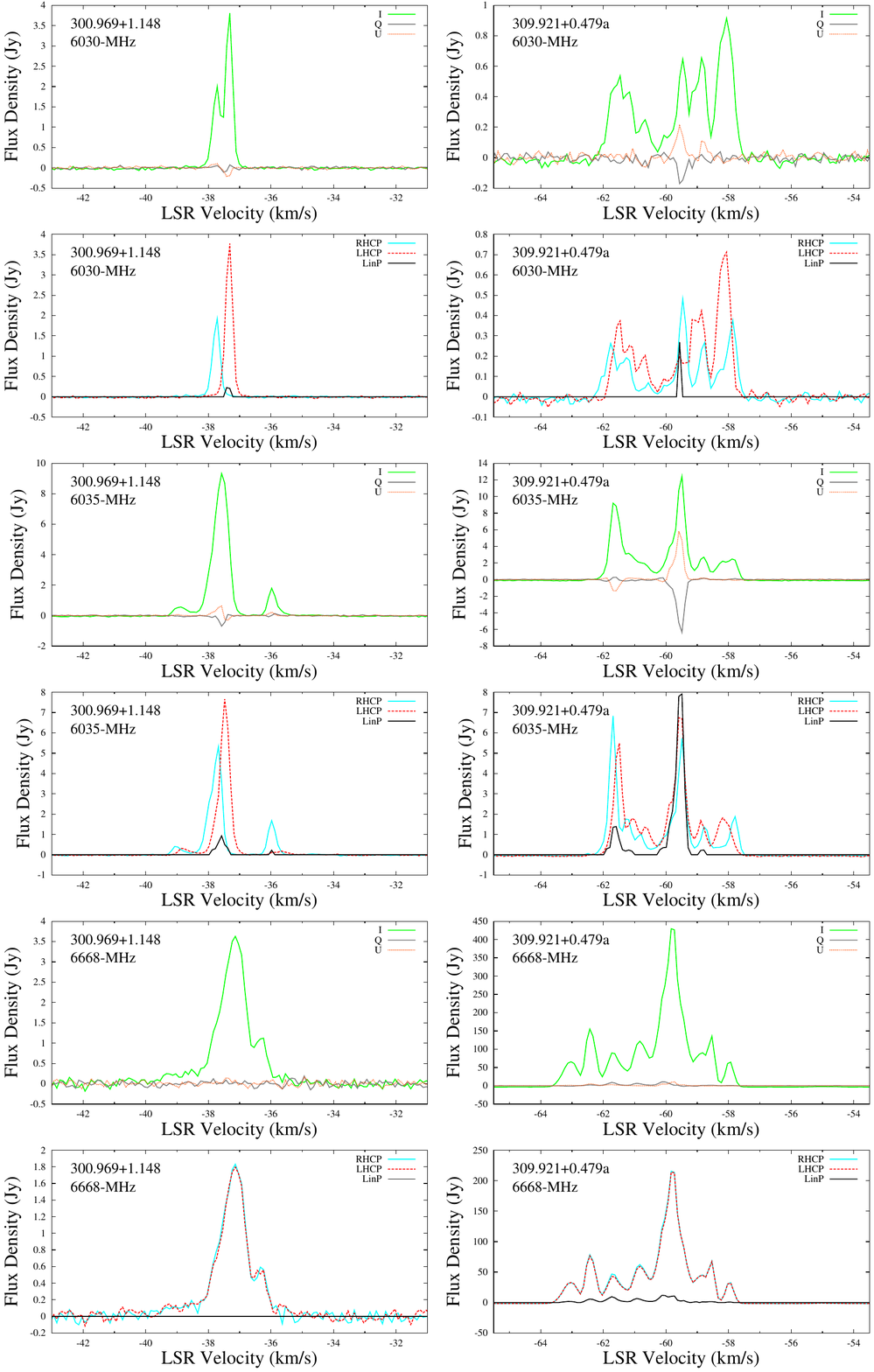}  
\caption{\small continued.}
\label{spectra}
\end{center}
\end{figure*}

\begin{figure*}
\begin{center}
\addtocounter{figure}{-1}
\renewcommand{\baselinestretch}{1.1}
\includegraphics[width=17cm]{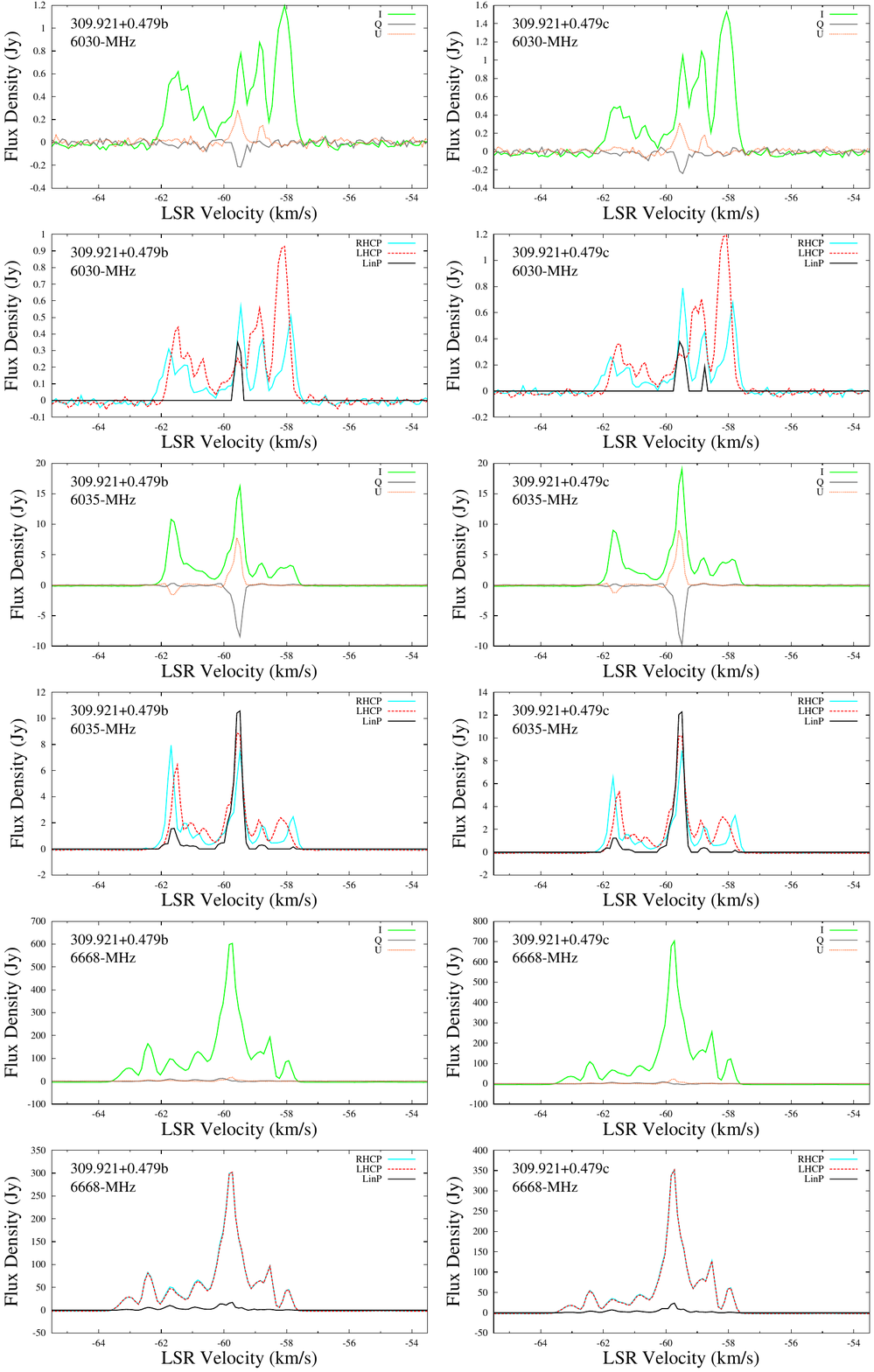}  
\caption{\small continued.}
\label{spectra}
\end{center}
\end{figure*}

\clearpage

\begin{figure*}
\begin{center}
\addtocounter{figure}{-1}
\renewcommand{\baselinestretch}{1.1}
\includegraphics[width=17cm]{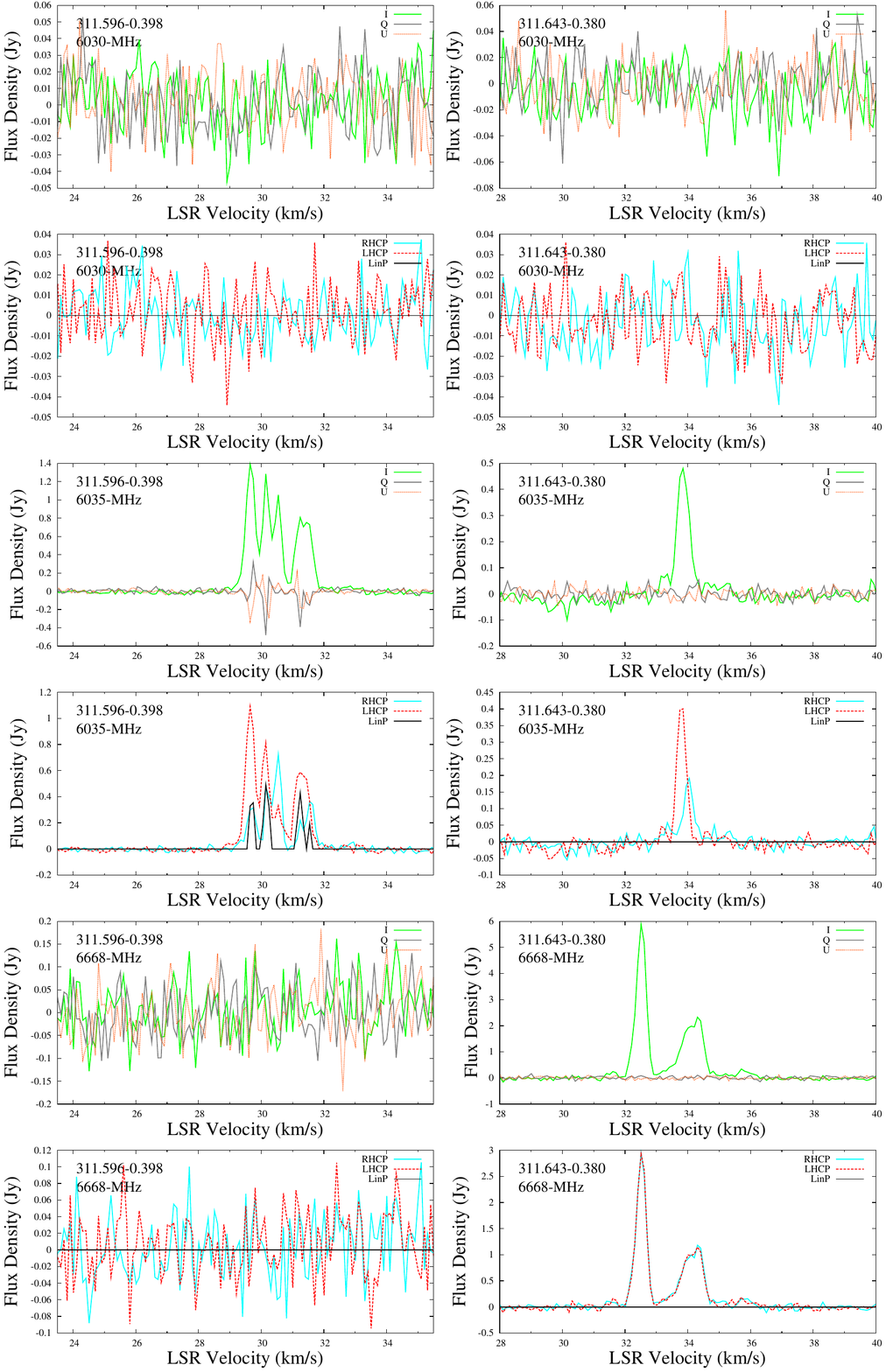}  
\caption{\small continued.}
\label{spectra}
\end{center}
\end{figure*}

\begin{figure*}
\begin{center}
\addtocounter{figure}{-1}
\renewcommand{\baselinestretch}{1.1}
\includegraphics[width=17cm]{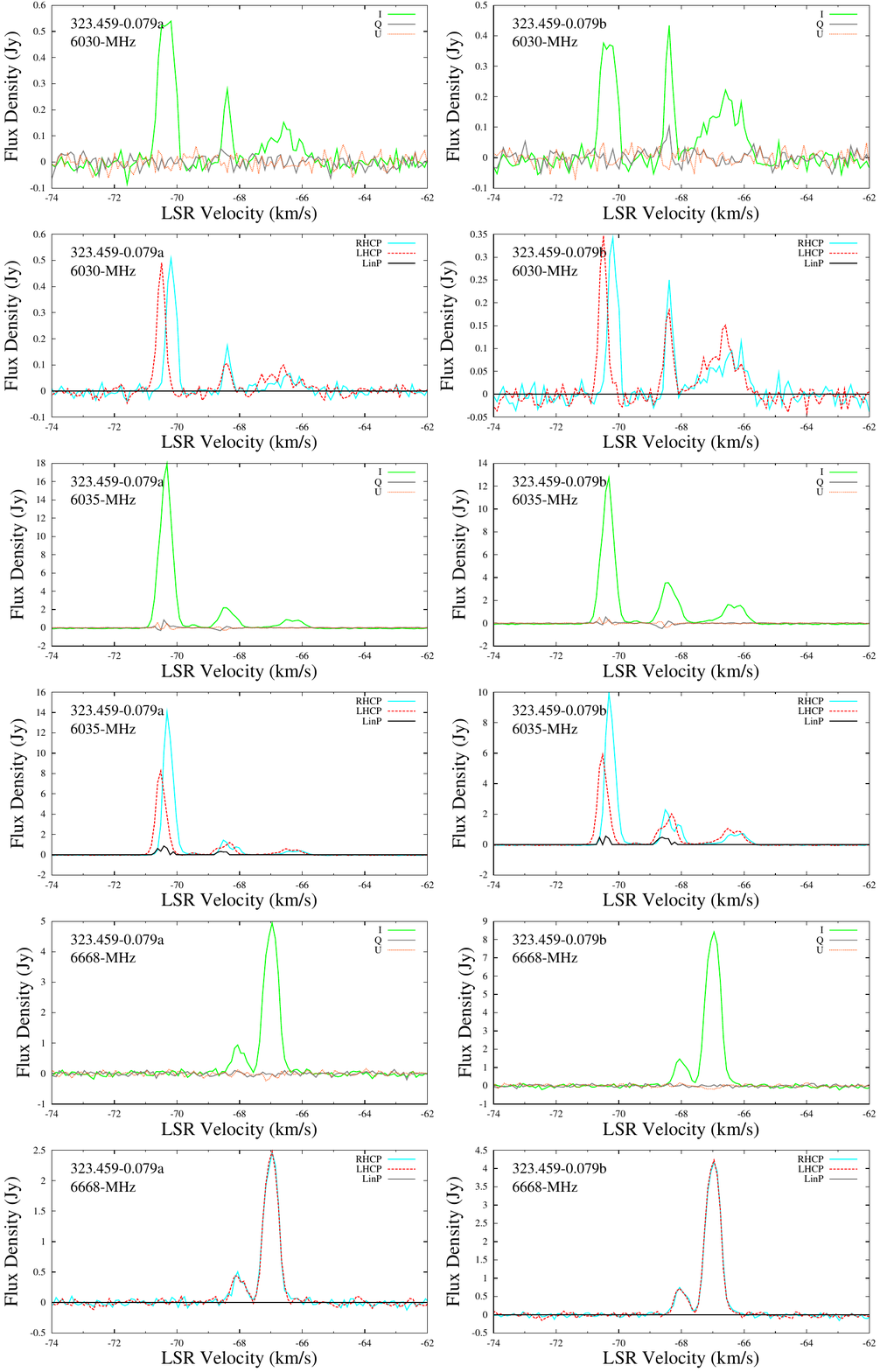}  
\caption{\small continued.}
\label{spectra}
\end{center}
\end{figure*}

\begin{figure*}
\begin{center}
\addtocounter{figure}{-1}
\renewcommand{\baselinestretch}{1.1}
\includegraphics[width=17cm]{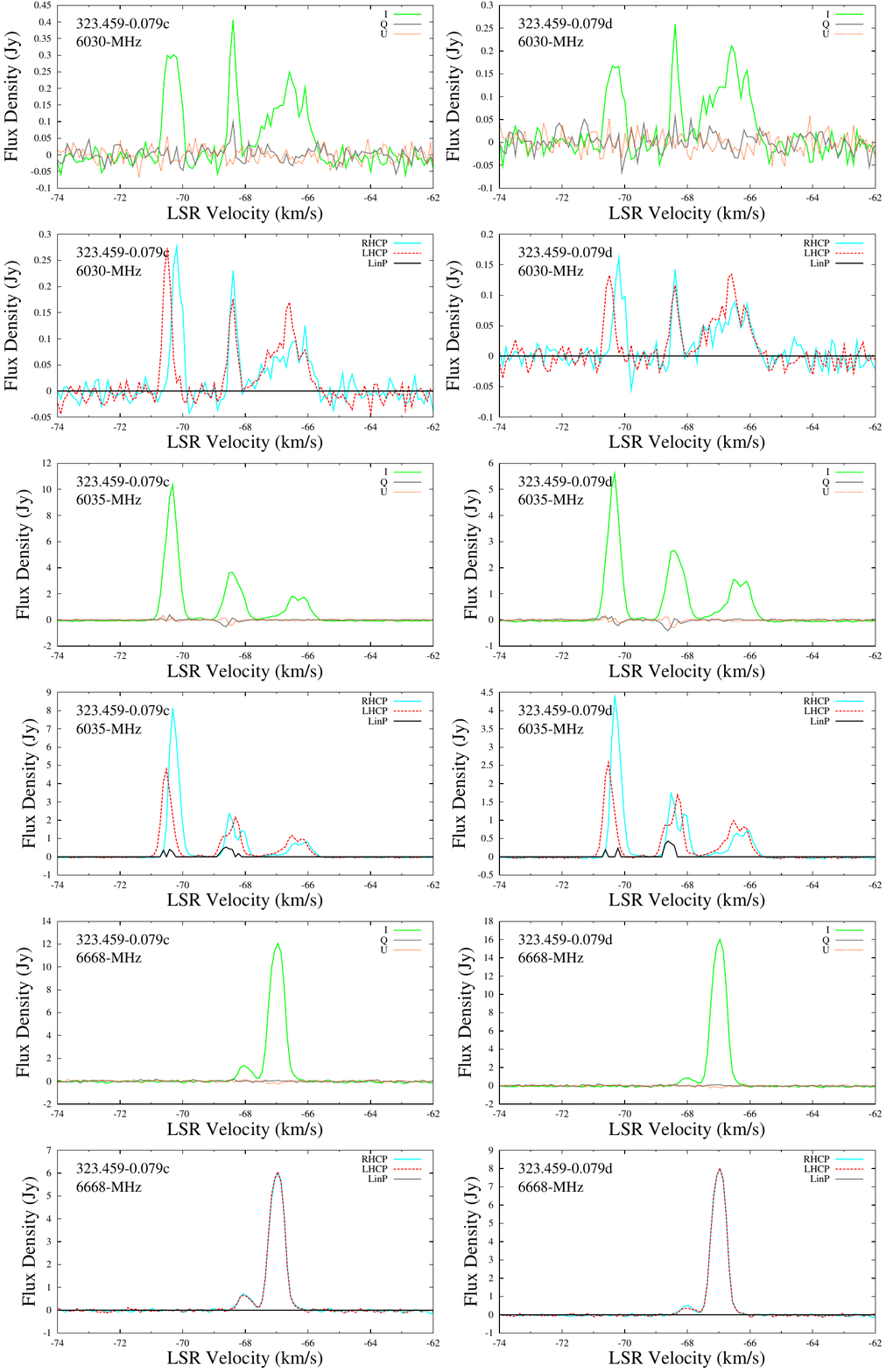}  
\caption{\small continued.}
\label{spectra}
\end{center}
\end{figure*}

\begin{figure*}
\begin{center}
\addtocounter{figure}{-1}
\renewcommand{\baselinestretch}{1.1}
\includegraphics[width=17cm]{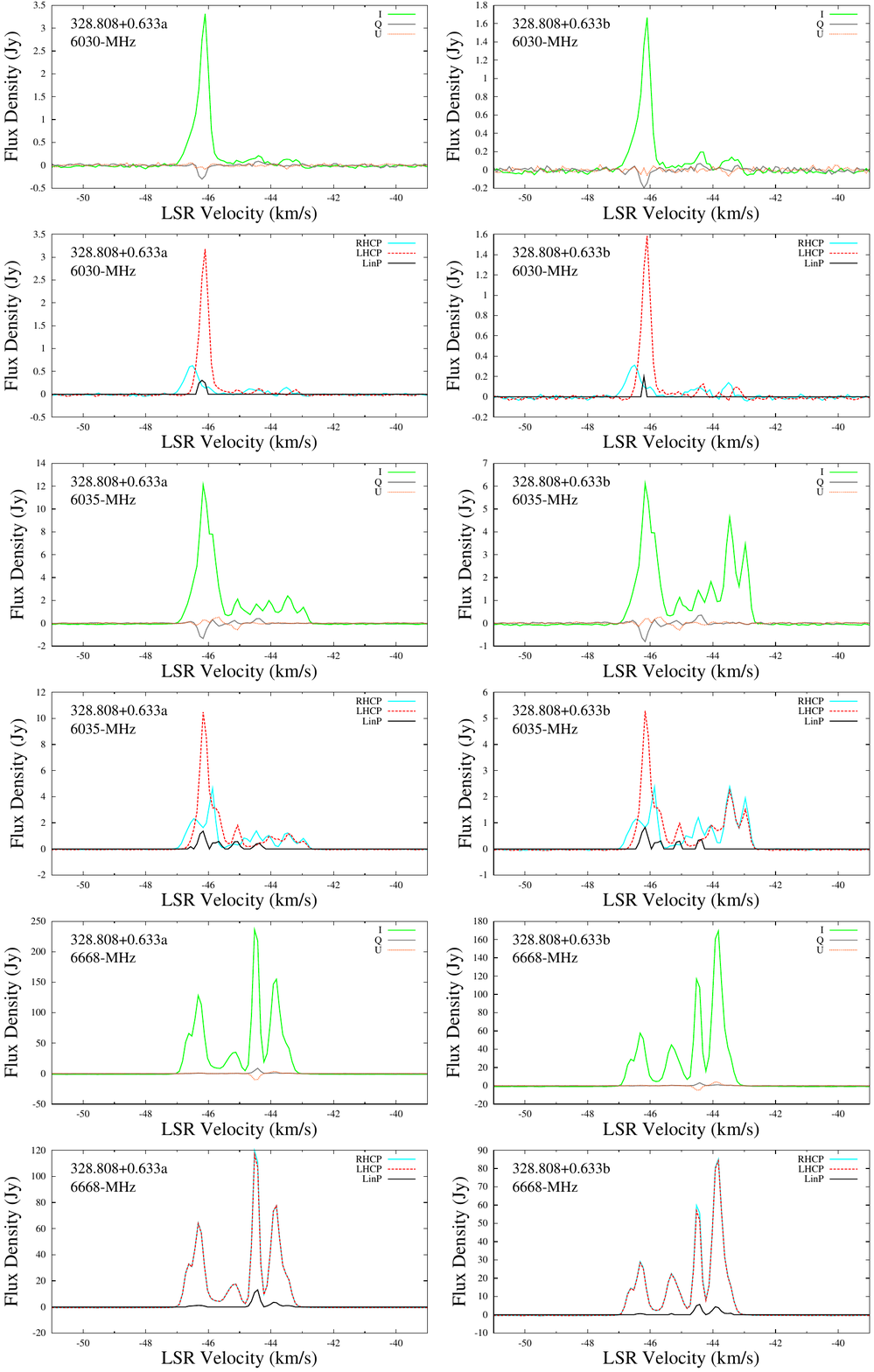}  
\caption{\small continued.}
\label{spectra}
\end{center}
\end{figure*}

\begin{figure*}
\begin{center}
\addtocounter{figure}{-1}
\renewcommand{\baselinestretch}{1.1}
\includegraphics[width=17cm]{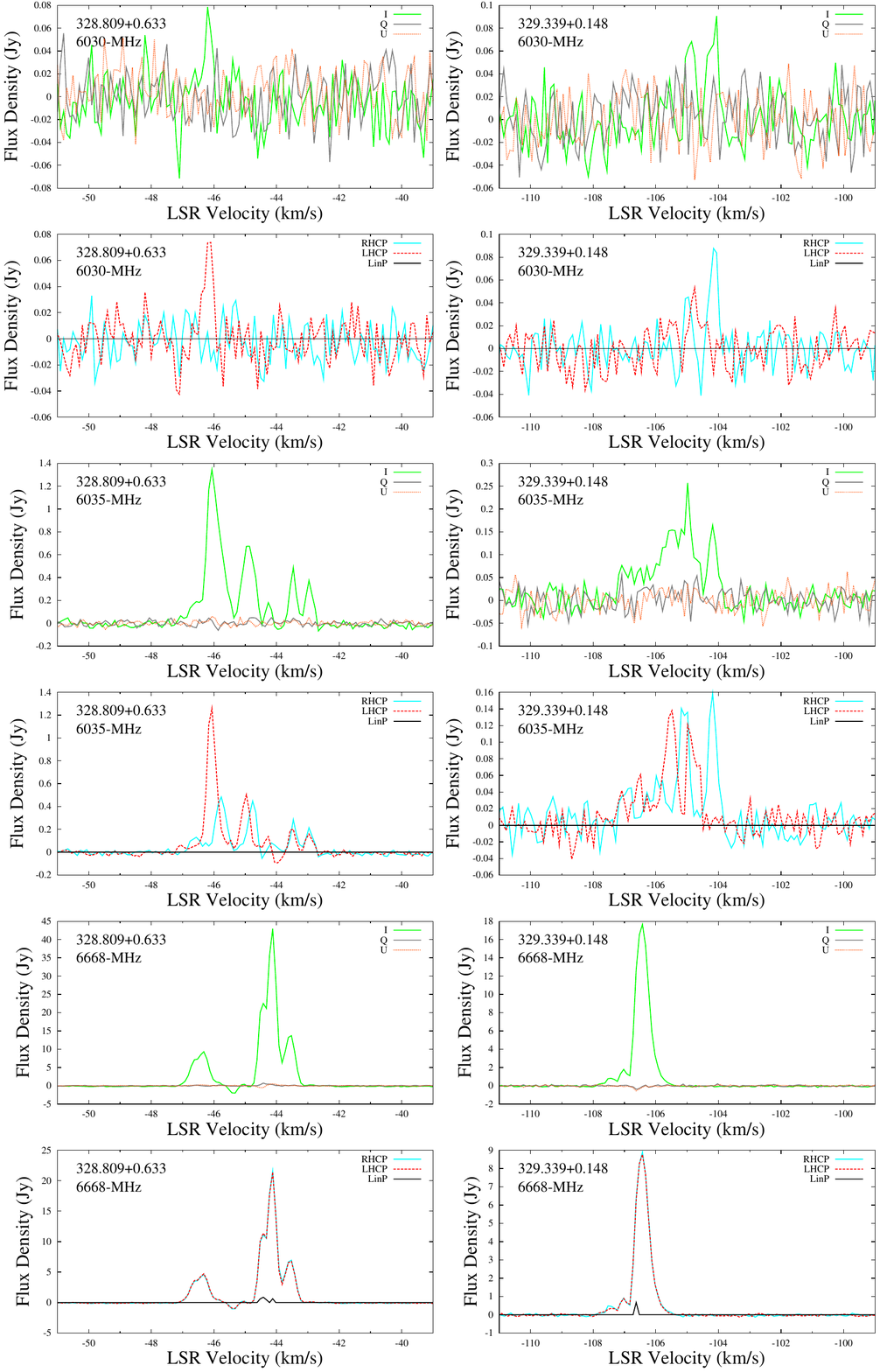}  
\caption{\small continued.}
\label{spectra}
\end{center}
\end{figure*}

\begin{figure*}
\begin{center}
\addtocounter{figure}{-1}
\renewcommand{\baselinestretch}{1.1}
\includegraphics[width=17cm]{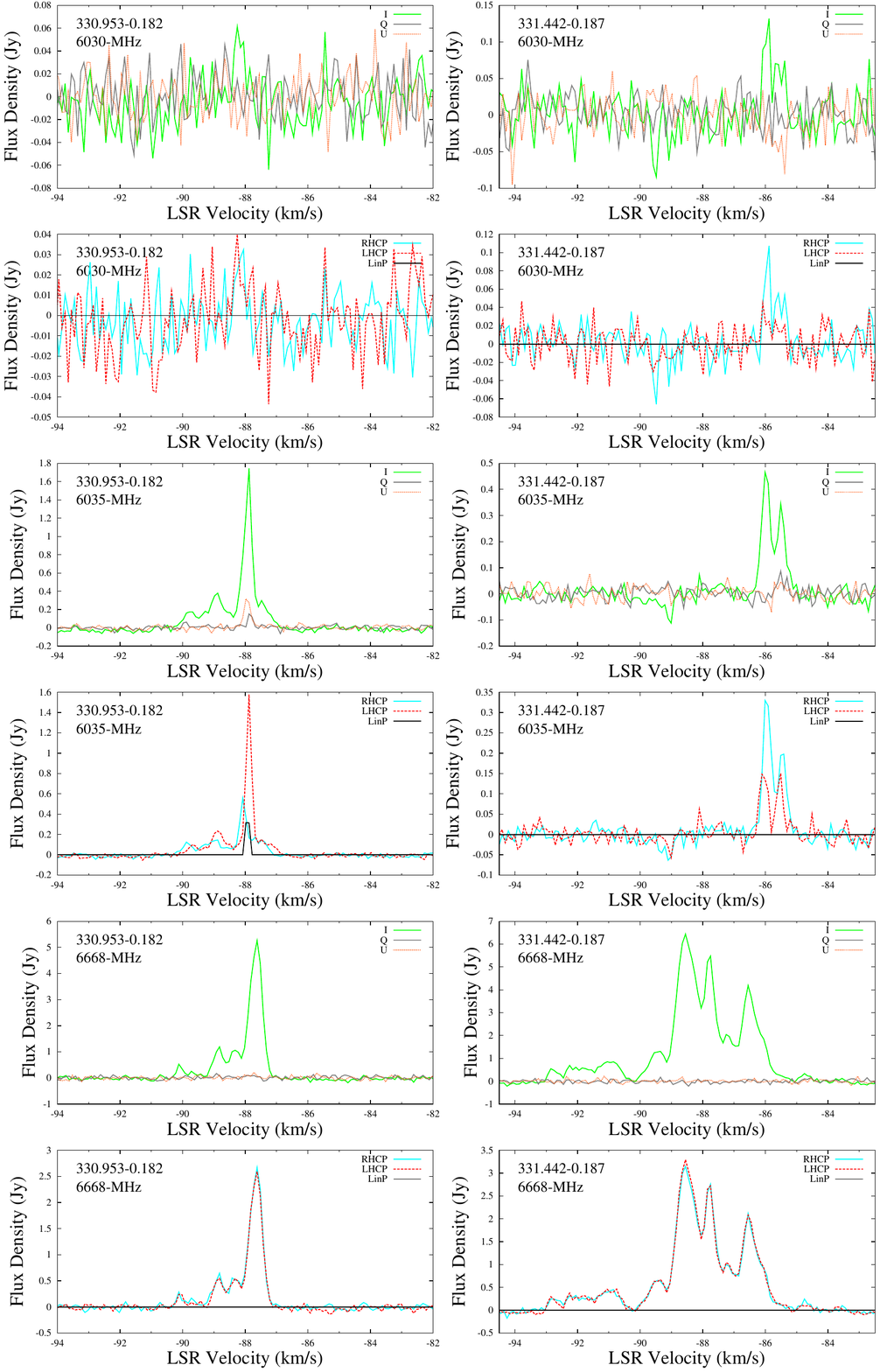}  
\caption{\small continued.}
\label{spectra}
\end{center}
\end{figure*}

\begin{figure*}
\begin{center}
\addtocounter{figure}{-1}
\renewcommand{\baselinestretch}{1.1}
\includegraphics[width=17cm]{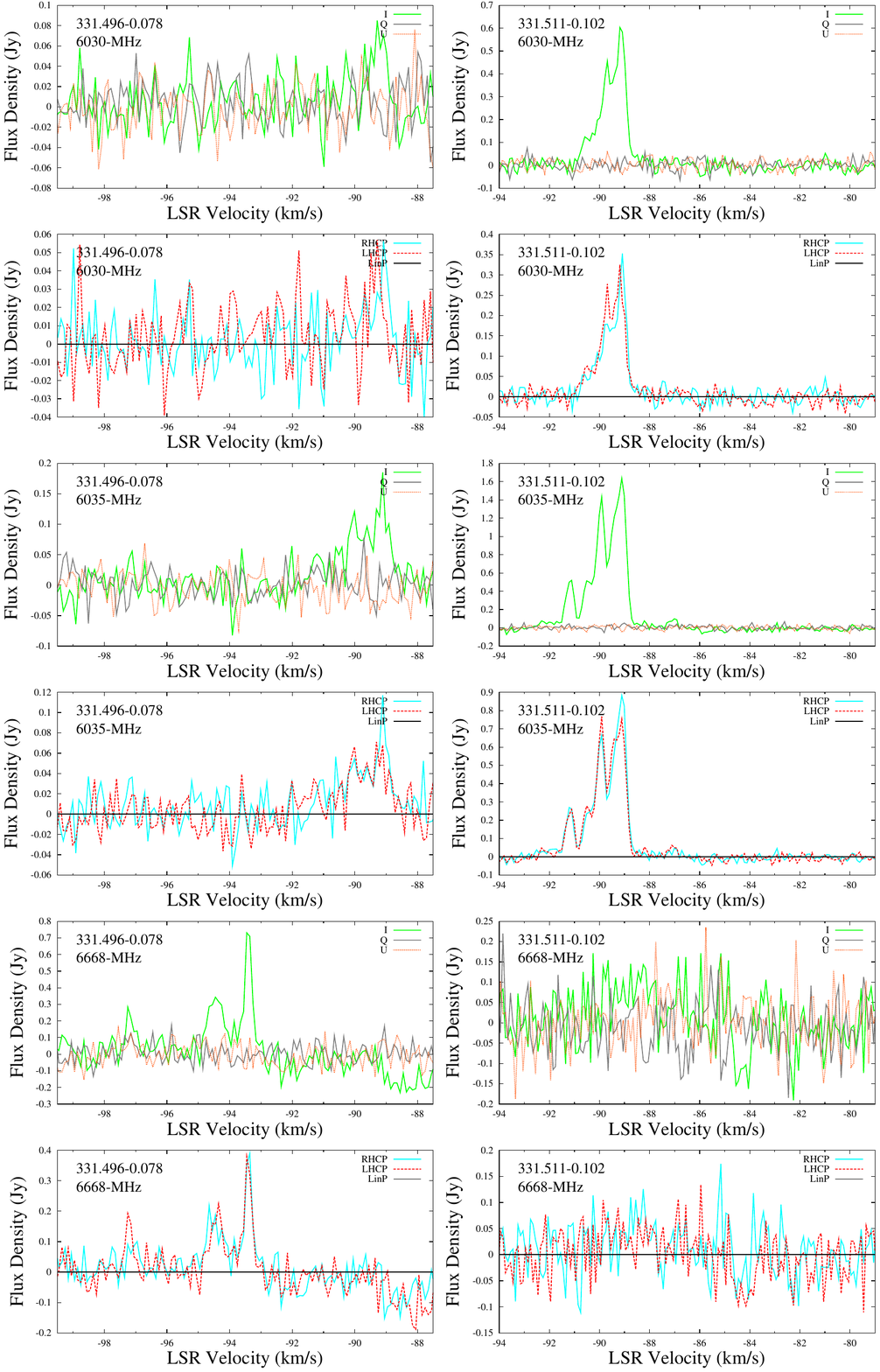}  
\caption{\small continued.}
\label{spectra}
\end{center}
\end{figure*}

\begin{figure*}
\begin{center}
\addtocounter{figure}{-1}
\renewcommand{\baselinestretch}{1.1}
\includegraphics[width=17cm]{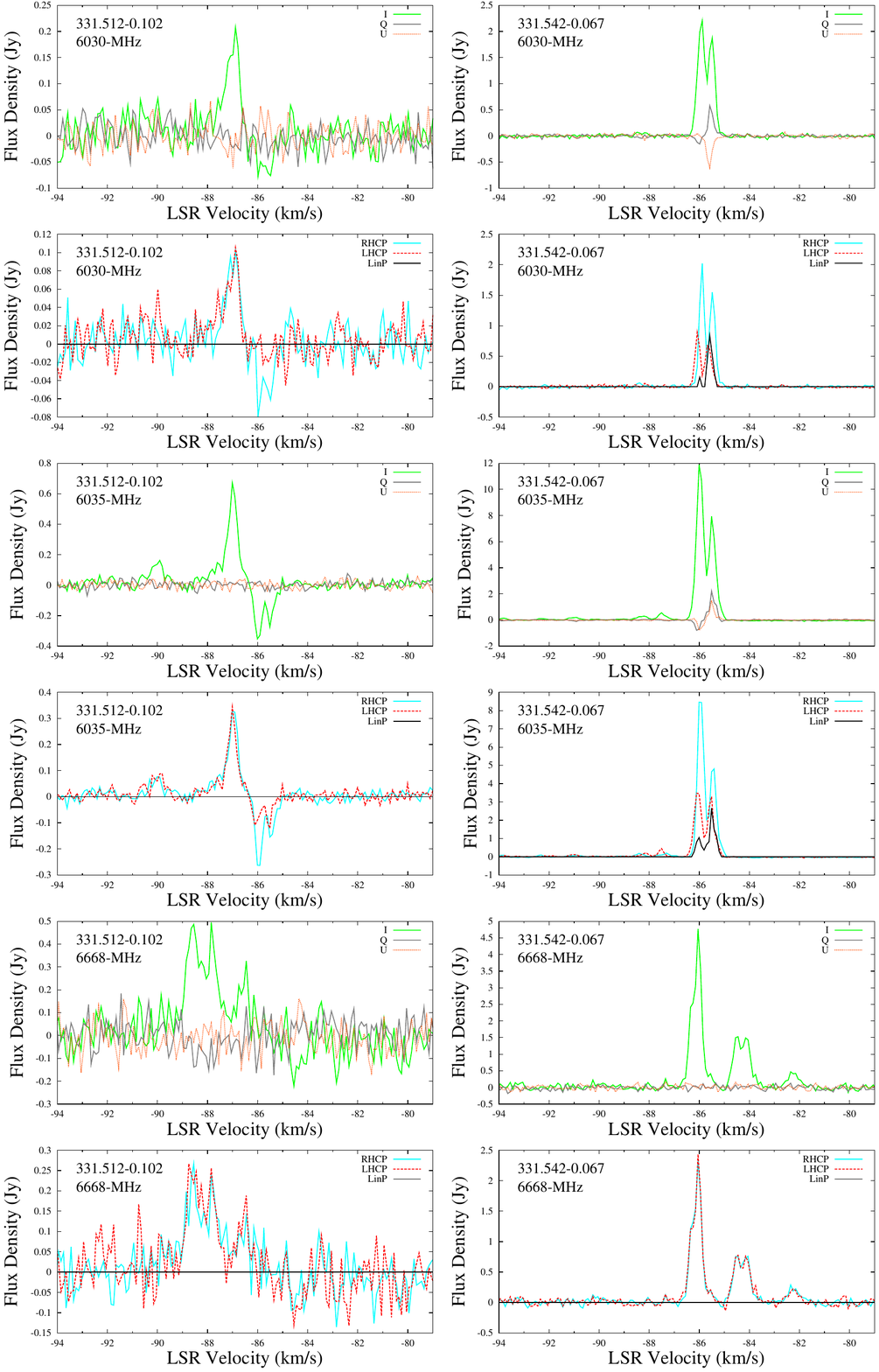}  
\caption{\small continued.}
\label{spectra}
\end{center}
\end{figure*}

\clearpage

\begin{figure*}
\begin{center}
\addtocounter{figure}{-1}
\renewcommand{\baselinestretch}{1.1}
\includegraphics[width=17cm]{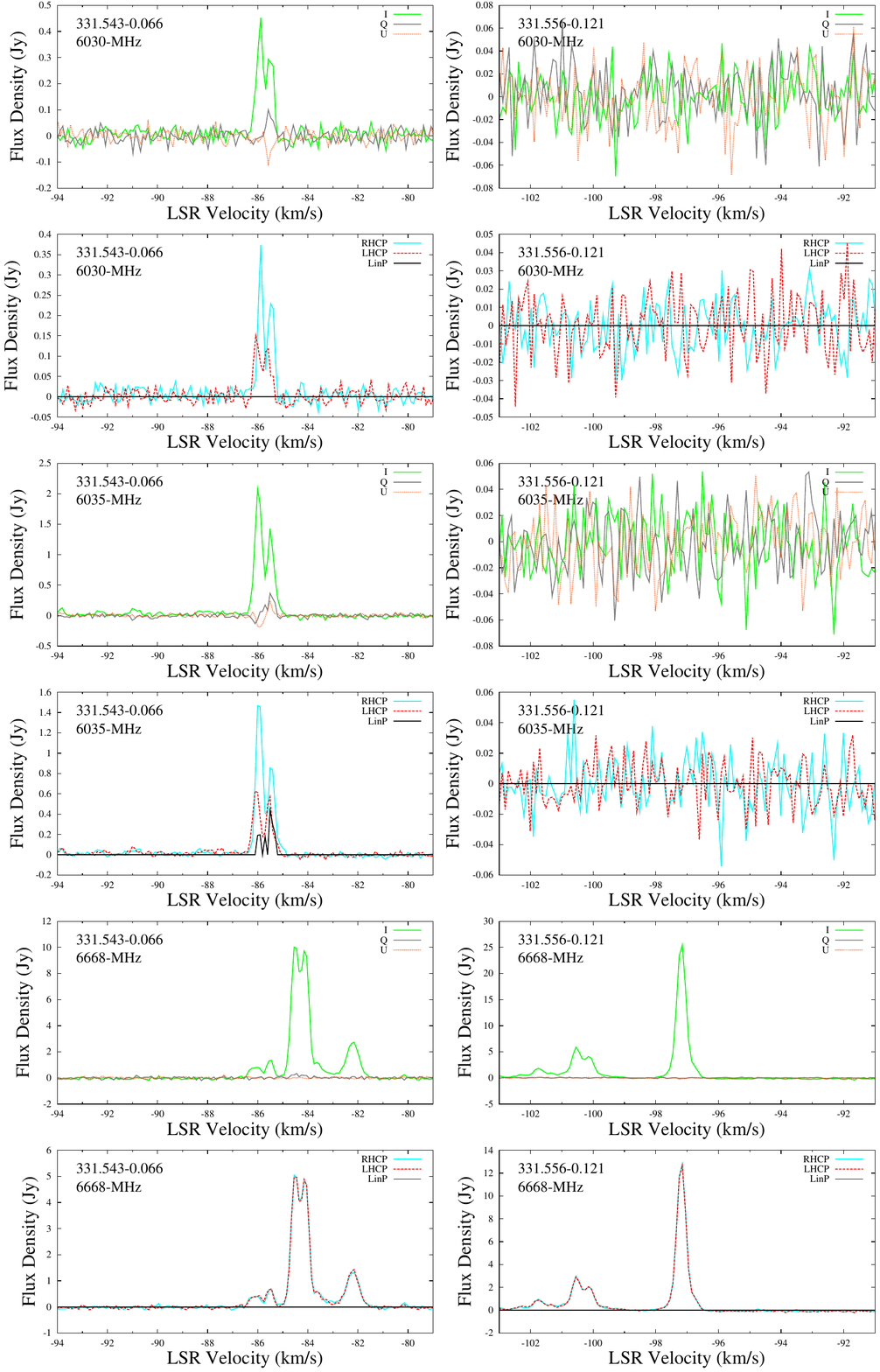}  
\caption{\small continued.}
\label{spectra}
\end{center}
\end{figure*}

\begin{figure*}
\begin{center}
\addtocounter{figure}{-1}
\renewcommand{\baselinestretch}{1.1}
\includegraphics[width=17cm]{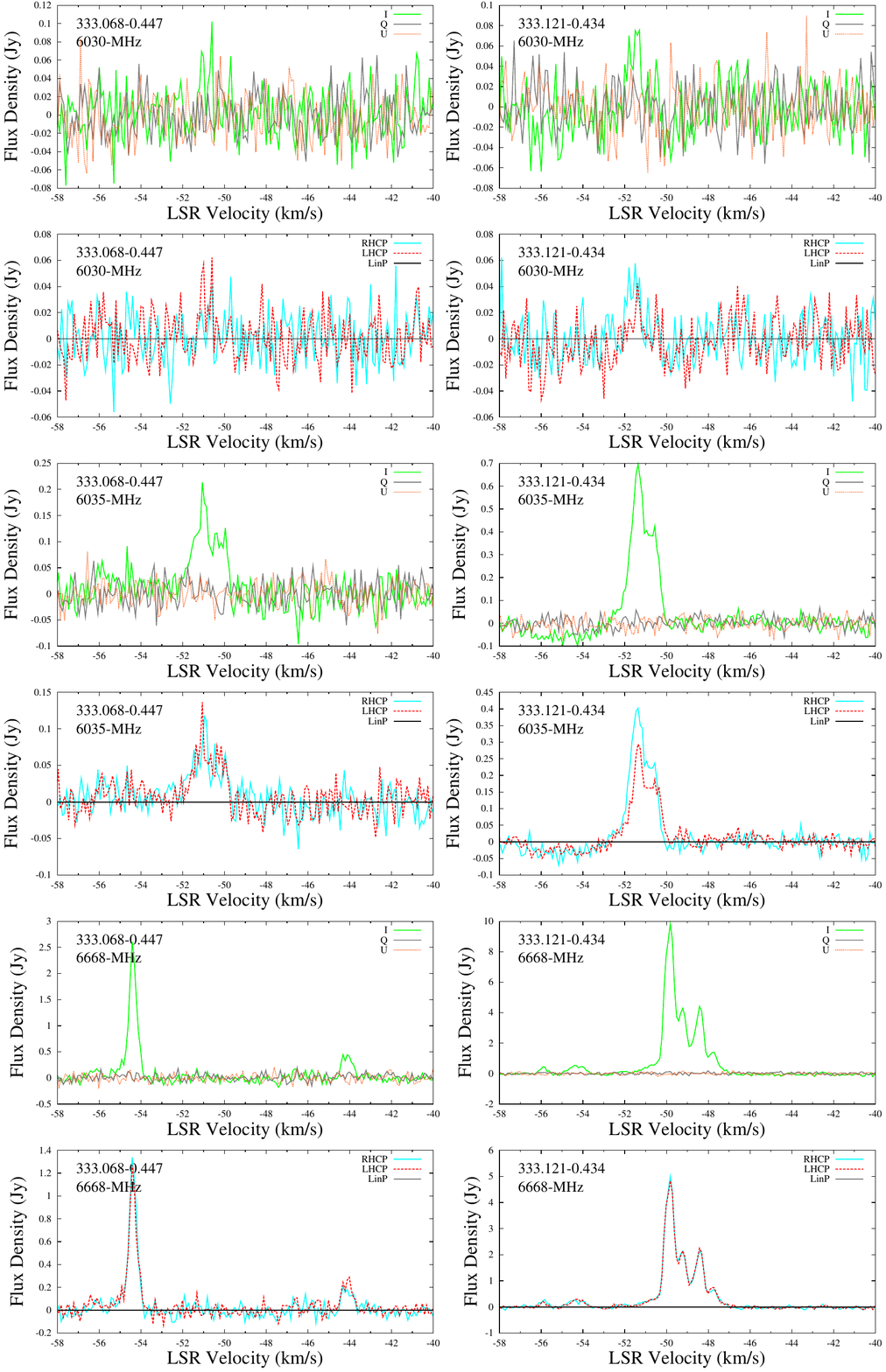}  
\caption{\small continued.}
\label{spectra}
\end{center}
\end{figure*}

\begin{figure*}
\begin{center}
\addtocounter{figure}{-1}
\renewcommand{\baselinestretch}{1.1}
\includegraphics[width=17cm]{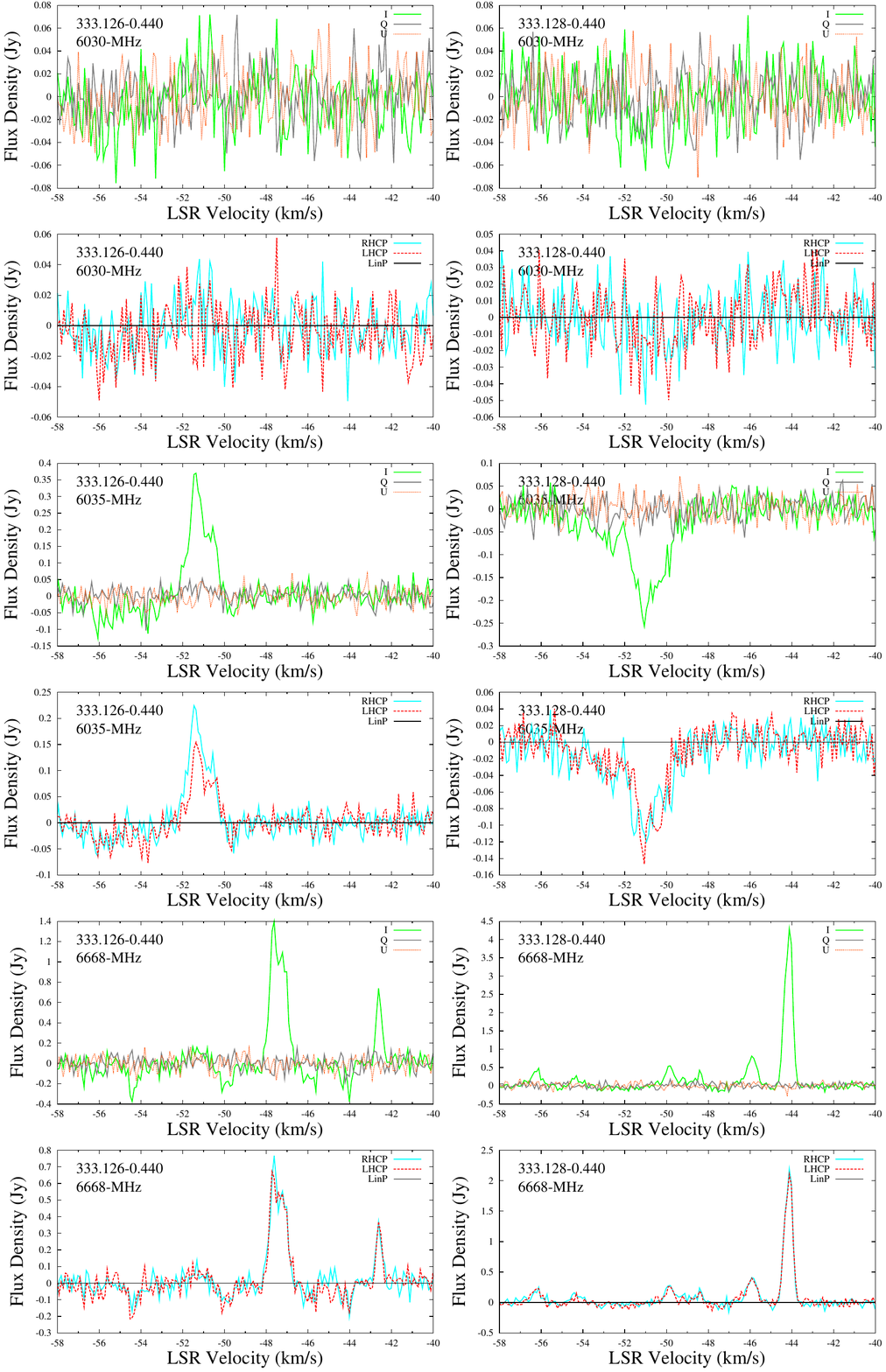}  
\caption{\small continued.}
\label{spectra}
\end{center}
\end{figure*}

\begin{figure*}
\begin{center}
\addtocounter{figure}{-1}
\renewcommand{\baselinestretch}{1.1}
\includegraphics[width=17cm]{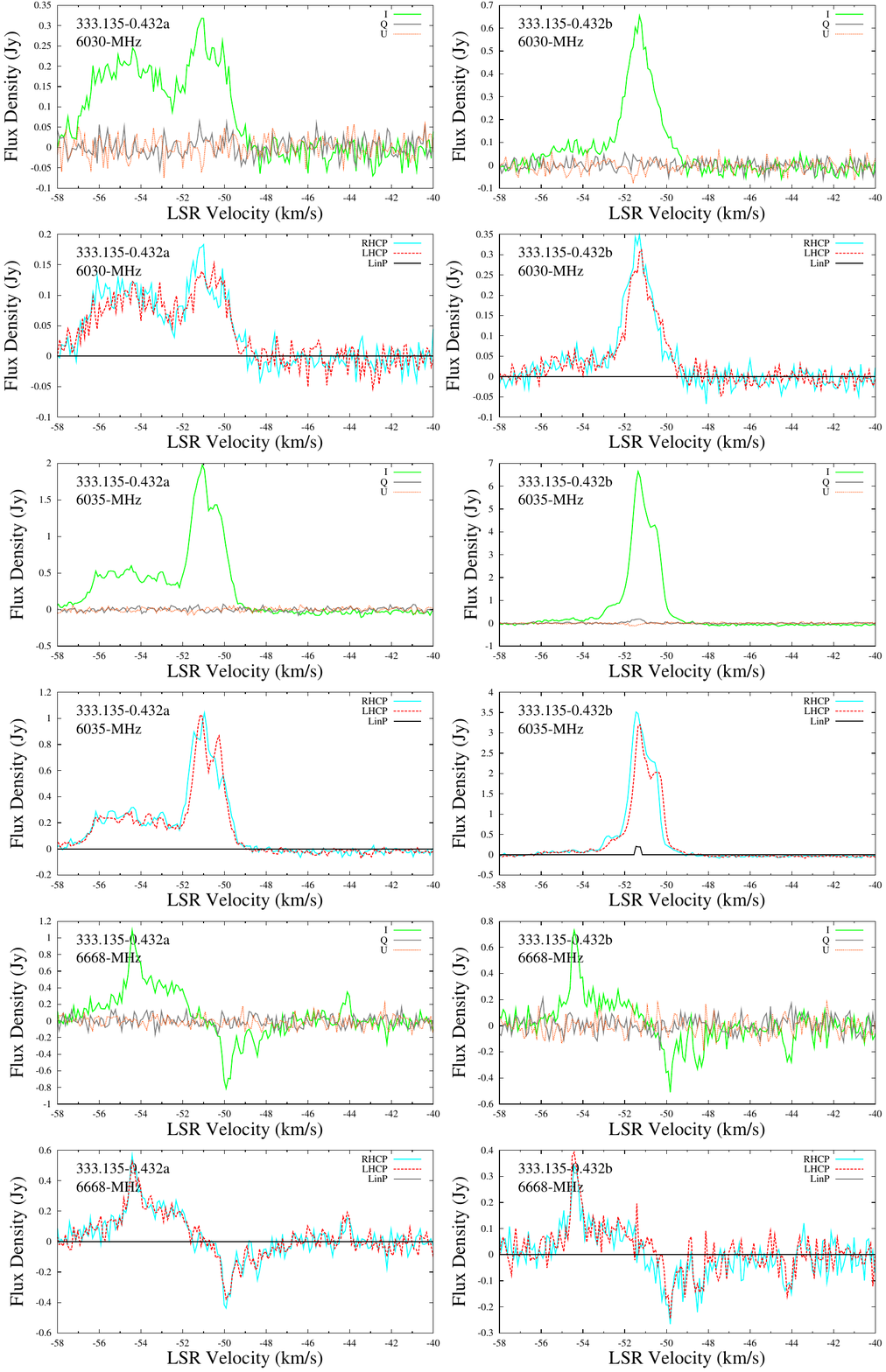}  
\caption{\small continued.}
\label{spectra}
\end{center}
\end{figure*}

\begin{figure*}
\begin{center}
\addtocounter{figure}{-1}
\renewcommand{\baselinestretch}{1.1}
\includegraphics[width=17cm]{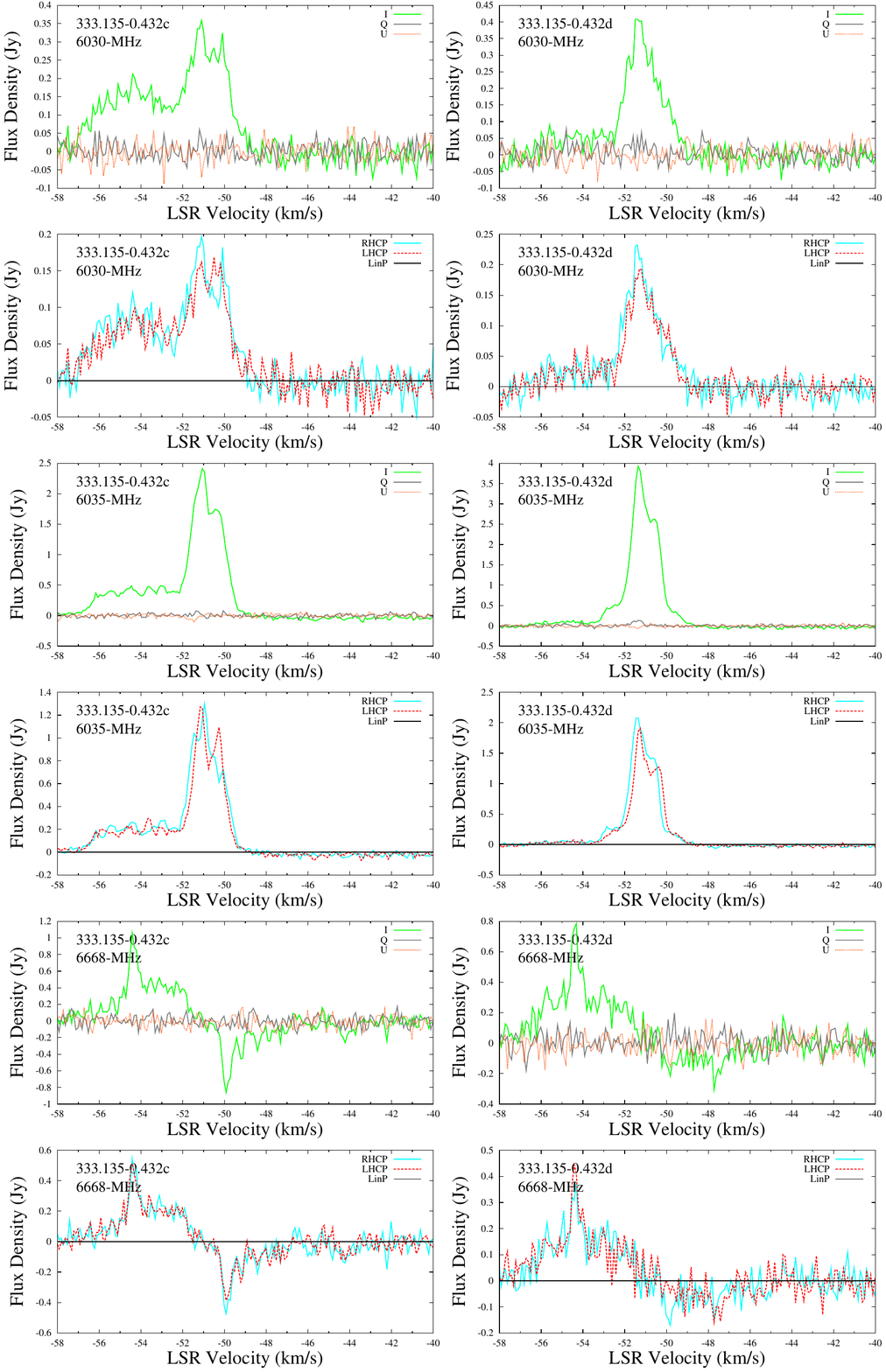}  
\caption{\small continued.}
\label{spectra}
\end{center}
\end{figure*}

\begin{figure*}
\begin{center}
\addtocounter{figure}{-1}
\renewcommand{\baselinestretch}{1.1}
\includegraphics[width=17cm]{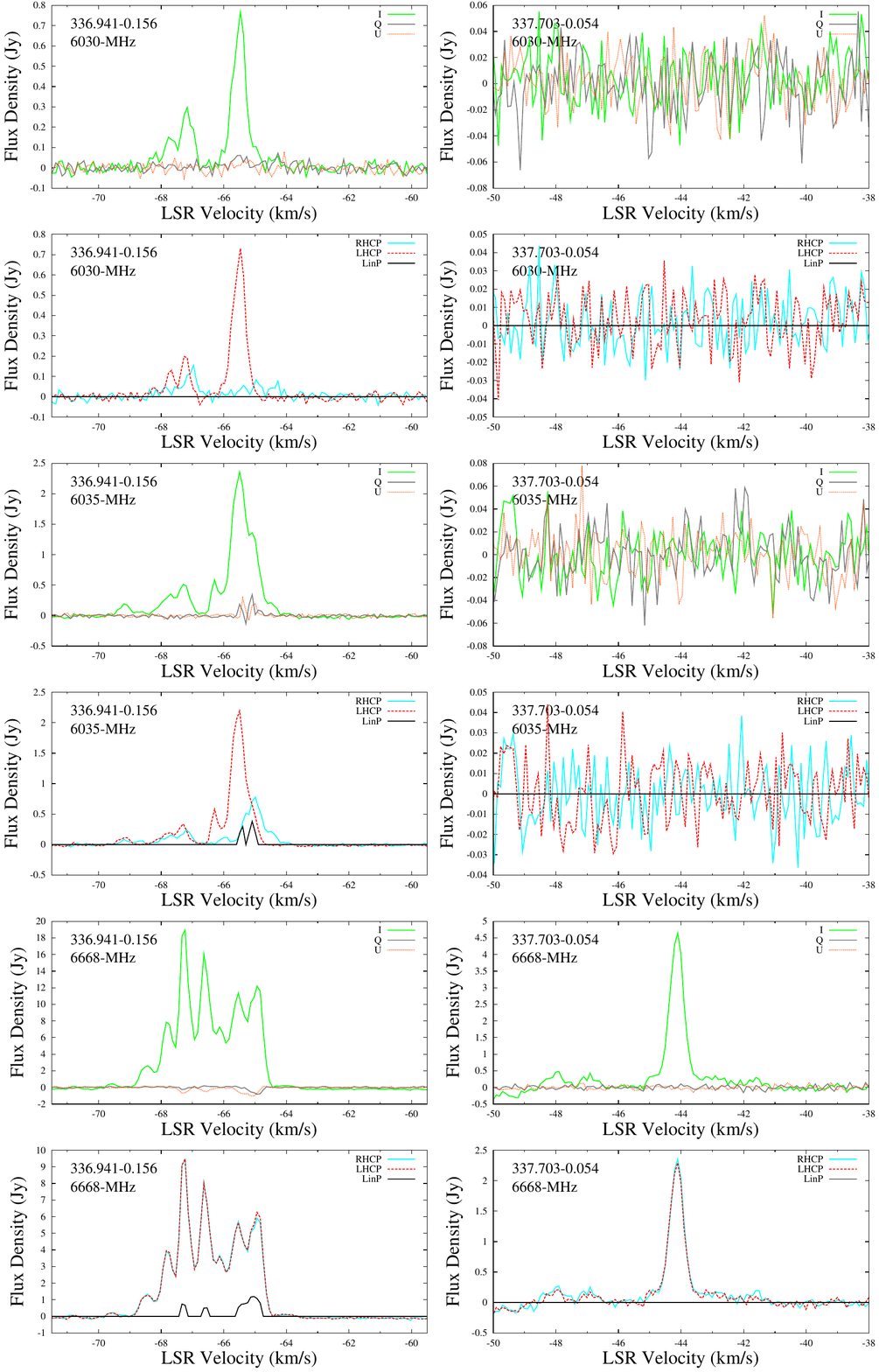}  
\caption{\small continued.}
\label{spectra}
\end{center}
\end{figure*}

\begin{figure*}
\begin{center}
\addtocounter{figure}{-1}
\renewcommand{\baselinestretch}{1.1}
\includegraphics[width=17cm]{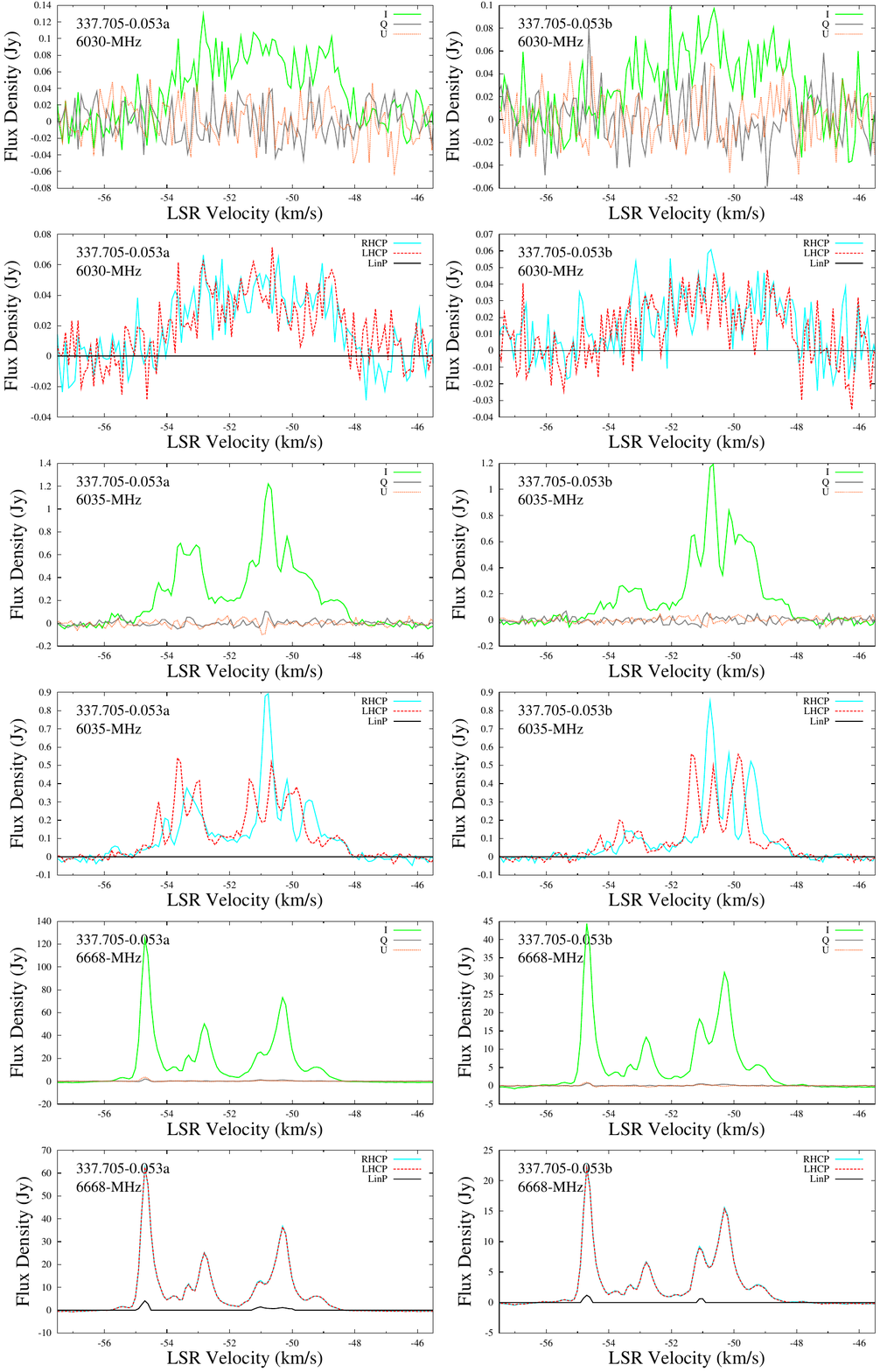}  
\caption{\small continued.}
\label{spectra}
\end{center}
\end{figure*}

\begin{figure*}
\begin{center}
\addtocounter{figure}{-1}
\renewcommand{\baselinestretch}{1.1}
\includegraphics[width=17cm]{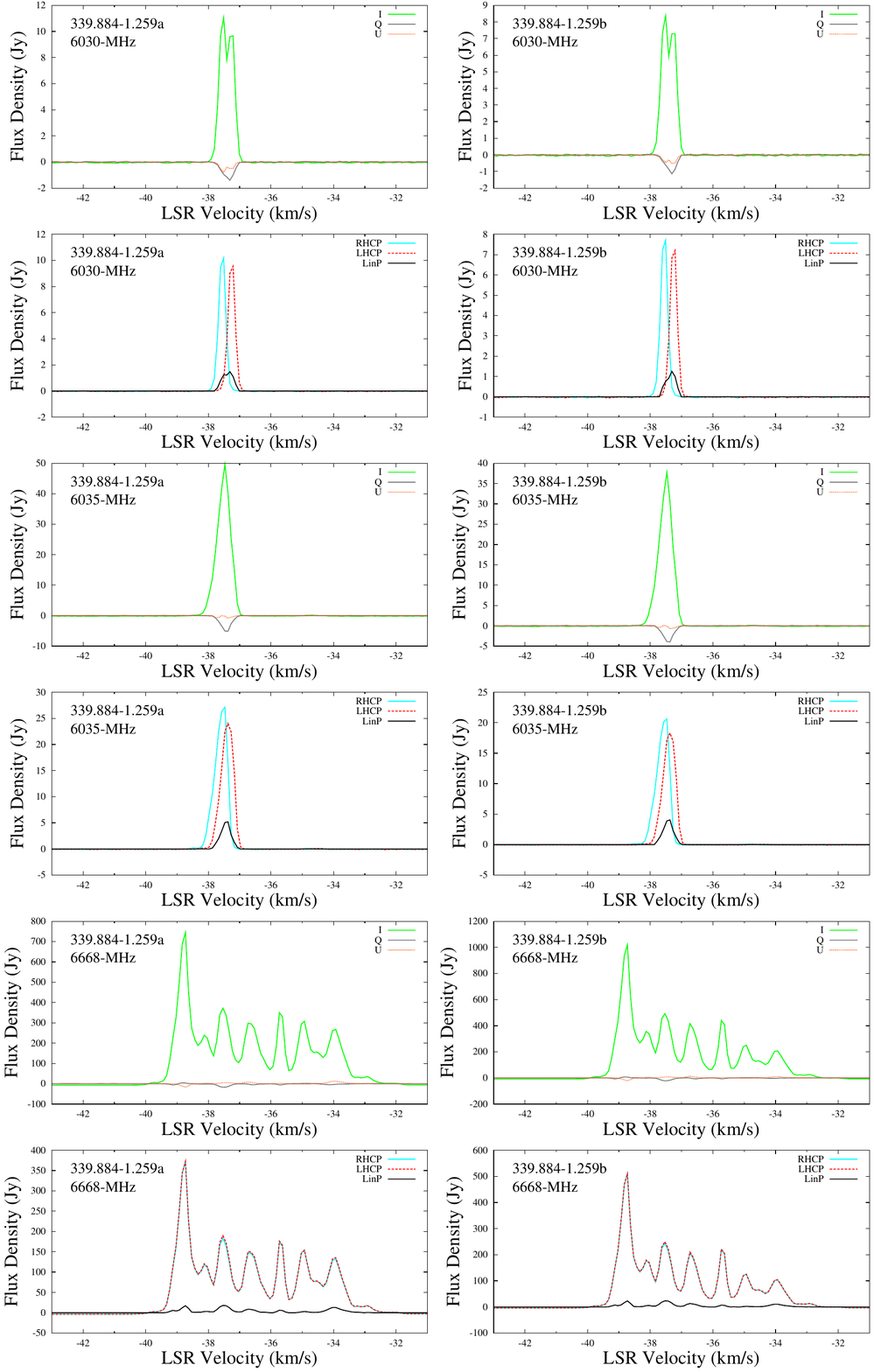}  
\caption{\small continued.}
\label{spectra}
\end{center}
\end{figure*}

\begin{figure*}
\begin{center}
\addtocounter{figure}{-1}
\renewcommand{\baselinestretch}{1.1}
\includegraphics[width=17cm]{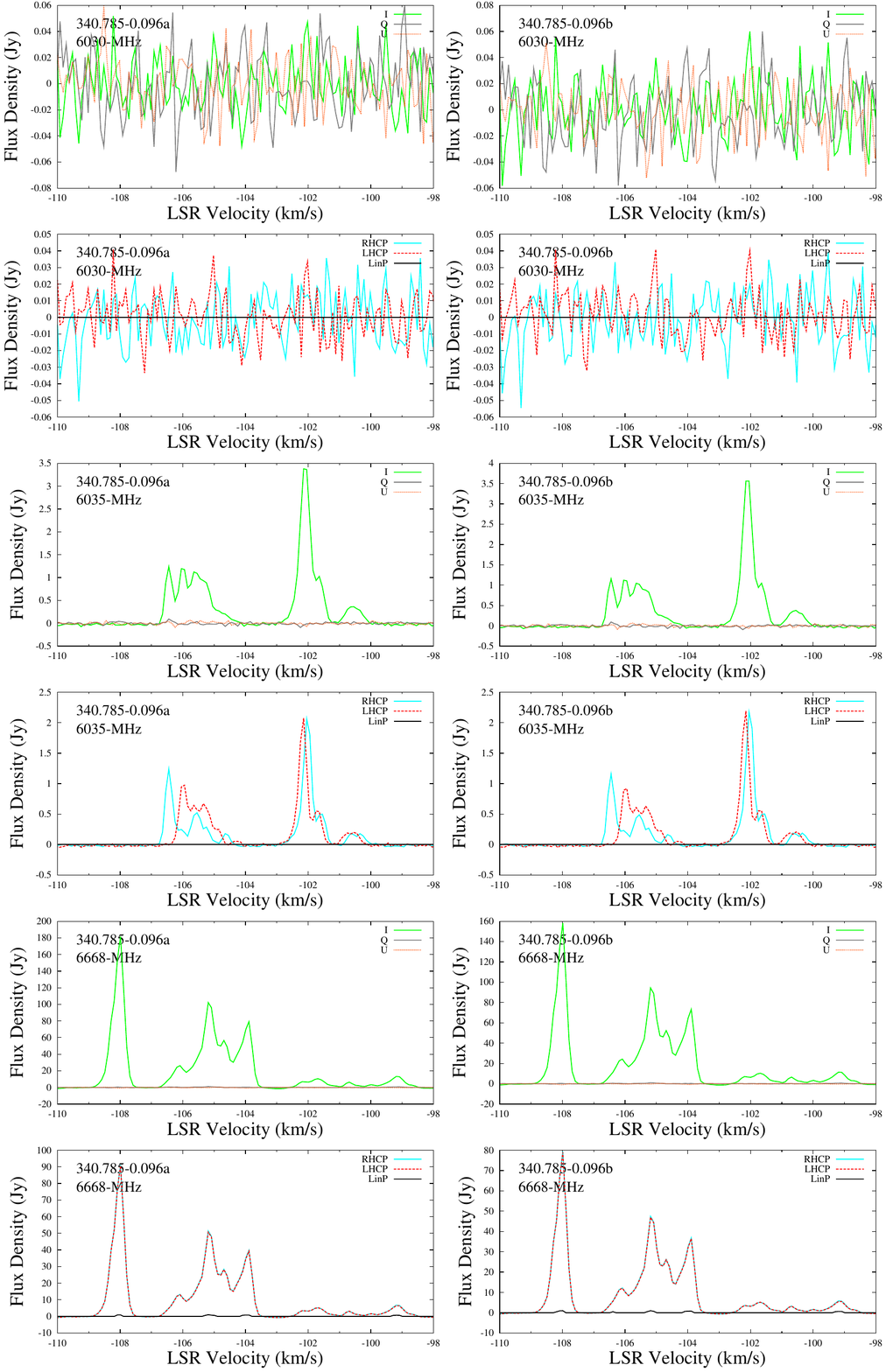}  
\caption{\small continued.}
\label{spectra}
\end{center}
\end{figure*}

\begin{figure*}
\begin{center}
\addtocounter{figure}{-1}
\renewcommand{\baselinestretch}{1.1}
\includegraphics[width=17cm]{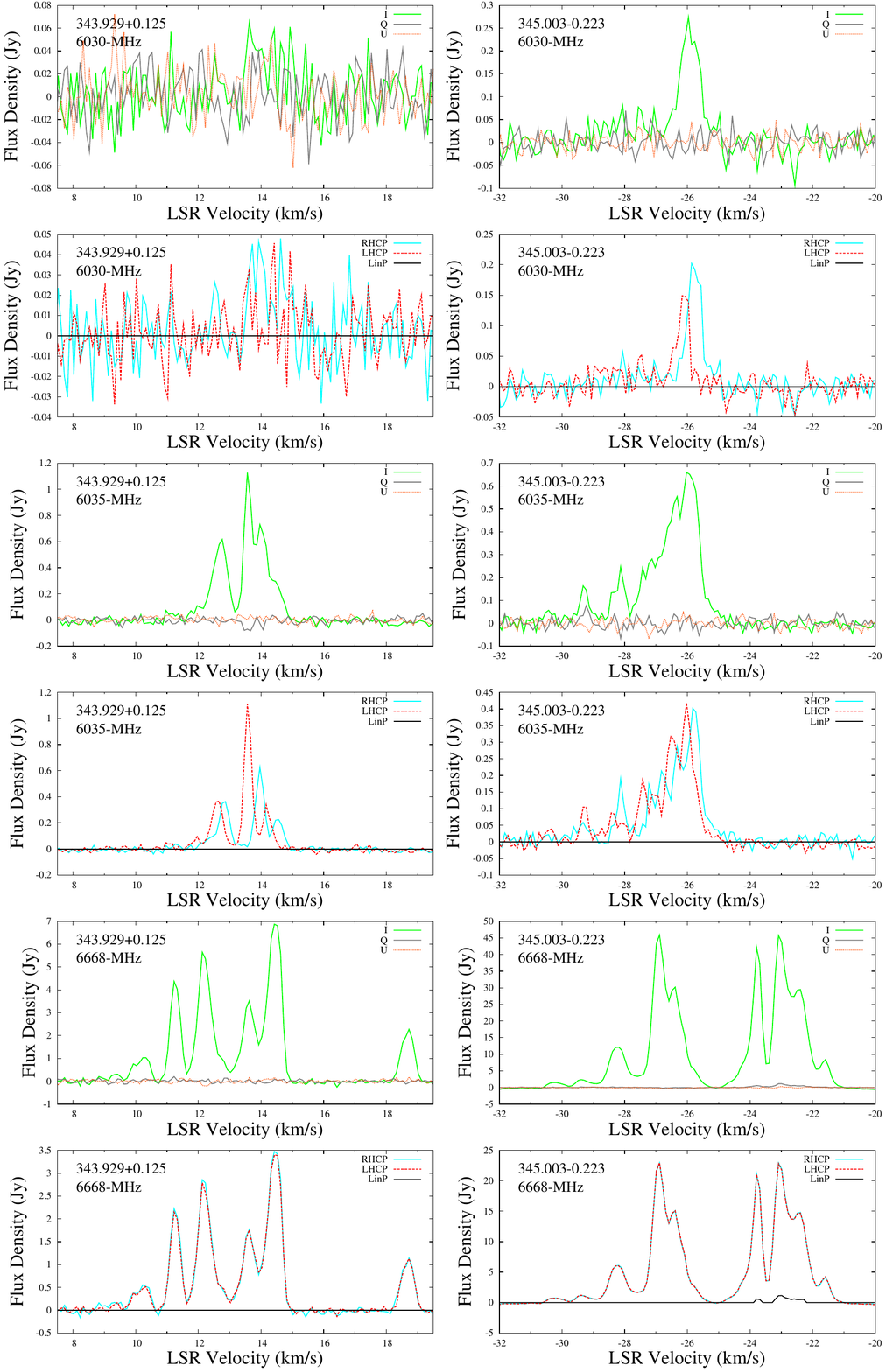}  
\caption{\small continued.}
\label{spectra}
\end{center}
\end{figure*}

\begin{figure*}
\begin{center}
\addtocounter{figure}{-1}
\renewcommand{\baselinestretch}{1.1}
\includegraphics[width=17cm]{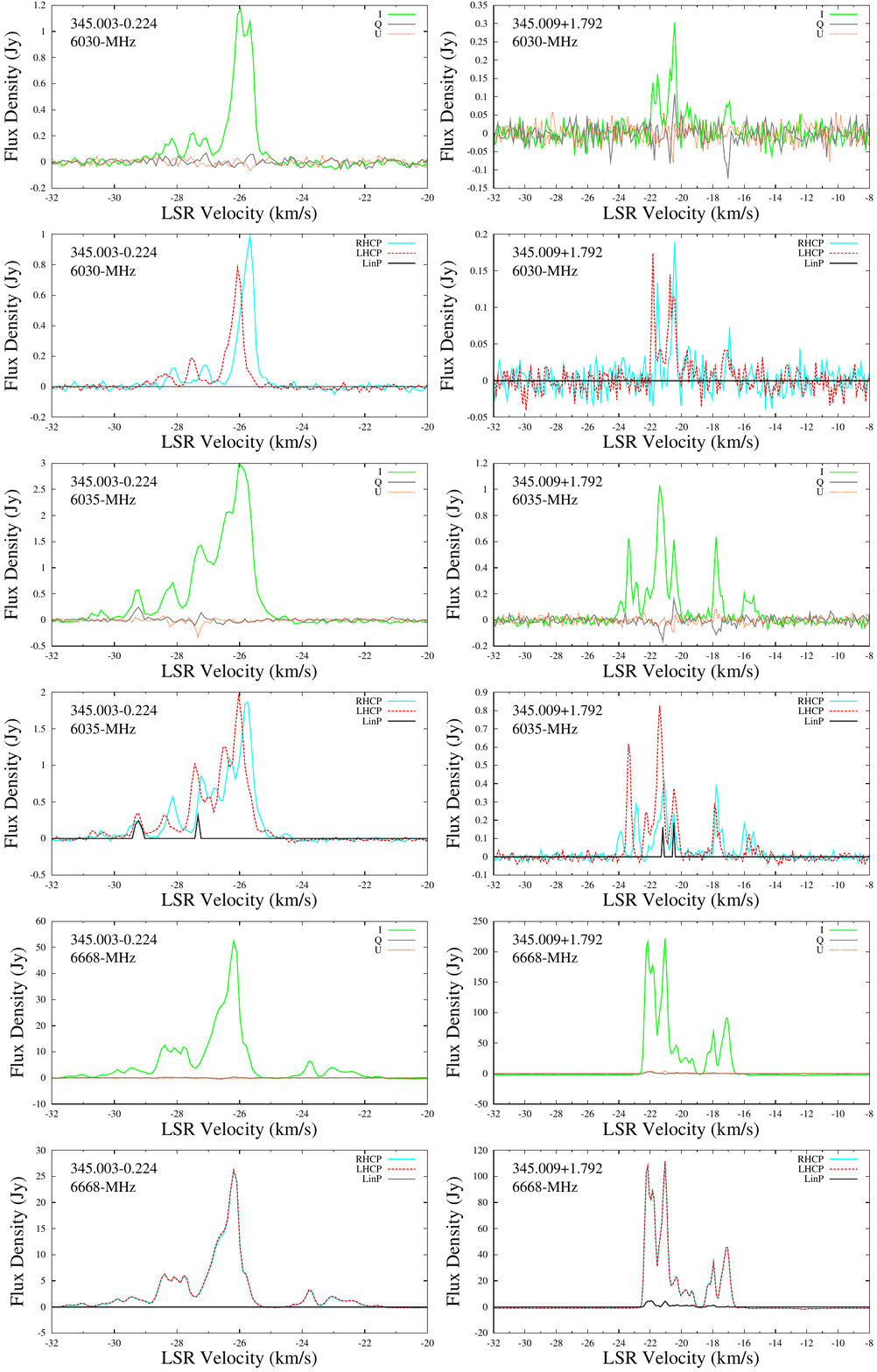}  
\caption{\small continued.}
\label{spectra}
\end{center}
\end{figure*}

\begin{figure*}
\begin{center}
\addtocounter{figure}{-1}
\renewcommand{\baselinestretch}{1.1}
\includegraphics[width=17cm]{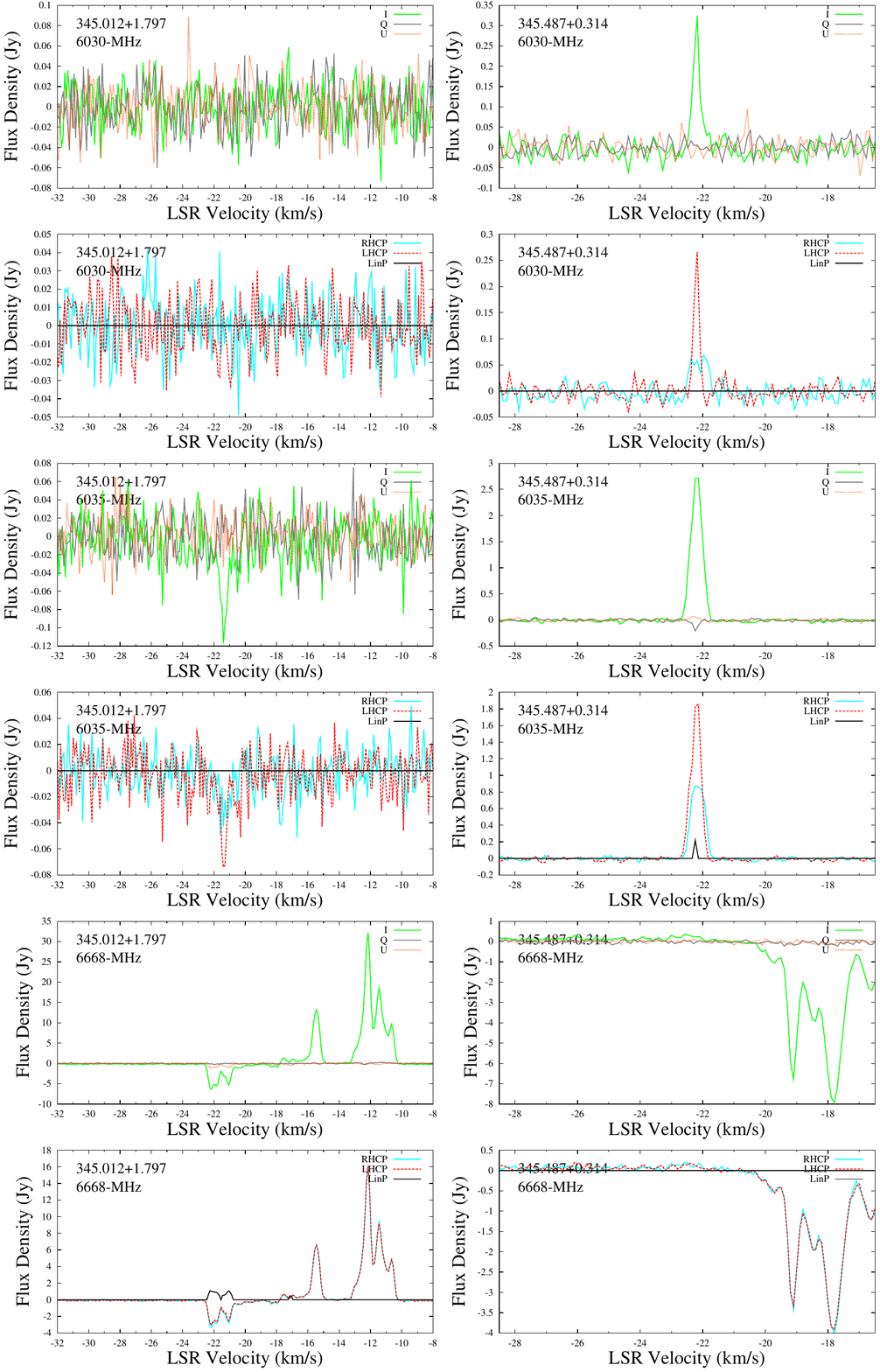}  
\caption{\small continued.}
\label{spectra}
\end{center}
\end{figure*}

\begin{figure*}
\begin{center}
\addtocounter{figure}{-1}
\renewcommand{\baselinestretch}{1.1}
\includegraphics[width=17cm]{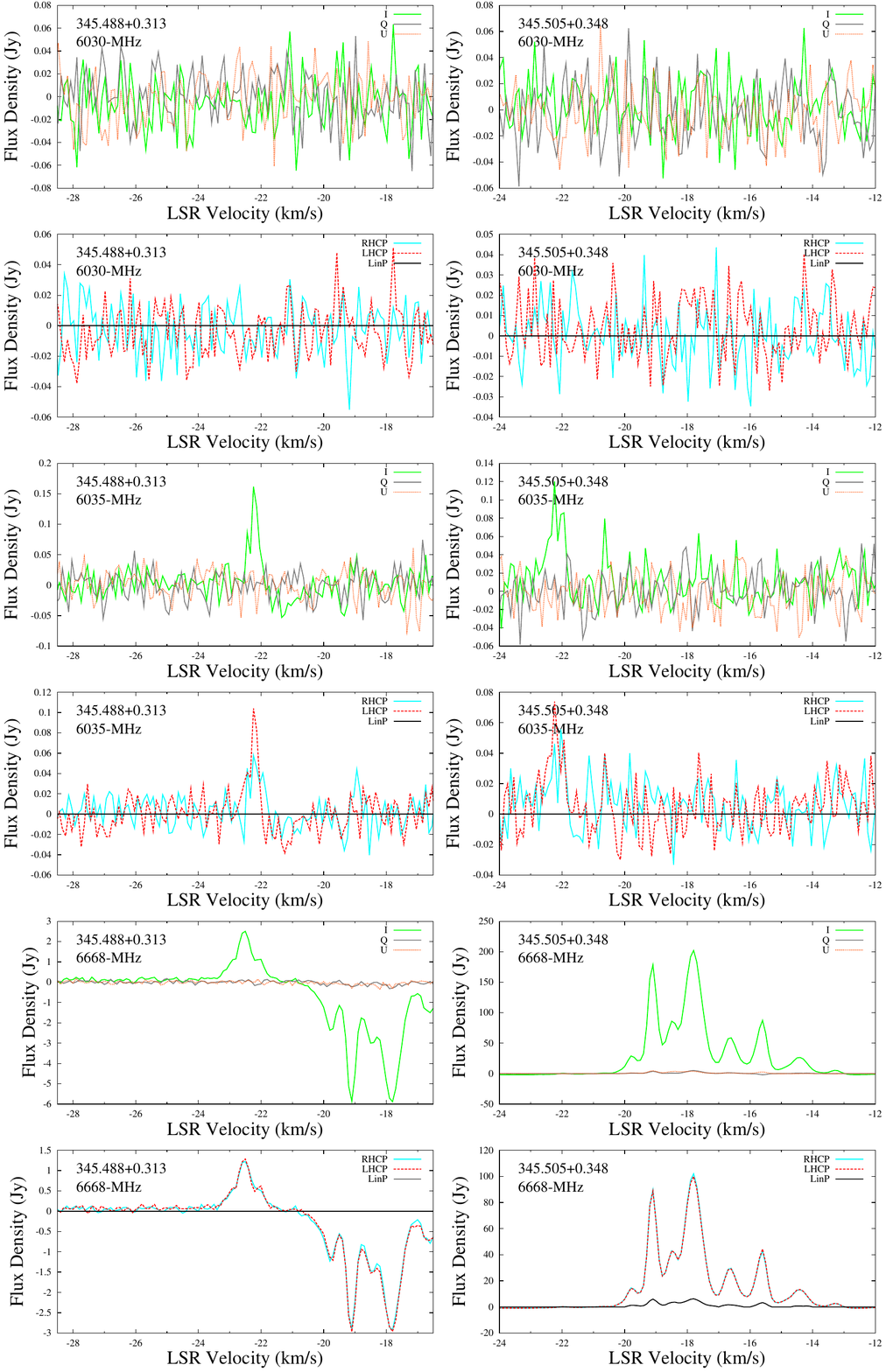}  
\caption{\small continued.}
\label{spectra}
\end{center}
\end{figure*}

\clearpage

\begin{figure*}
\begin{center}
\addtocounter{figure}{-1}
\renewcommand{\baselinestretch}{1.1}
\includegraphics[width=17cm]{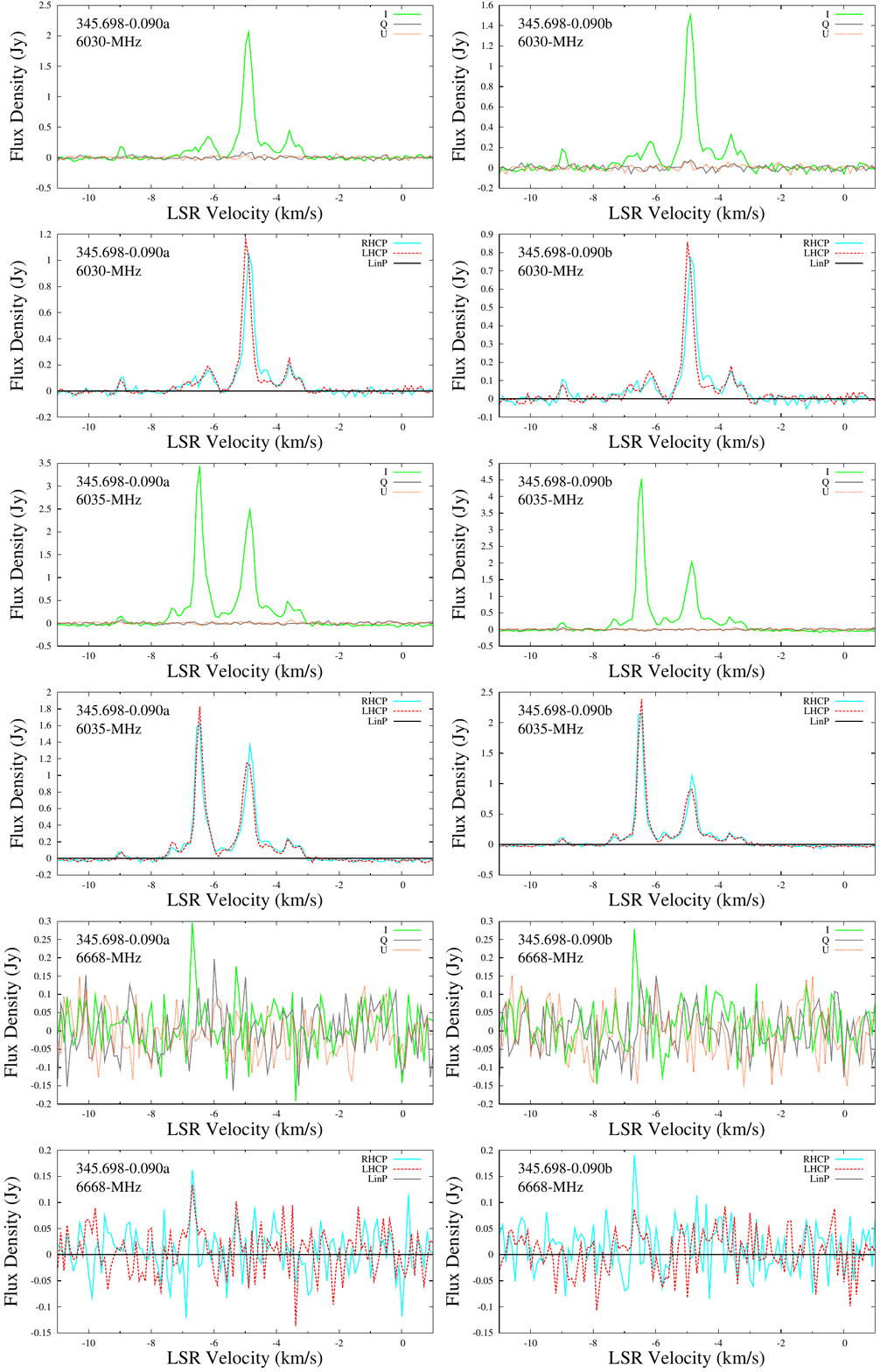}  
\caption{\small continued.}
\label{spectra}
\end{center}
\end{figure*}

\begin{figure*}
\begin{center}
\addtocounter{figure}{-1}
\renewcommand{\baselinestretch}{1.1}
\includegraphics[width=17cm]{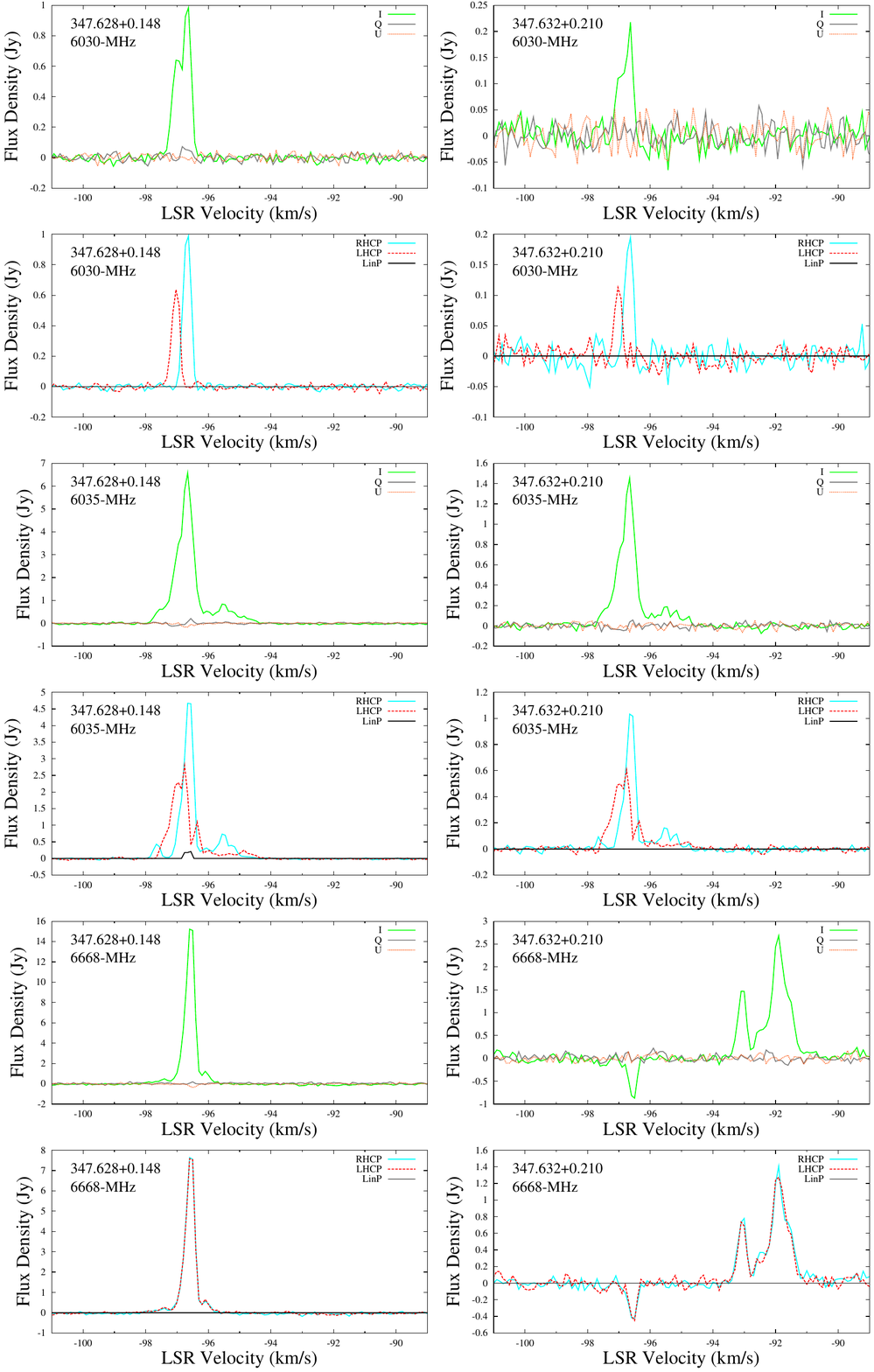}  
\caption{\small continued.}
\label{spectra}
\end{center}
\end{figure*}

\begin{figure*}
\begin{center}
\addtocounter{figure}{-1}
\renewcommand{\baselinestretch}{1.1}
\includegraphics[width=17cm]{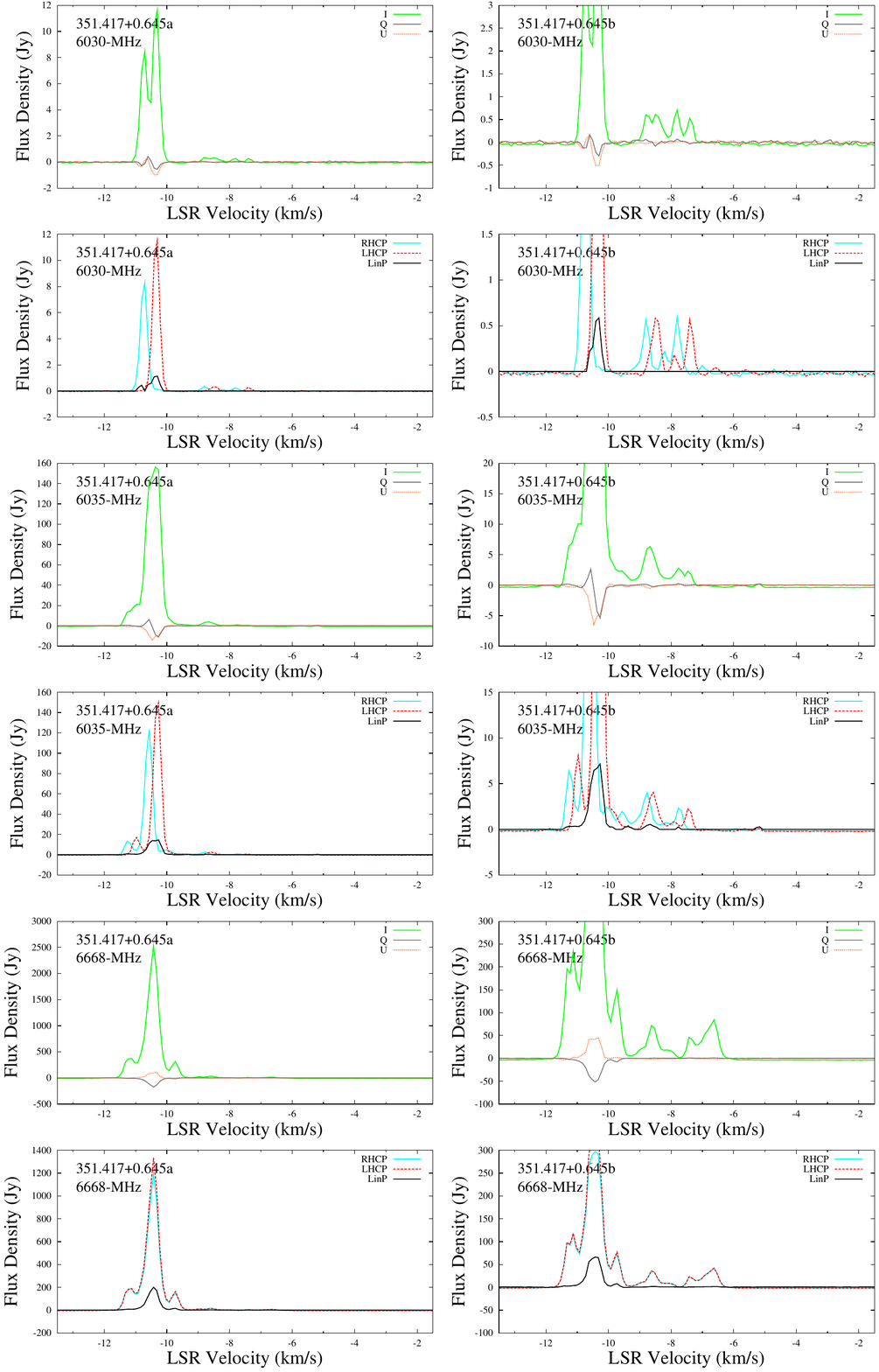}  
\caption{\small continued.}
\label{spectra}
\end{center}
\end{figure*}

\clearpage

\begin{figure*}
\begin{center}
\addtocounter{figure}{-1}
\renewcommand{\baselinestretch}{1.1}
\includegraphics[width=17cm]{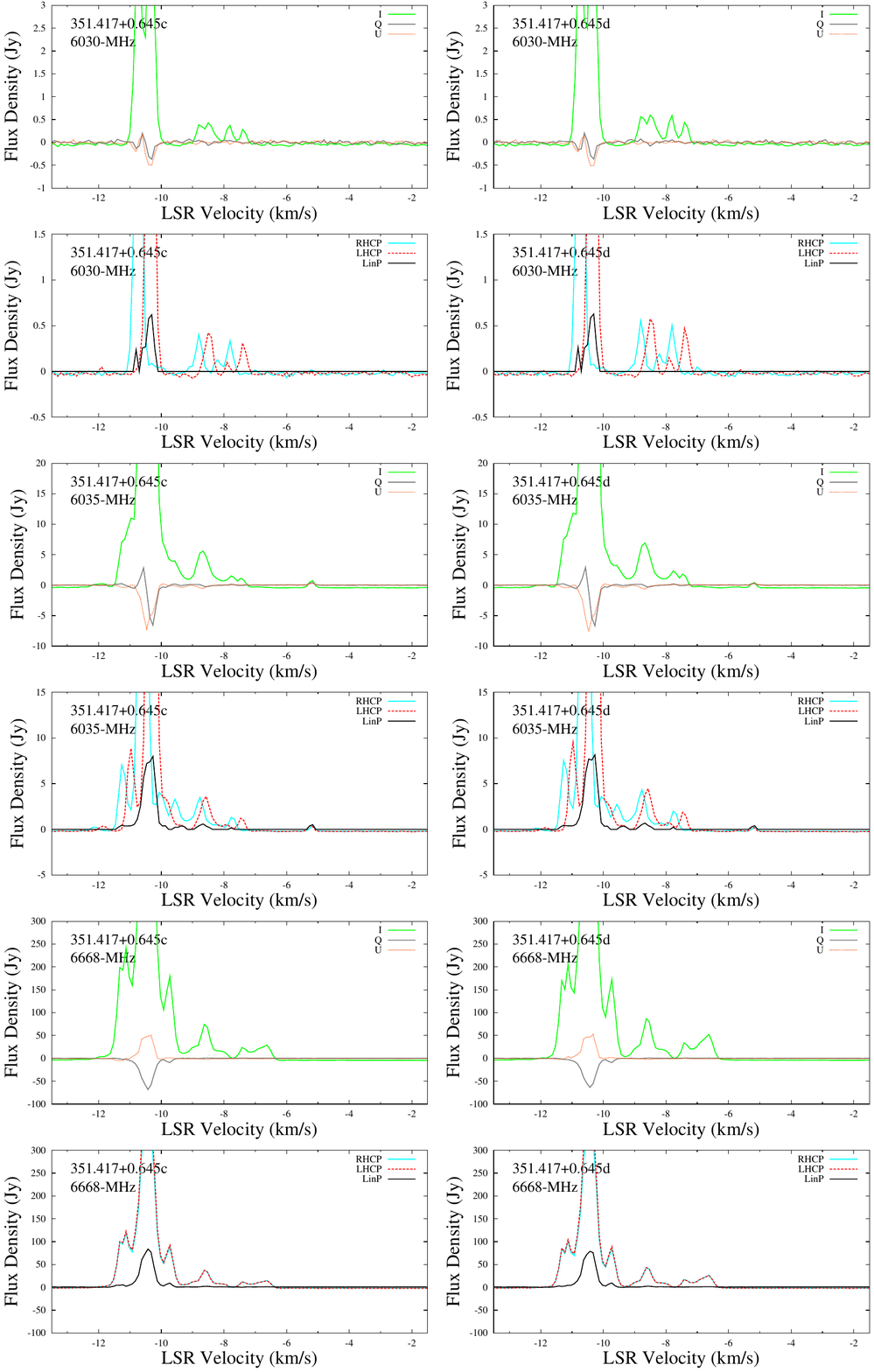}  
\caption{\small continued.}
\label{spectra}
\end{center}
\end{figure*}

\begin{figure*}
\begin{center}
\addtocounter{figure}{-1}
\renewcommand{\baselinestretch}{1.1}
\includegraphics[width=17cm]{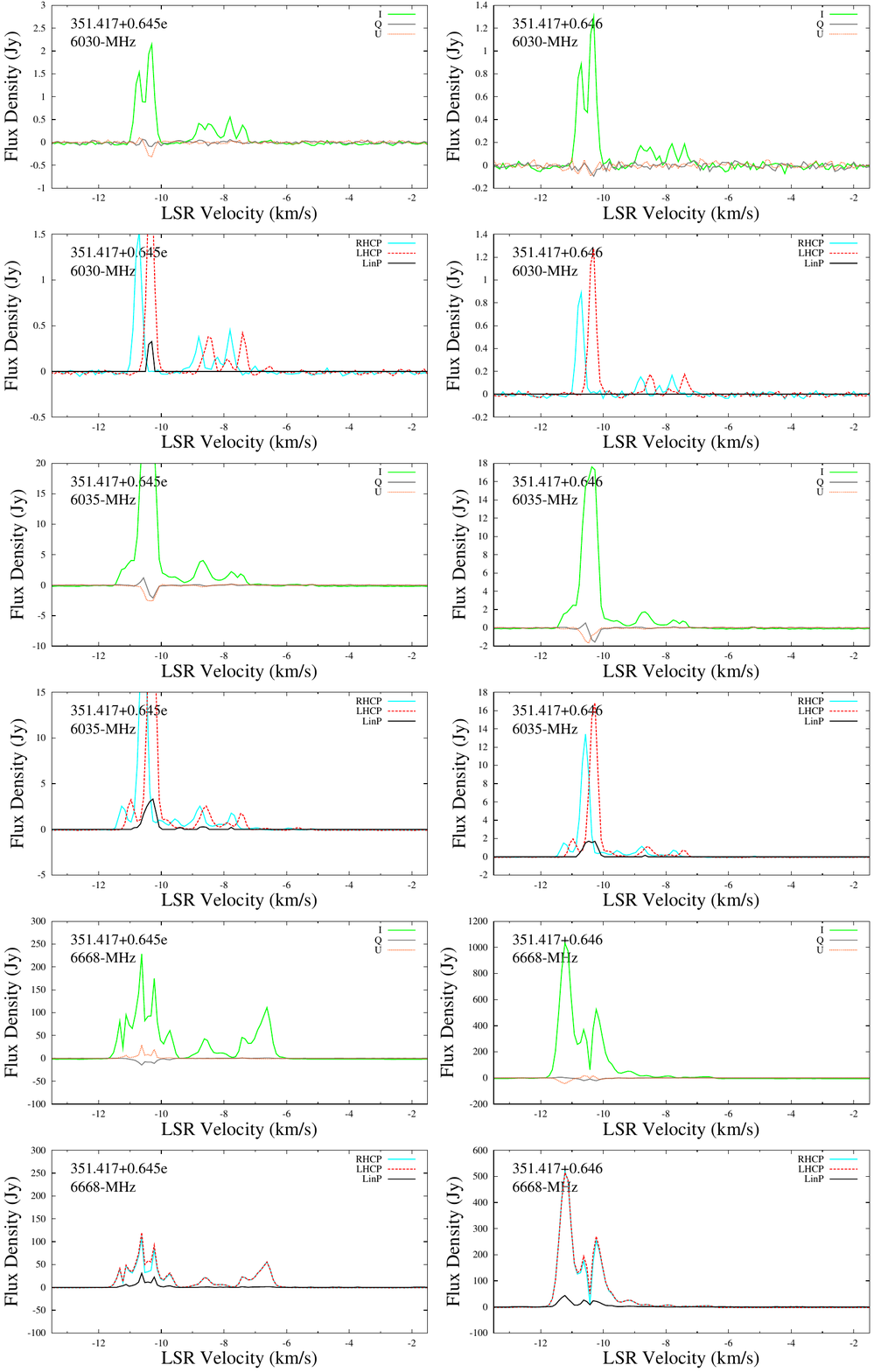}  
\caption{\small continued.}
\label{spectra}
\end{center}
\end{figure*}

\begin{figure*}
\begin{center}
\addtocounter{figure}{-1}
\renewcommand{\baselinestretch}{1.1}
\includegraphics[width=17cm]{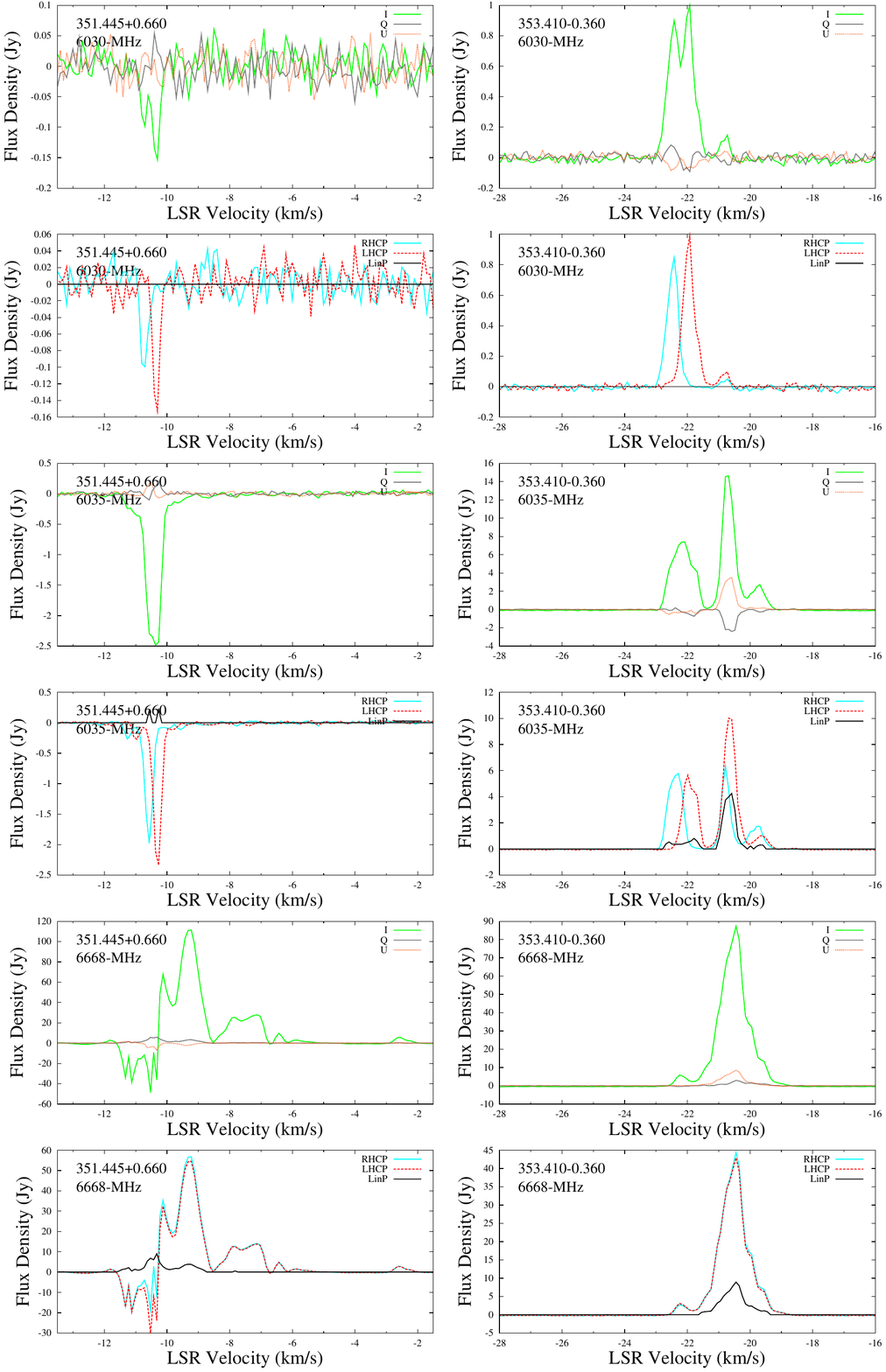}  
\caption{\small continued.}
\label{spectra}
\end{center}
\end{figure*}

\clearpage

\begin{figure*}
\begin{center}
\addtocounter{figure}{-1}
\renewcommand{\baselinestretch}{1.1}
\includegraphics[width=17cm]{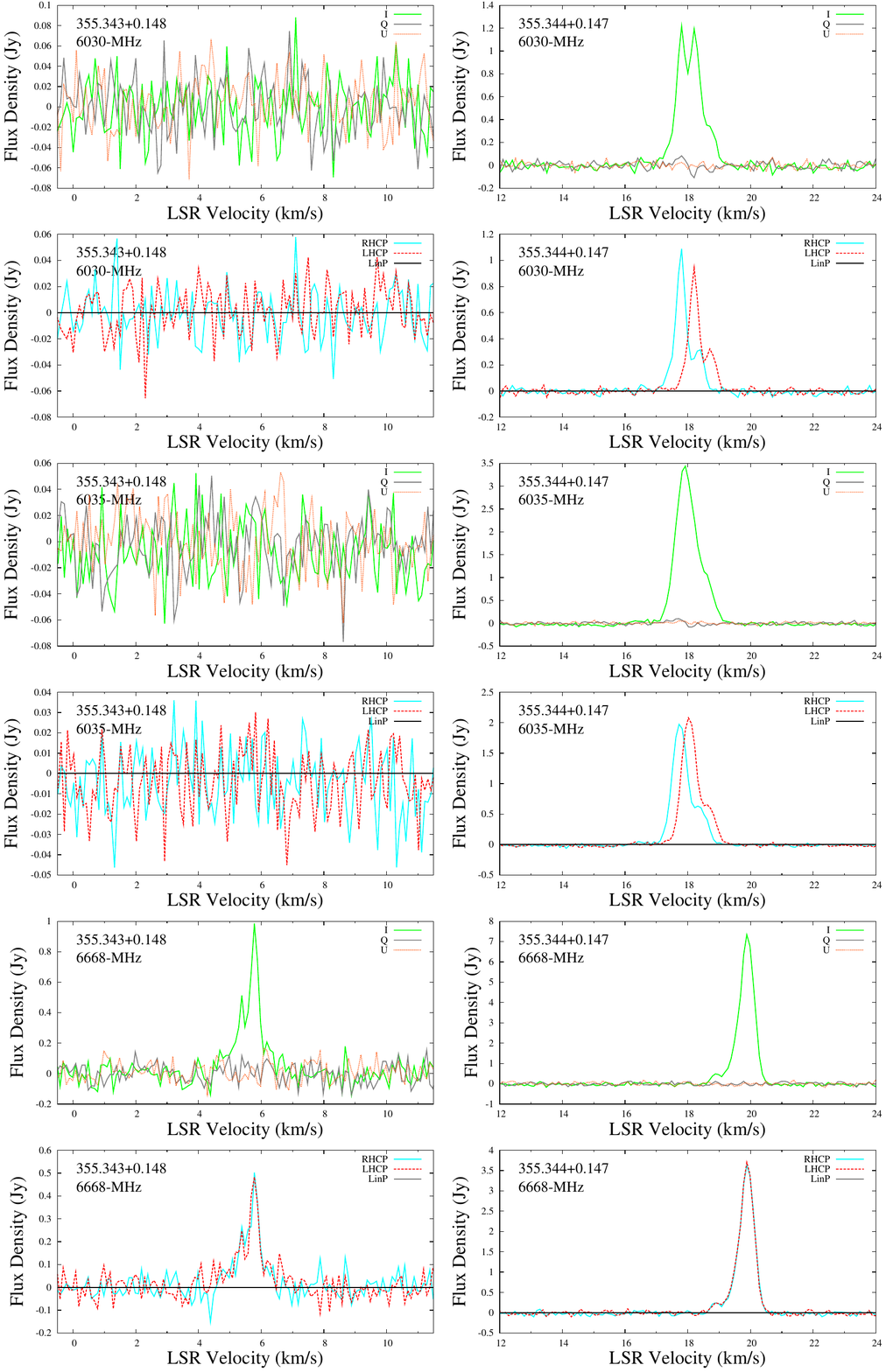}  
\caption{\small continued.}
\label{spectra}
\end{center}
\end{figure*}

\begin{figure*}
\begin{center}
\addtocounter{figure}{-1}
\renewcommand{\baselinestretch}{1.1}
\includegraphics[width=17cm]{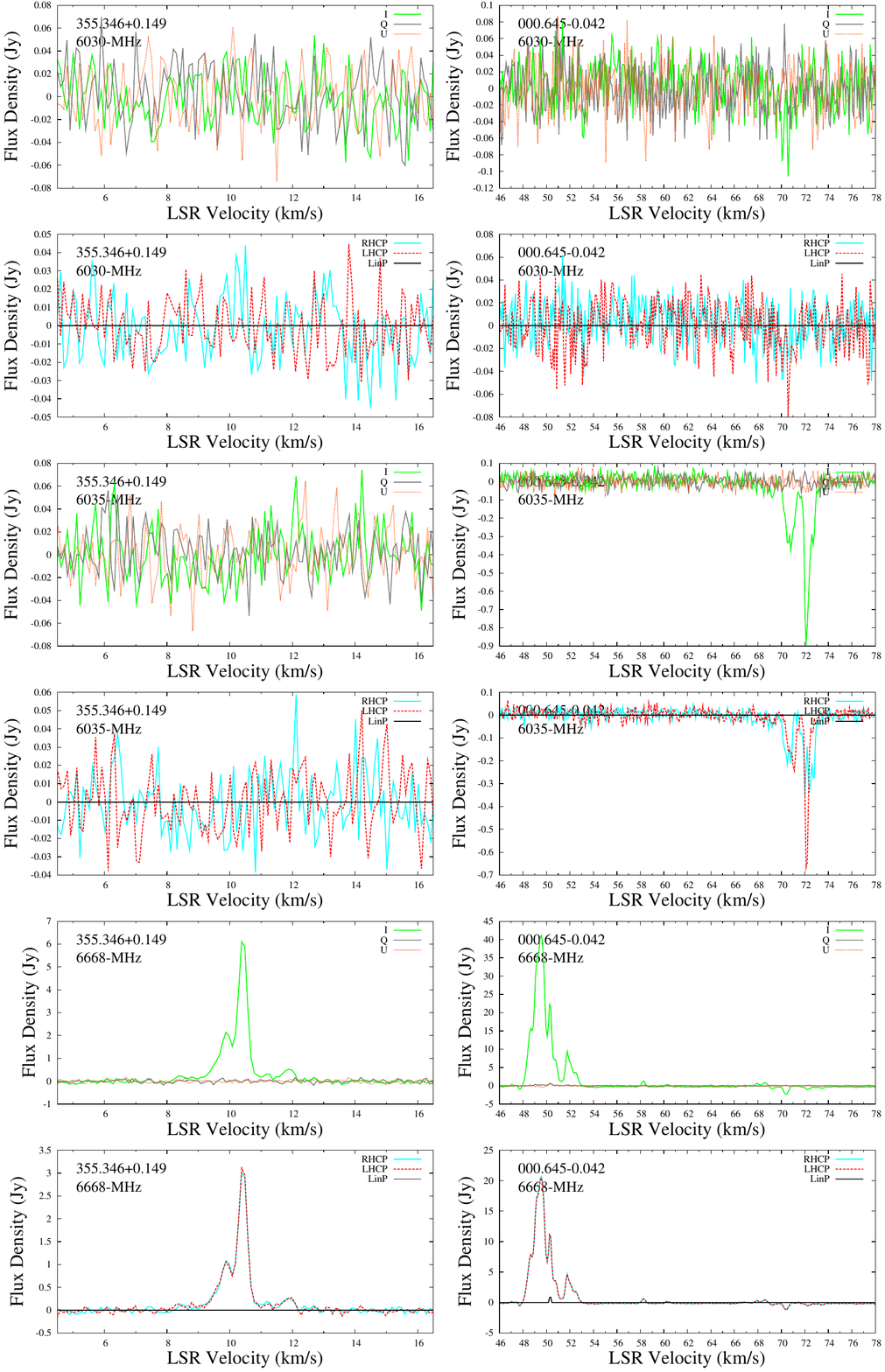}  
\caption{\small continued.}
\label{spectra}
\end{center}
\end{figure*}

\begin{figure*}
\begin{center}
\addtocounter{figure}{-1}
\renewcommand{\baselinestretch}{1.1}
\includegraphics[width=17cm]{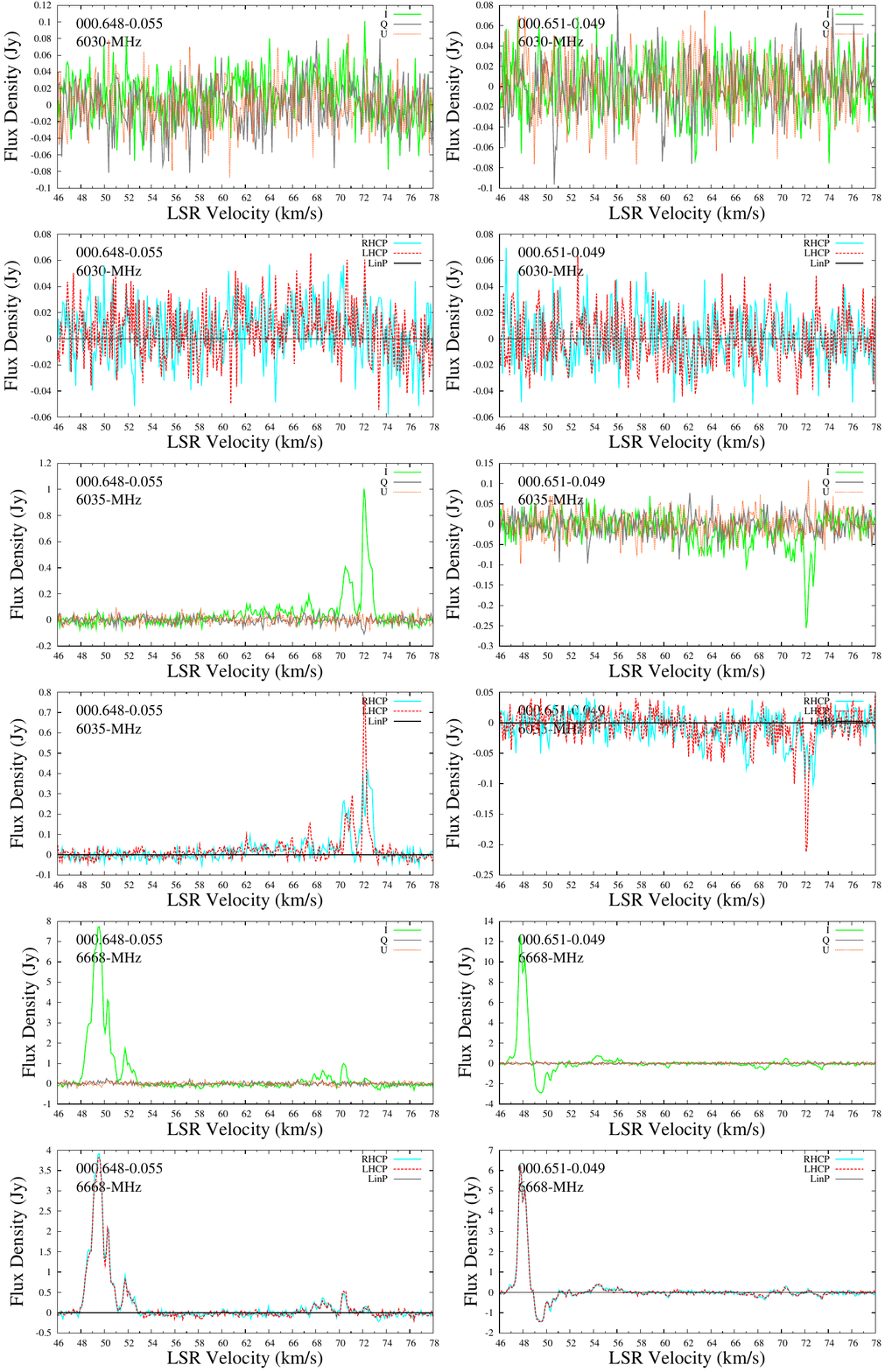}  
\caption{\small continued.}
\label{spectra}
\end{center}
\end{figure*}

\begin{figure*}
\begin{center}
\addtocounter{figure}{-1}
\renewcommand{\baselinestretch}{1.1}
\includegraphics[width=17cm]{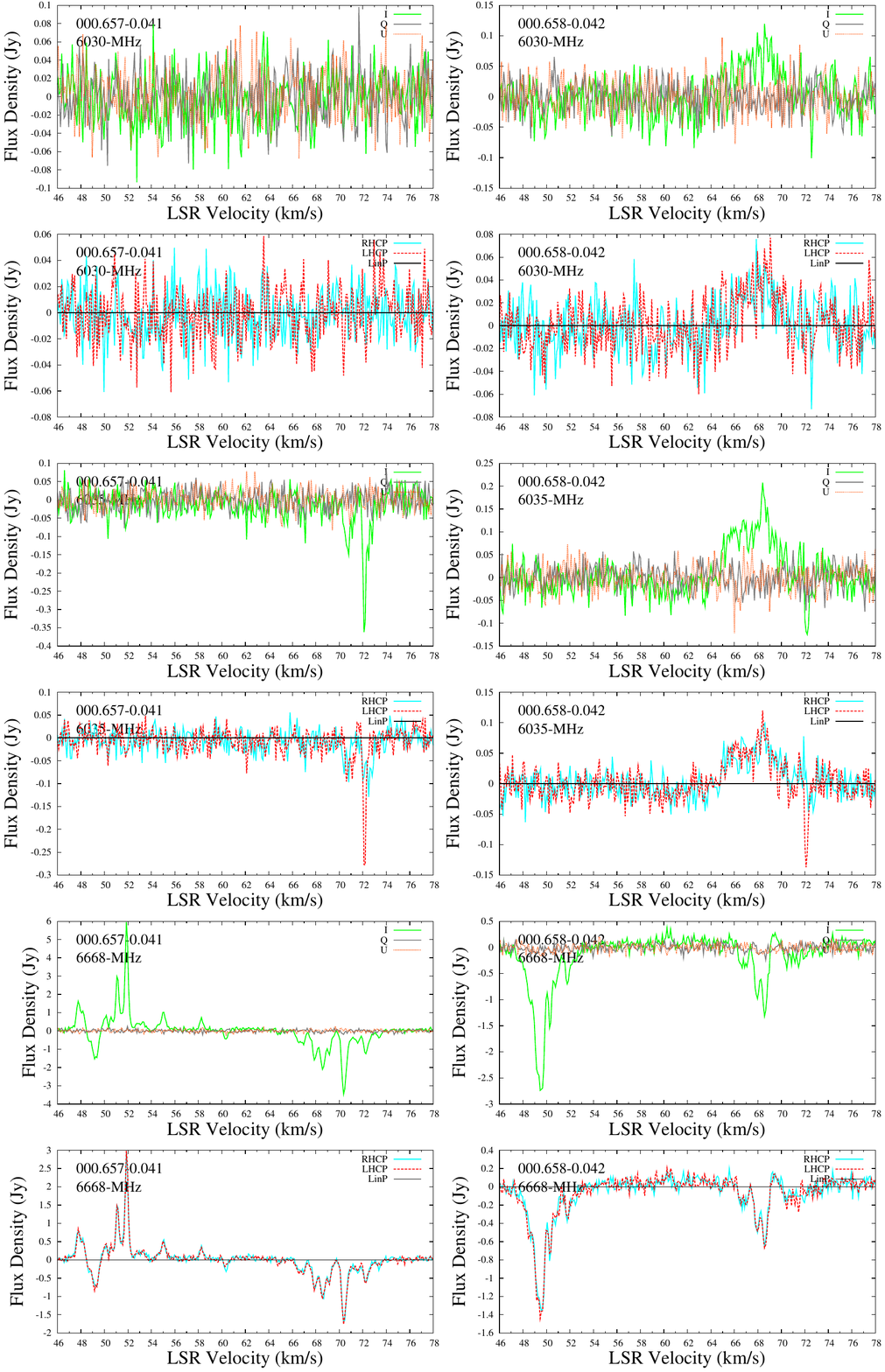}  
\caption{\small continued.}
\label{spectra}
\end{center}
\end{figure*}

\clearpage

\begin{figure*}
\begin{center}
\addtocounter{figure}{-1}
\renewcommand{\baselinestretch}{1.1}
\includegraphics[width=17cm]{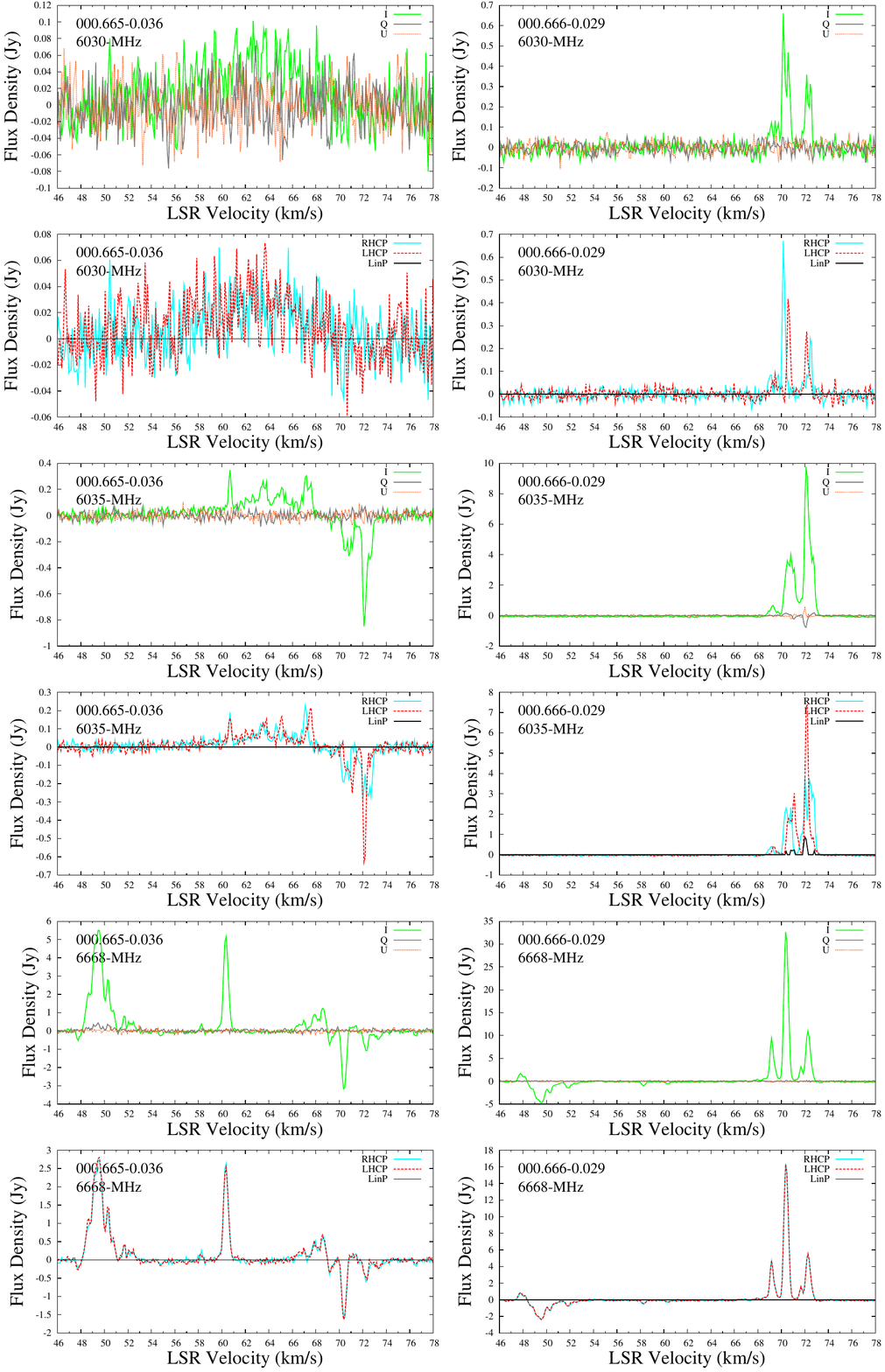}  
\caption{\small continued.}
\label{spectra}
\end{center}
\end{figure*}

\begin{figure*}
\begin{center}
\addtocounter{figure}{-1}
\renewcommand{\baselinestretch}{1.1}
\includegraphics[width=17cm]{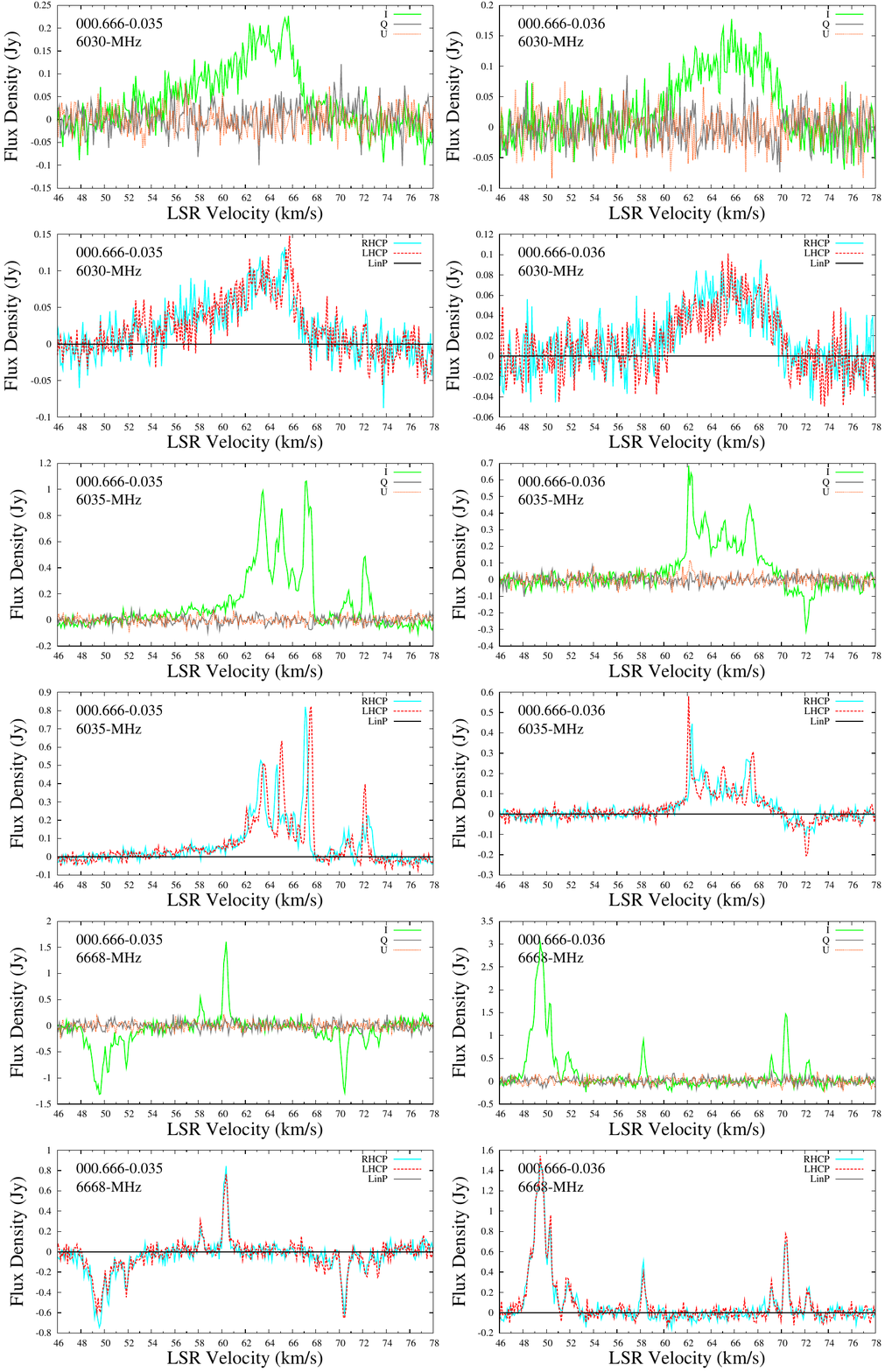}  
\caption{\small continued.}
\label{spectra}
\end{center}
\end{figure*}

\begin{figure*}
\begin{center}
\addtocounter{figure}{-1}
\renewcommand{\baselinestretch}{1.1}
\includegraphics[width=17cm]{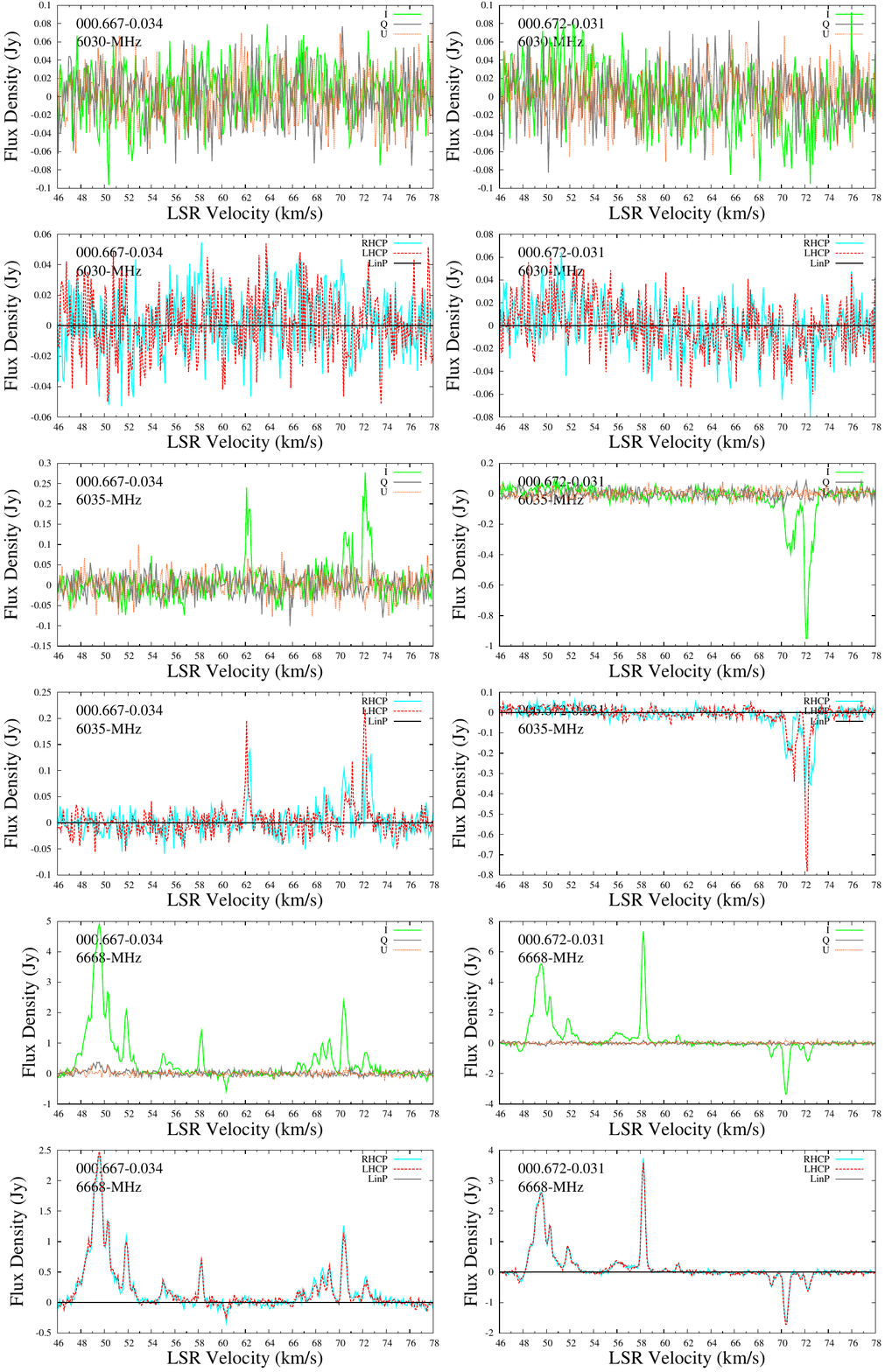}  
\caption{\small continued.}
\label{spectra}
\end{center}
\end{figure*}

\begin{figure*}
\begin{center}
\addtocounter{figure}{-1}
\renewcommand{\baselinestretch}{1.1}
\includegraphics[width=17cm]{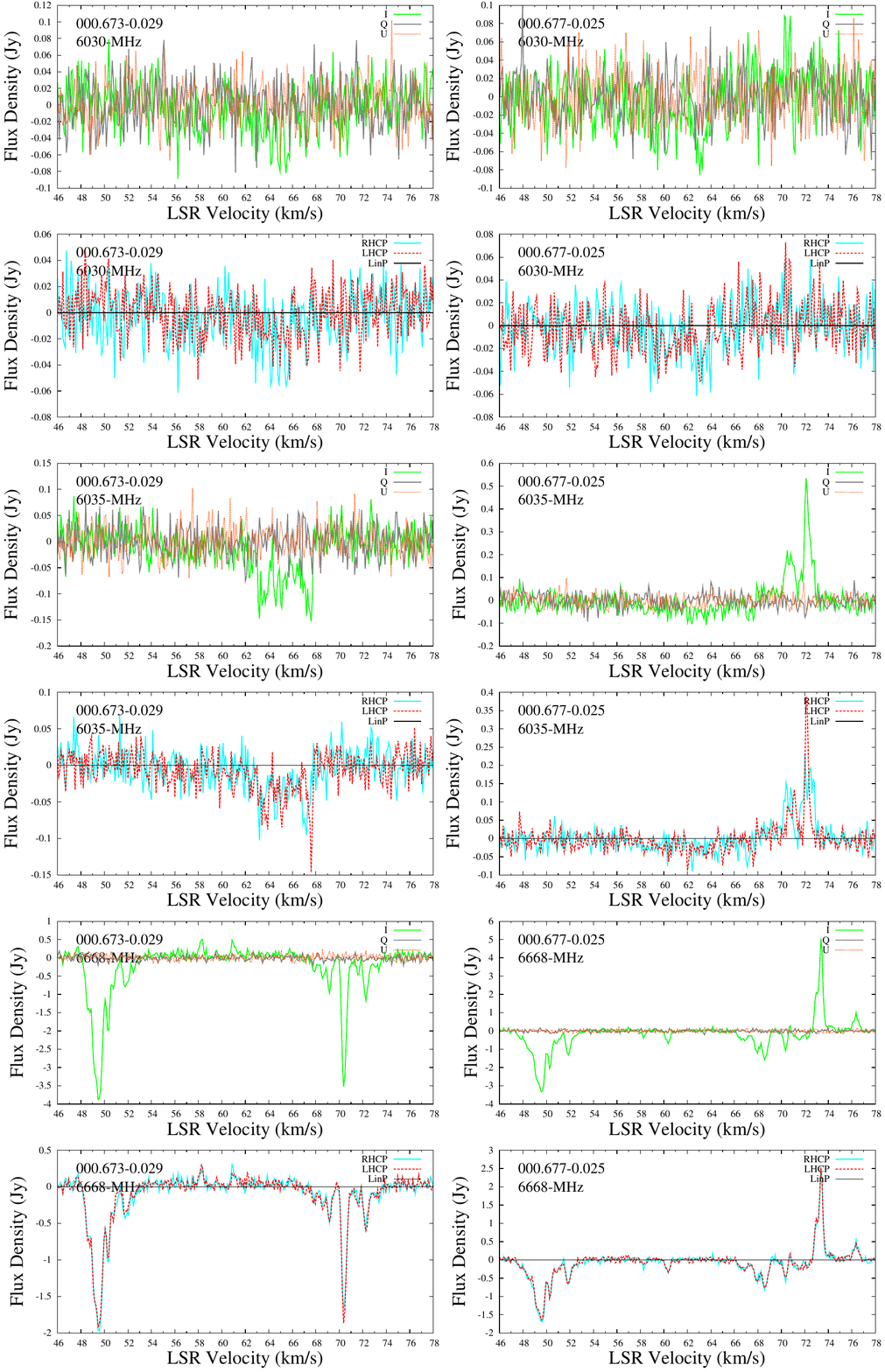}  
\caption{\small continued.}
\label{spectra}
\end{center}
\end{figure*}

\begin{figure*}
\begin{center}
\addtocounter{figure}{-1}
\renewcommand{\baselinestretch}{1.1}
\includegraphics[width=17cm]{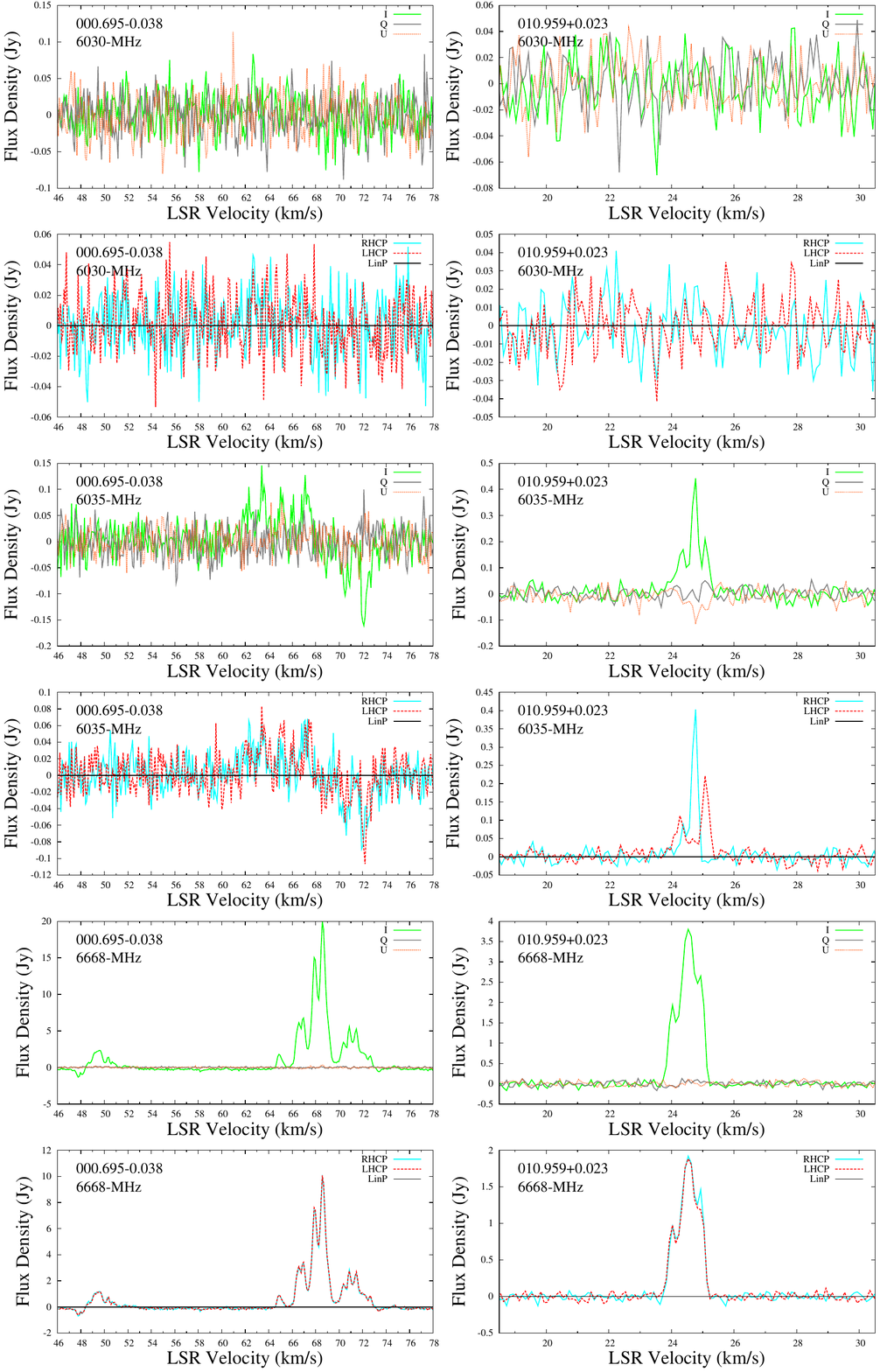}  
\caption{\small continued.}
\label{spectra}
\end{center}
\end{figure*}

\begin{figure*}
\begin{center}
\addtocounter{figure}{-1}
\renewcommand{\baselinestretch}{1.1}
\includegraphics[width=17cm]{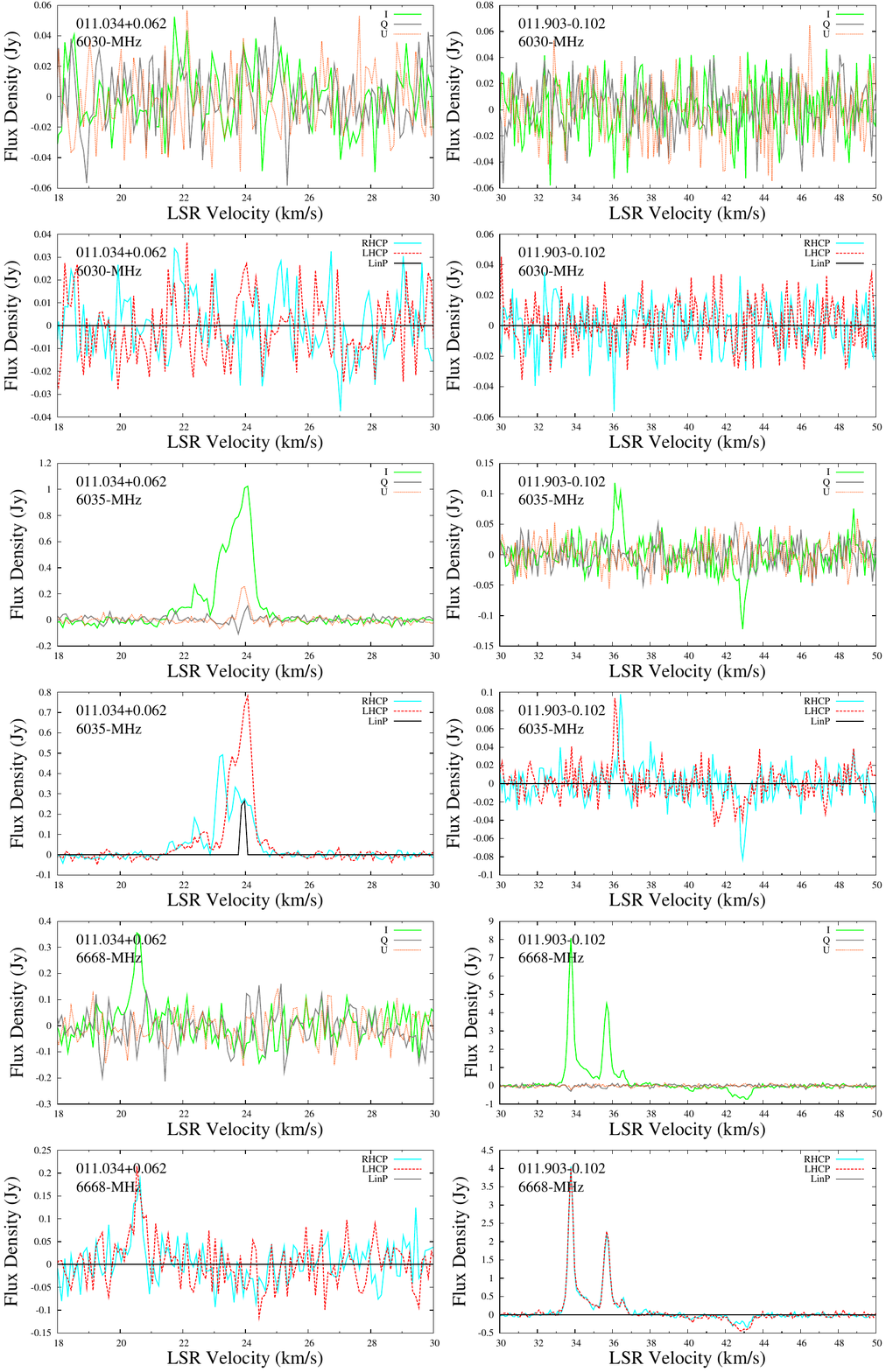}  
\caption{\small continued.}
\label{spectra}
\end{center}
\end{figure*}

\begin{figure*}
\begin{center}
\addtocounter{figure}{-1}
\renewcommand{\baselinestretch}{1.1}
\includegraphics[width=17cm]{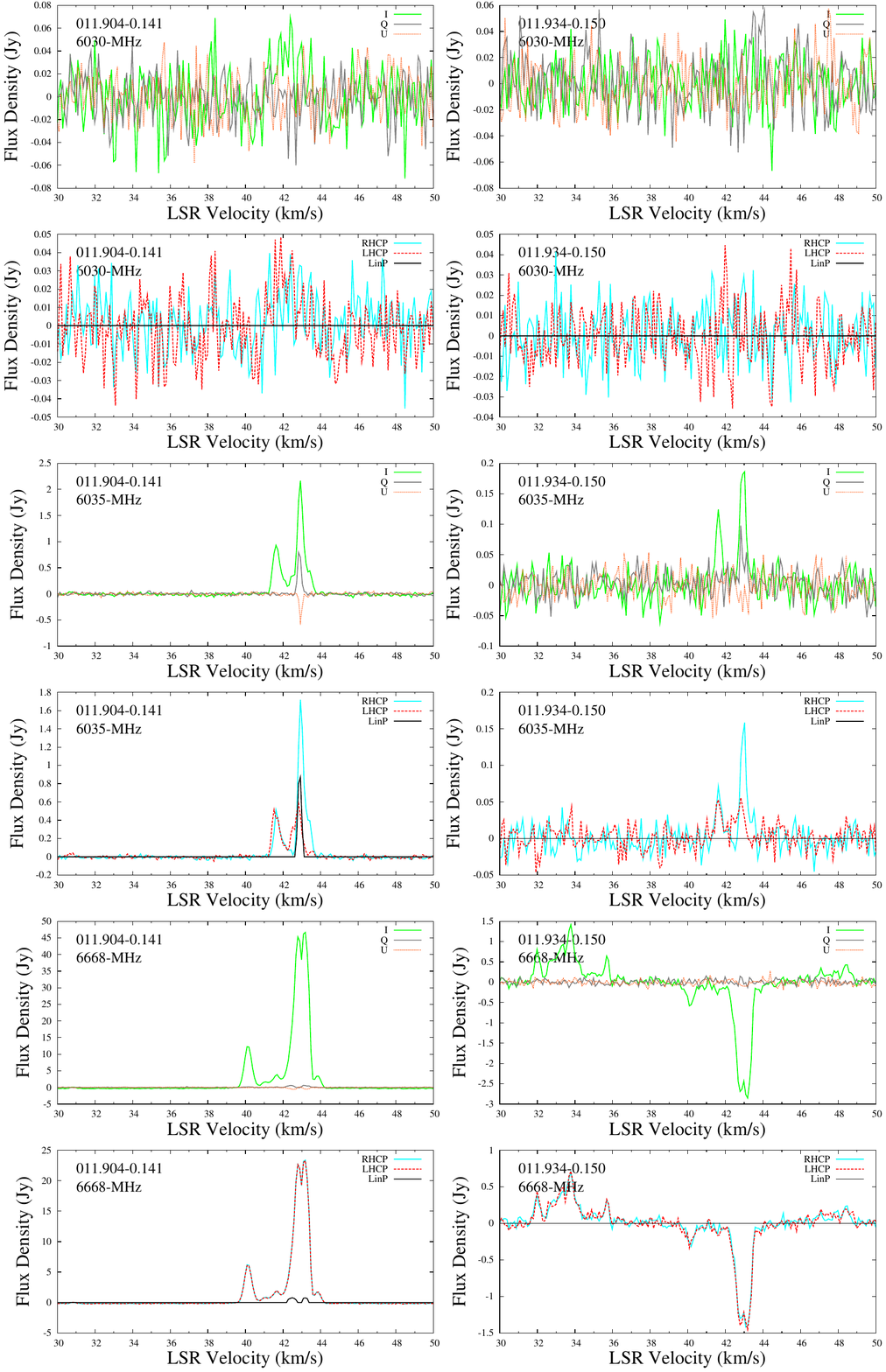}  
\caption{\small continued.}
\label{spectra}
\end{center}
\end{figure*}

\begin{figure*}
\begin{center}
\addtocounter{figure}{-1}
\renewcommand{\baselinestretch}{1.1}
\includegraphics[width=17cm]{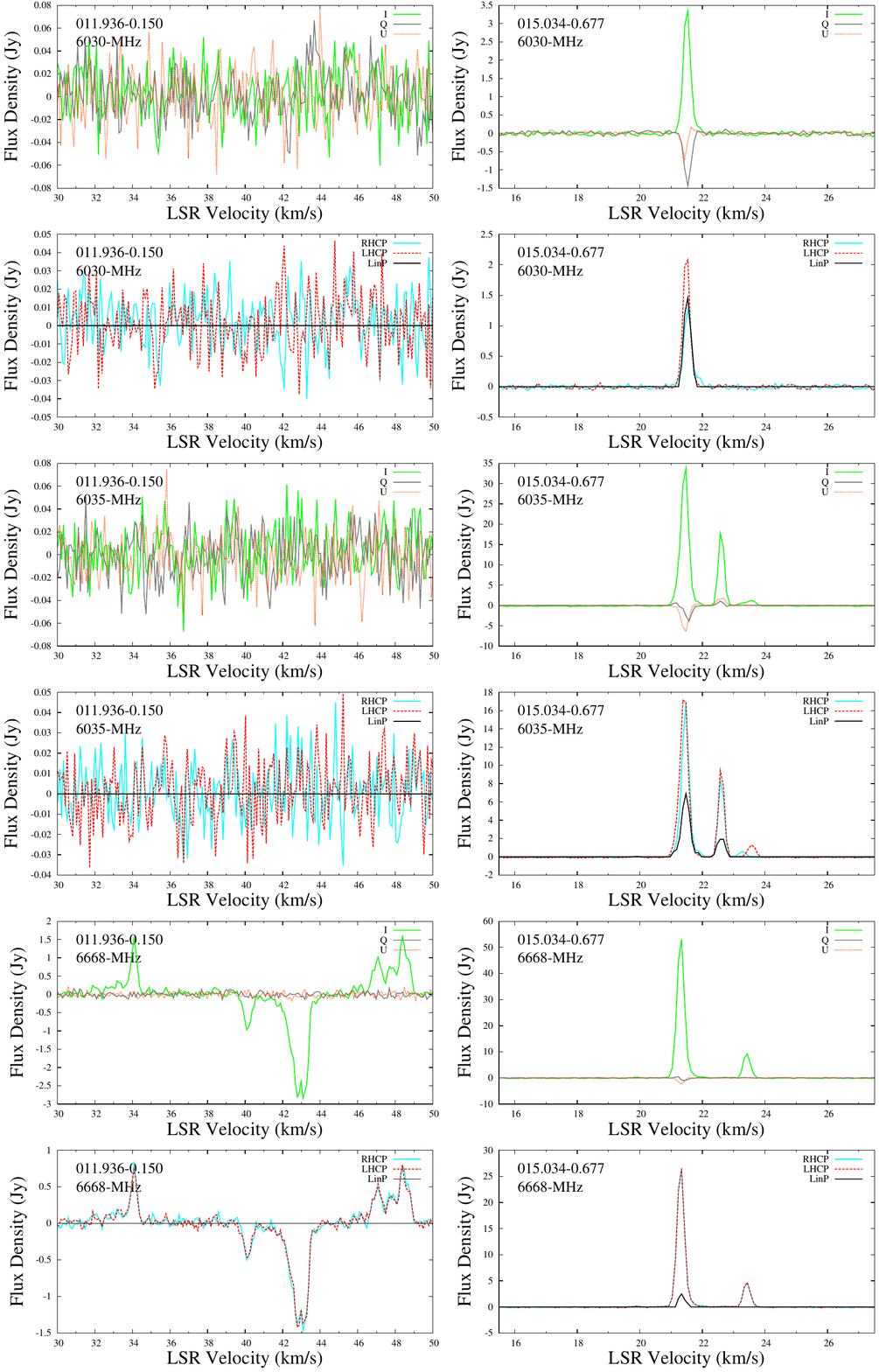}  
\caption{\small continued.}
\label{spectra}
\end{center}
\end{figure*}

\begin{figure*}
\begin{center}
\renewcommand{\baselinestretch}{1.1}
\includegraphics[width=17cm]{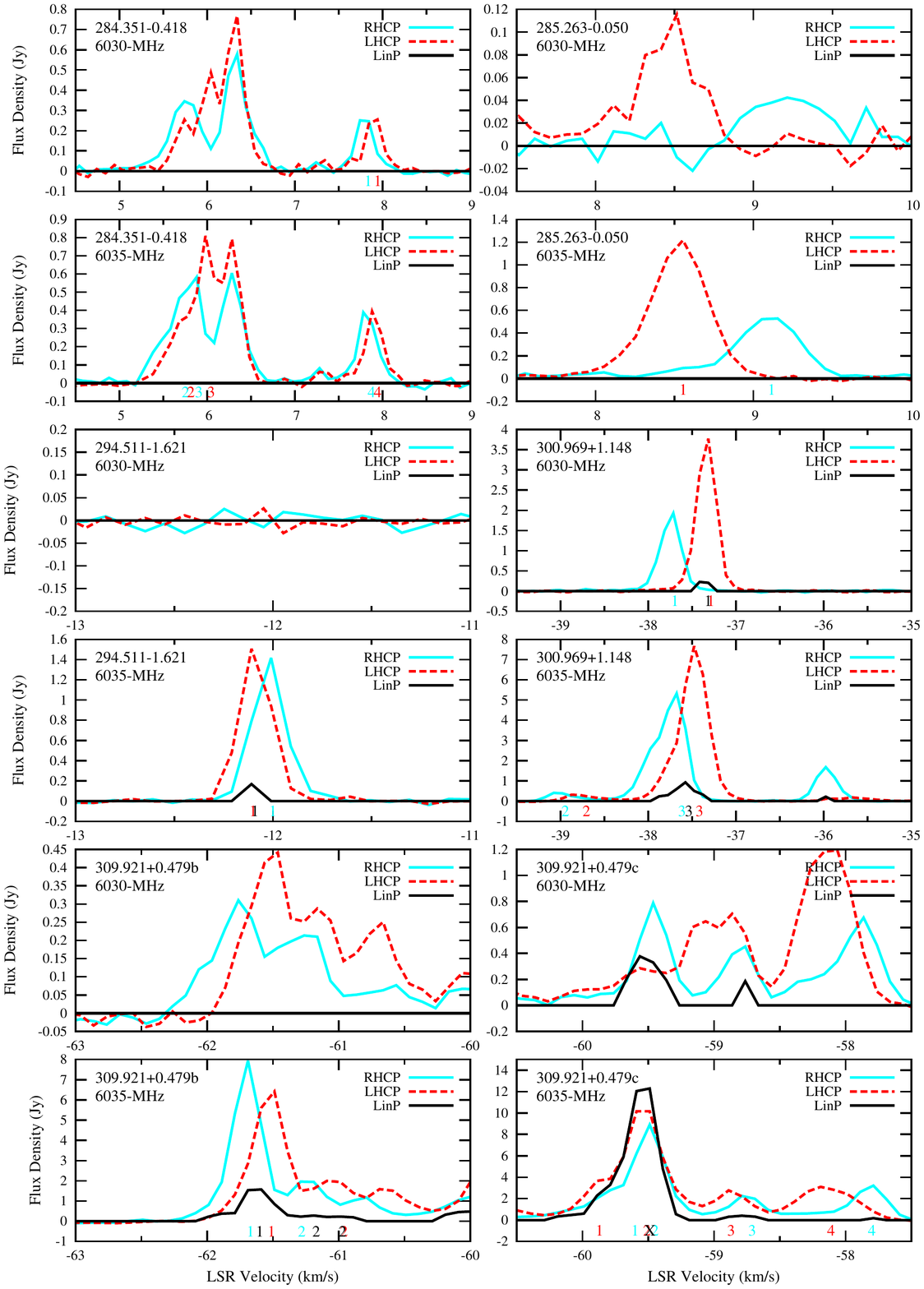}  
\caption{\small Spectra of Zeeman pattern features, showing (from top to bottom) 6030-MHz OH and 6035-MHz OH.}
\label{newspectra}
\end{center}
\end{figure*}

\begin{figure*}
\begin{center}
\addtocounter{figure}{-1}
\renewcommand{\baselinestretch}{1.1}
\includegraphics[width=17cm]{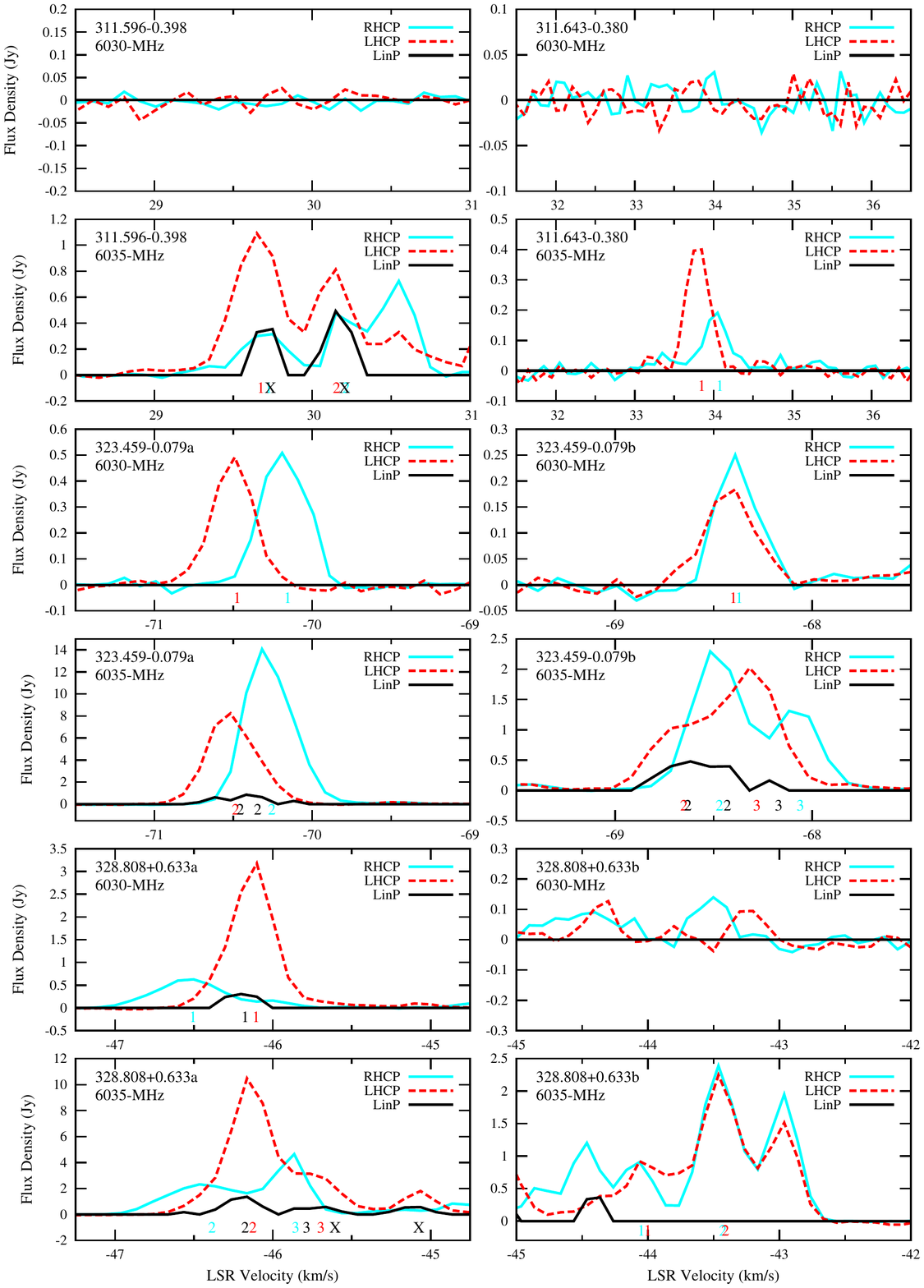}  
\caption{\small continued.}
\label{newspectra}
\end{center}
\end{figure*}

\begin{figure*}
\begin{center}
\addtocounter{figure}{-1}
\renewcommand{\baselinestretch}{1.1}
\includegraphics[width=17cm]{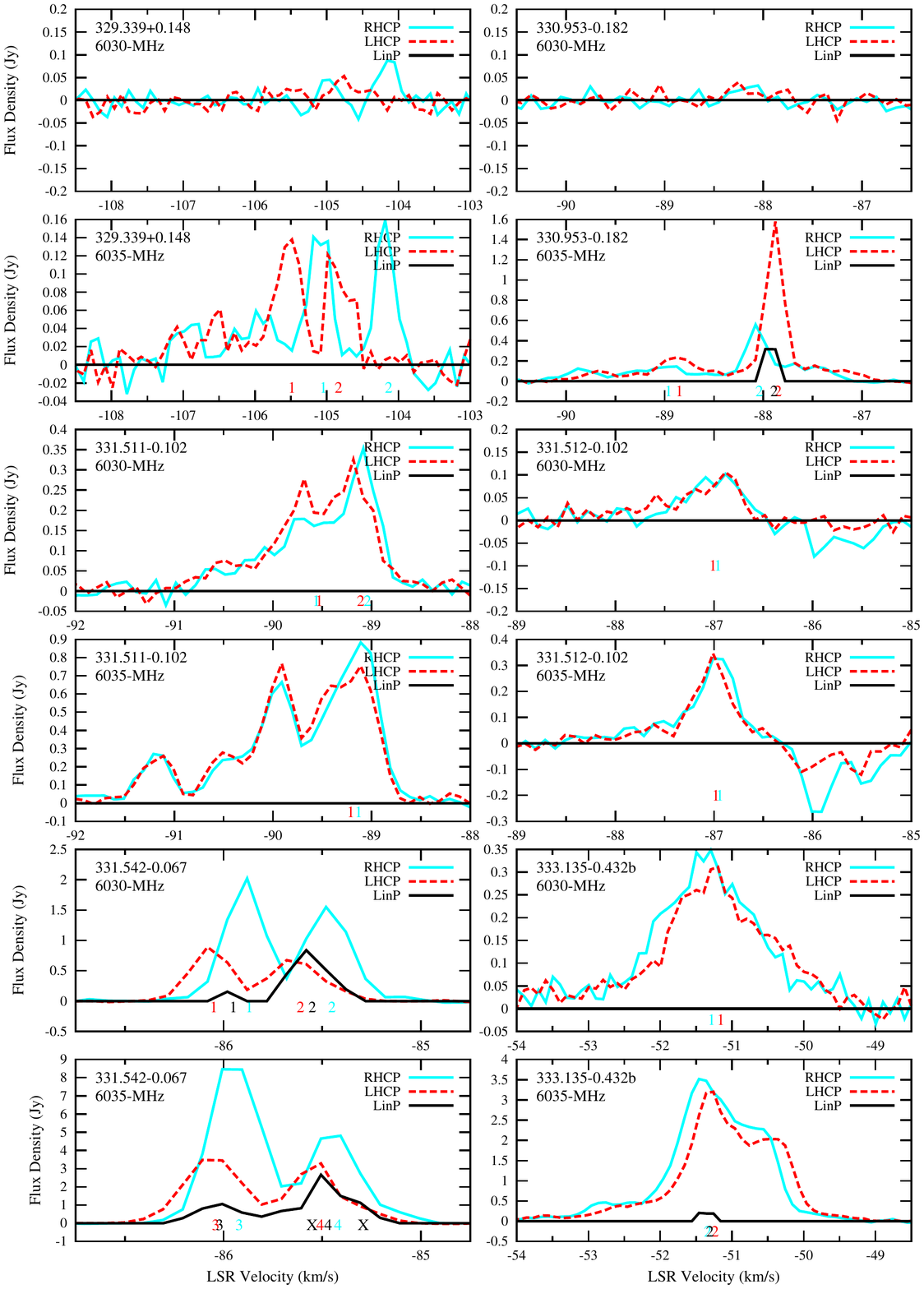}  
\caption{\small continued.}
\label{newspectra}
\end{center}
\end{figure*}

\begin{figure*}
\begin{center}
\addtocounter{figure}{-1}
\renewcommand{\baselinestretch}{1.1}
\includegraphics[width=17cm]{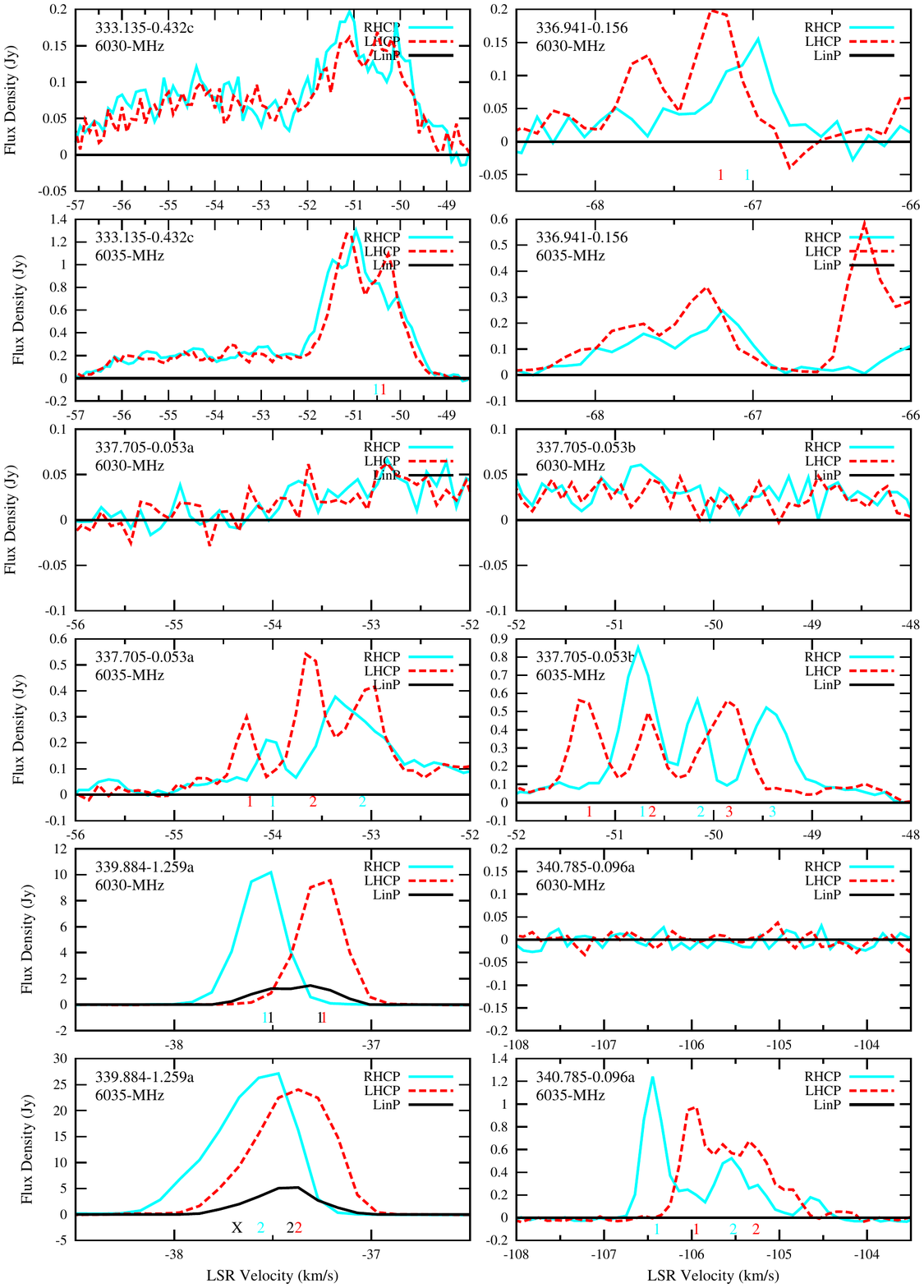}  
\caption{\small continued.}
\label{newspectra}
\end{center}
\end{figure*}

\begin{figure*}
\begin{center}
\addtocounter{figure}{-1}
\renewcommand{\baselinestretch}{1.1}
\includegraphics[width=17cm]{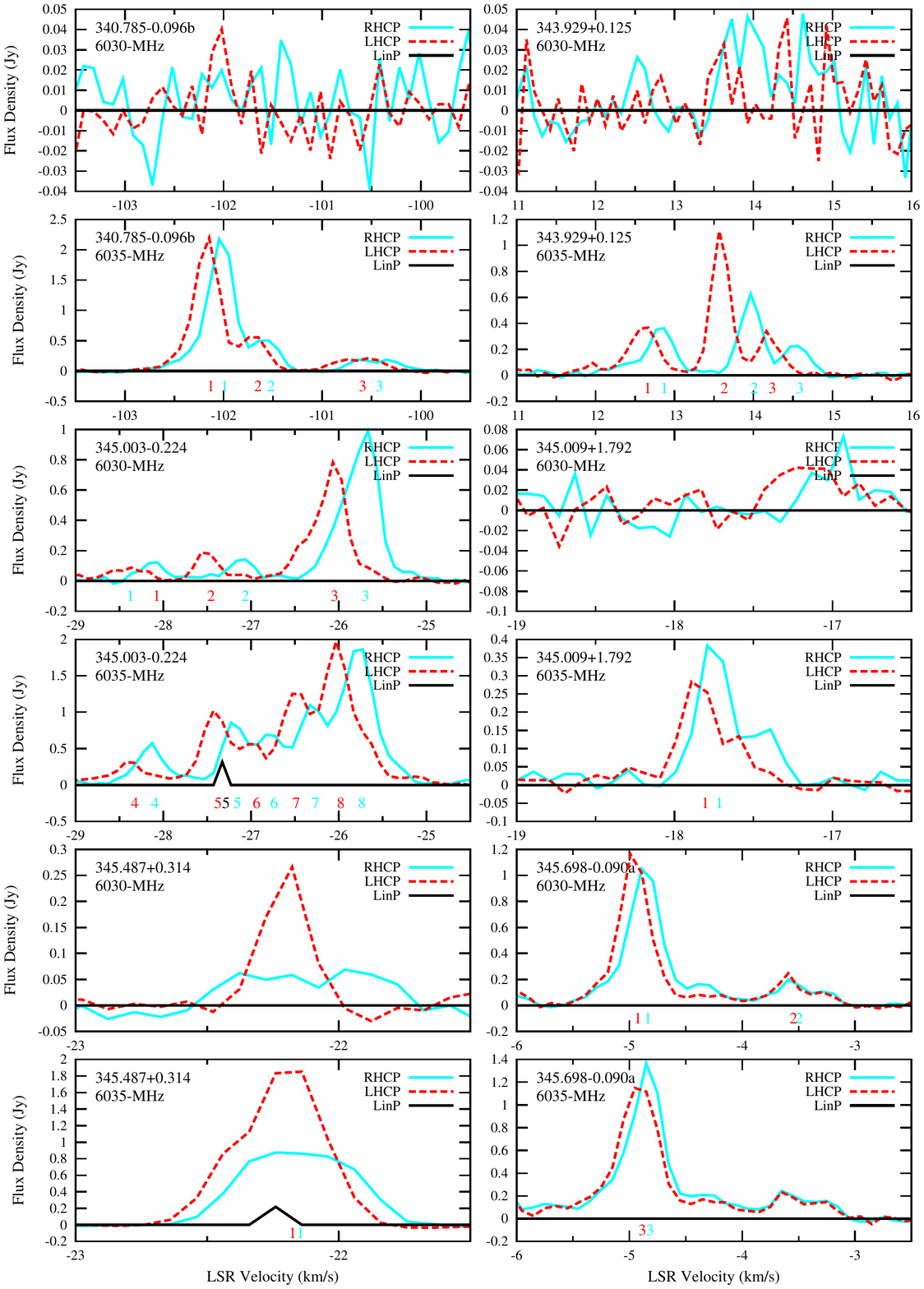}  
\caption{\small continued.}
\label{newspectra}
\end{center}
\end{figure*}

\begin{figure*}
\begin{center}
\addtocounter{figure}{-1}
\renewcommand{\baselinestretch}{1.1}
\includegraphics[width=17cm]{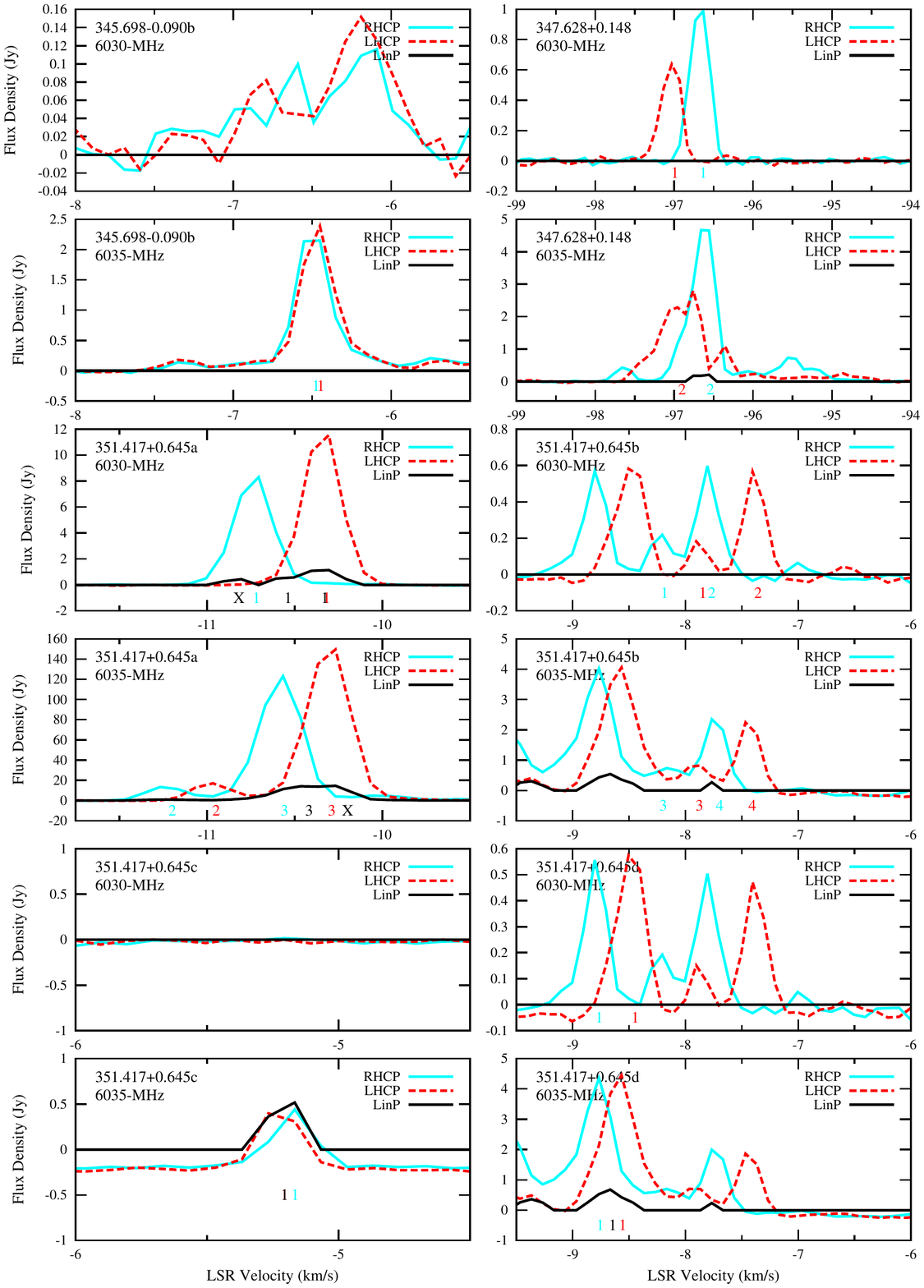}  
\caption{\small continued.}
\label{newspectra}
\end{center}
\end{figure*}

\begin{figure*}
\begin{center}
\addtocounter{figure}{-1}
\renewcommand{\baselinestretch}{1.1}
\includegraphics[width=17cm]{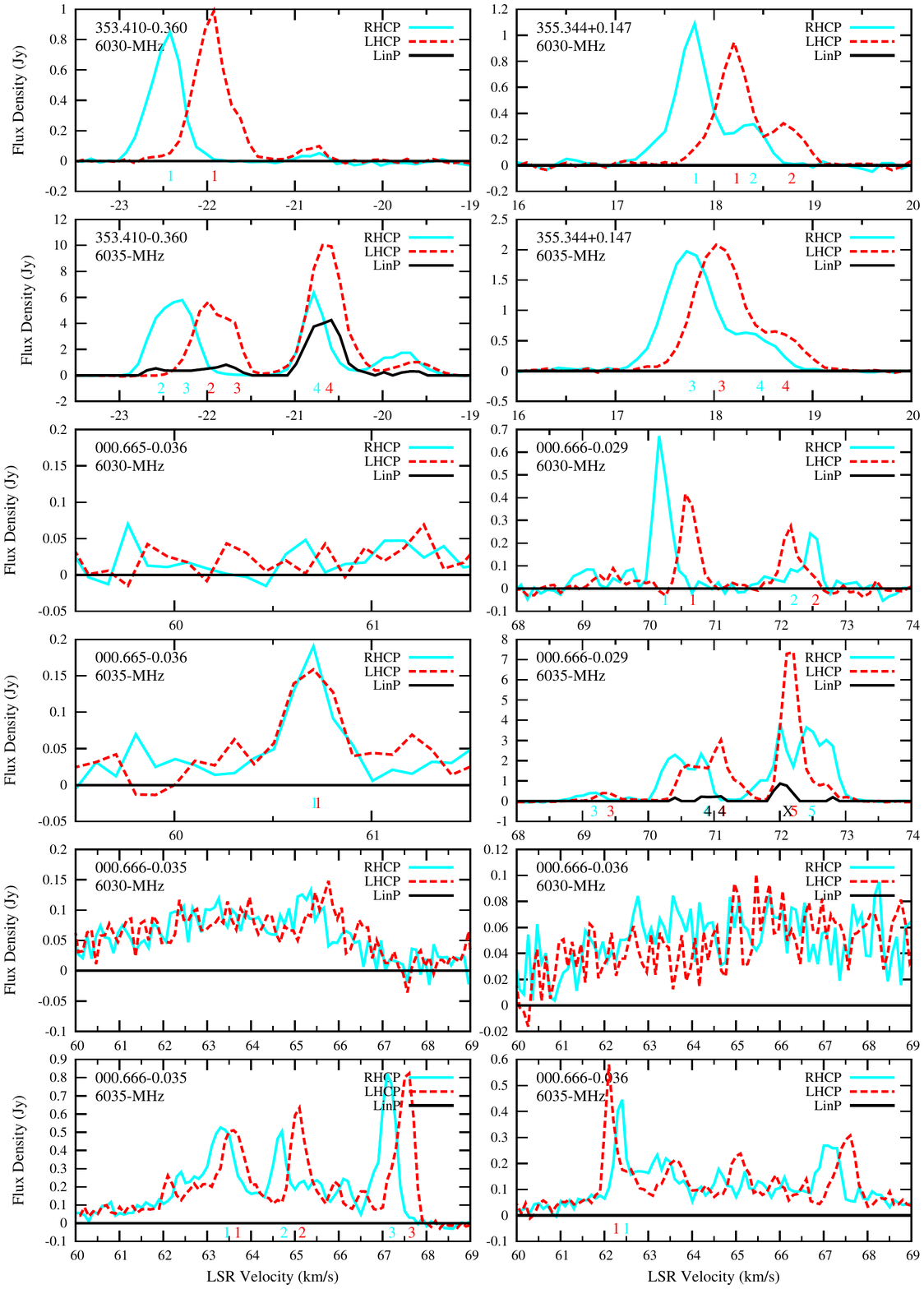}  
\caption{\small continued.}
\label{newspectra}
\end{center}
\end{figure*}

\begin{figure*}
\begin{center}
\addtocounter{figure}{-1}
\renewcommand{\baselinestretch}{1.1}
\includegraphics[width=17cm]{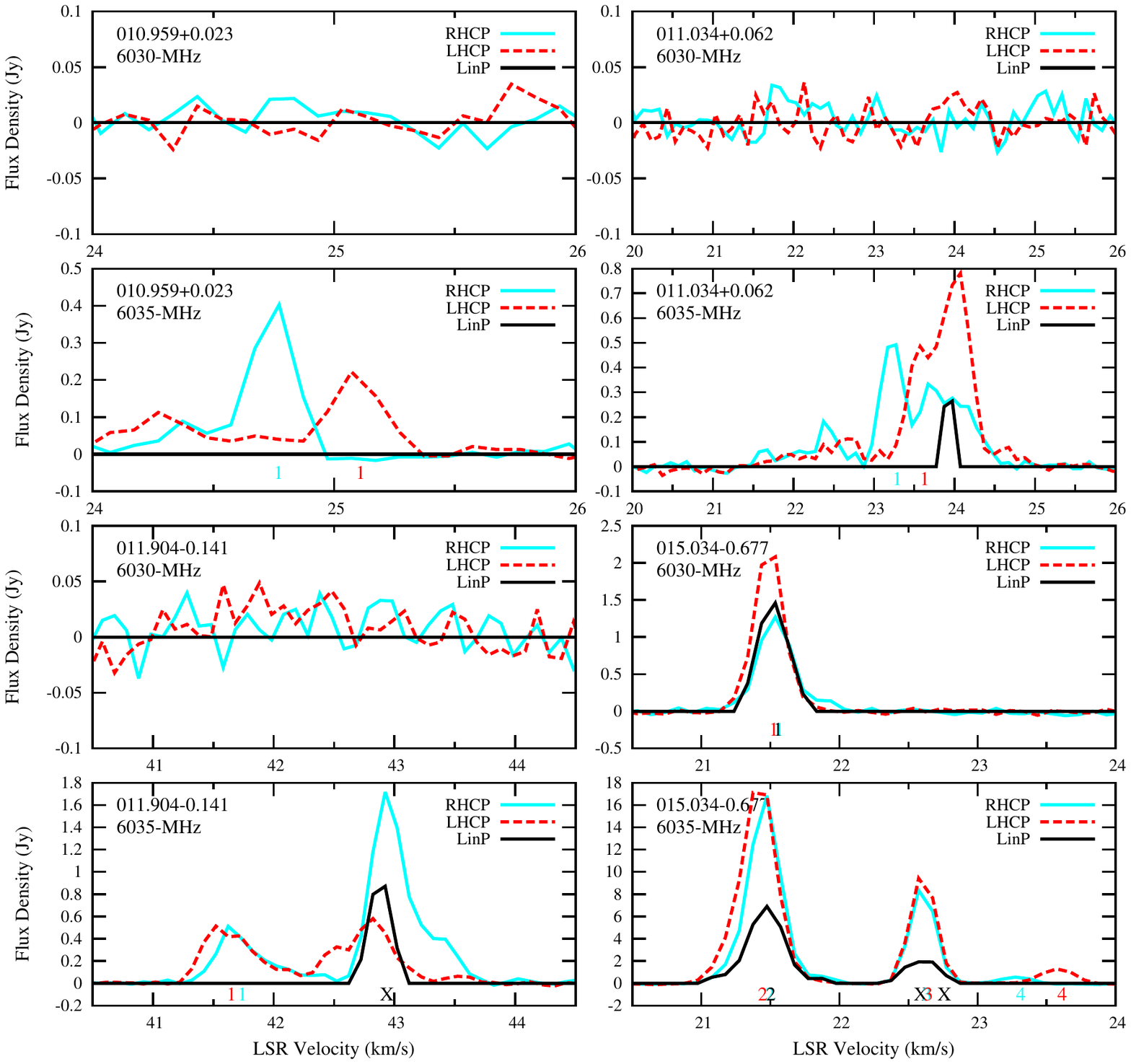}  
\caption{\small continued.}
\label{newspectra}
\end{center}
\end{figure*}

\label{lastpage}

\end{document}